%% file: main.tex
\documentclass[acmsmall,screen,nonacm]{acmart}

\input{utils/predefs}
\input{utils/style}

\title{A Nonlinear Abstract Domain for Bayesian Neural Networks}

\begin{document}

\author{Qiuhan Gu}
\orcid{0009-0002-3574-0538}
\affiliation{%
  \institution{University of Illinois Urbana-Champaign}
  \city{Urbana}
  \country{USA}
}
\email{qiuhang2@illinois.edu}

\author{Gagandeep Singh}
\orcid{0000-0002-9299-2961}
\affiliation{%
  \institution{University of Illinois Urbana-Champaign}
  \city{Urbana}
  \country{USA}
}
\email{ggnds@illinois.edu}

\input{tex/00_abstract}


\maketitle

\input{tex/01_intro}

\input{tex/02_pre}

\input{tex/03_overview}
\input{tex/04_concrete}

\input{tex/05_expohedron}

\input{tex/08_eval}
\input{tex/09_related}
\input{tex/10_conclusion}

\clearpage
\bibliographystyle{ACM-Reference-Format}
\bibliography{references}
\clearpage
\input{tex/appendix}

\end{document}

%% file: utils/predefs.tex
\usepackage{booktabs}   
\usepackage{subcaption} 

\usepackage{enumitem}
\usepackage{xcolor}
\usepackage{listings}
\newtheorem{theorem}{Theorem}
\newtheorem{definition}{Definition}[section]
\newtheorem{lemma}{Lemma}
\newtheorem{proposition}{Proposition}

\theoremstyle{remark}

\usepackage{wrapfig}
\usepackage{longtable}
\usepackage{pifont}
\newcommand{\xmark}{\ding{55}}

\newcommand{\lstref}[1]{\hyperref[lst:#1]{Listing~\ref*{lst:#1}}}
\newcommand{\figref}[1]{\hyperref[fig:#1]{Fig.~\ref*{fig:#1}}}
\newcommand{\figlabel}[1]{\label{fig:#1}}
\newcommand{\secref}[1]{\hyperref[sec:#1]{\S\ref*{sec:#1}}}
\newcommand{\seclabel}[1]{\label{sec:#1}}
\newcommand{\tabref}[1]{\hyperref[tab:#1]{Table~\ref*{tab:#1}}}
\newcommand{\tablabel}[1]{\label{tab:#1}}
\newcommand{\ruleref}[1]{\hyperref[rule:#1]{Rule~\ref*{rule:#1}}}

\newcommand{\defref}[1]{\hyperref[def:#1]{Definition~\ref*{def:#1}}}
\newcommand{\deflabel}[1]{\label{def:#1}}
\newcommand{\lemmaref}[1]{\hyperref[lemma:#1]{Lemma~\ref*{lemma:#1}}}
\newcommand{\lemmalabel}[1]{\label{lemma:#1}}
\newcommand{\theoremref}[1]{\hyperref[theorem:#1]{Theorem~\ref*{theorem:#1}}}
\newcommand{\theoremlabel}[1]{\label{theorem:#1}}
\newcommand{\eqnref}[1]{\hyperref[eq:#1]{Equation~\ref*{eq:#1}}}
\newcommand{\eqnlabel}[1]{\label{eq:#1}}
\newcommand{\algref}[1]{\hyperref[alg:#1]{Algorithm~\ref*{alg:#1}}}

\newcommand{\semref}[1]{\hyperref[rule:#1]{\textsc{#1}}}

\newcommand{\boxref}[1]{\hyperref[box:#1]{Note~\ref*{box:#1}}}

\newcommand{\appref}[1]{\hyperref[alg:#1]{Appendix~\ref*{alg:#1}}}
\newcommand{\applabel}[1]{\label{alg:#1}}

\usepackage{xspace}

\usepackage{algorithm}
\usepackage[noend]{algpseudocode}

\usepackage{tikz}
\usetikzlibrary{arrows.meta,positioning}

\usepackage{multirow}

\usepackage{pifont}

\newcommand{\cmark}{\ding{108}}      
\newcommand{\pcmark}{\ding{109}}     

%% file: utils/style.tex
\definecolor{diagramcolor}{rgb}{0.17,0.37,0.69}
\definecolor{keywords}{rgb}{0.05,0.05,0.9}
\definecolor{typewords}{rgb}{0,0.5,0}
\definecolor{greencomments}{rgb}{0,0.5,0}
\definecolor{turqusnumbers}{rgb}{0.17,0.57,0.69}
\definecolor{redstrings}{rgb}{0.5,0,0}
\definecolor{codegreen}{rgb}{0,0.6,0}
\definecolor{codegray}{rgb}{0.5,0.5,0.5}
\definecolor{codepurple}{rgb}{0.58,0,0.82}
\definecolor{backcolour}{RGB}{250,250,250}

\lstdefinelanguage{PPL}
{
    morekeywords={
        condition, gaussian, normal, N, sample, observe, Laplace
    },
    morekeywords=[2]{Bool, Int, Real},
    keywordstyle=\bfseries\color{keywords},
    keywordstyle=[2]{\bfseries\color{typewords}},
    sensitive=false,
    morecomment=[l][\color{greencomments}]{//},
    morestring=[b]",
    stringstyle=\color{redstrings},
    literate={~}{{$\sim$}}1
}

%% file: tex/00_abstract.tex
\begin{abstract}
Bayesian neural networks (BNNs) explicitly model epistemic uncertainty by leveraging probability distributions over network parameters, making them better suited for risk-aware decision-making. Yet soundly propagating uncertainty through the network and capturing the entire output distribution to support predictive uncertainty quantification remains a fundamental challenge. Existing approaches are either sampling-based, providing no formal guarantees, or propagate only selected statistical properties, such as expectations and entropy, limiting their applicability to general probabilistic queries. To address these challenges,  
we present \textbf{the first nonlinear abstract domain for soundly approximating complete probability distributions of BNNs}. 
The abstract domain builds upon a novel marginal collecting concrete semantics that soundly abstracts the classical joint-measure semantics. 
It combines support partitioning with exponential-affine bounds to represent probability measures, capturing both discrete and continuous components of mixed distributions as well as nonlinear density shapes over bounded and unbounded supports. We develop sound abstract transformers that can efficiently handle high-dimensional affine transformations and nonlinear ReLU activations, while remaining highly parallelizable. We further extend the framework beyond BNNs to support bounded-loop probabilistic programs with Bayesian inference to demonstrate its generality. We implement the resulting abstract interpreter as a GPU-accelerated analyzer. Experimental results show that the framework significantly improves precision over existing probabilistic reasoning approaches while soundly capturing the complete output distribution, demonstrating strong efficiency and scalability.
\end{abstract}

%% file: tex/01_intro.tex
\section{Introduction}
Deep neural networks (DNNs) are increasingly deployed in safety-critical applications, such as autonomous driving and medical diagnosis~\cite{llm1, llm2, llm3, llm4}. However, conventional deterministic DNNs rely on fixed network parameters and produce point predictions, therefore lacking principled quantification of epistemic uncertainty arising from limited training data~\cite{epistemic, quantification2,10.1007/978-3-031-71162-6_17,xu2025supportneedcertifiedvae,uncertaintydeepneural}, making it difficult to assess the reliability of their outputs~\cite{unreliable1, unreliable2, unreliable3}. 
Bayesian neural networks (BNNs)~\cite{BNN1} address this limitation by modeling posterior distributions over network parameters, thereby capturing and propagating parameter uncertainty through the network to the predictions~\cite{epistemic}.
One of the central goals of BNN analysis is therefore to quantify the resulting predictive uncertainty by reasoning about the output probability distributions, which enables downstream applications such as risk assessment, uncertainty-aware decision-making, and active learning~\cite{uncertaintyapplication3,uncertaintyapplication2,uncertaintyapplication1}.
However, the probabilistic nature of BNNs makes this task substantially more challenging~\cite{monograph1,ugo,Corina}. 
Traditional symbolic approaches, such as PSI~\cite{psi}, aim to compute exact and simplified analytical forms, but they are not applicable to BNNs, as intermediate neuron distributions often lack closed-form representations. Moreover, repeated compositions of high-dimensional affine transformations and nonlinear activations quickly make exact symbolic reasoning intractable. 
To address this issue, many researchers have been working on approximate probabilistic reasoning techniques, including both sampling-based methods~\cite{sample1, sample2, sample3, sample4} and methods that compute only partial distributional information~\cite{makesureyouareunsure, expectation1, expectation2, entropy}. Sampling-based methods approximate the output distribution by drawing a finite number of weight samples, reducing the BNN to a deterministic neural network, to which existing DNN reasoning techniques can be applied. However, the resulting guarantees depend heavily on the sampled realizations and fail to capture the entire output distribution, especially rare events that occur with low probability. Methods that track partial distributional information, on the other hand, avoid explicitly constructing the output distribution by propagating only selected quantities of interest, such as expectations, variance, and entropy. As a result, the resulting analysis is tailored to specific goals and cannot be generalized to answer arbitrary probabilistic queries.
Therefore, without a sufficiently expressive and sound representation, the rich probabilistic information provided by BNNs cannot be fully leveraged with formal guarantees. 
Such limitations can have devastating consequences in real-world applications. 
For example, the work~\cite{cancer} uses BNNs to guide cancer drug dose selection by estimating the probability that predicted drug toxicity exceeds a prescribed threshold, where a slight underestimation of the actual probability could lead to false negatives and potentially life-threatening treatment decisions. 
Consequently, soundly capturing the complete output distribution remains a fundamental challenge. Taking a step further, an even greater challenge is performing fine-grained analysis of predictive uncertainty at intermediate neurons to understand how epistemic uncertainty propagates through the network and identify where it becomes amplified, which also provides deeper insights into model interpretability~\cite{uncertainty}.


Abstract interpretation~\cite{AI, AI2,roberto2,roberto1} can enable compositional, sound probabilistic reasoning by constructing tractable abstractions while preserving soundness guarantees. Existing popular abstract domains, such as the Interval domain, the Polyhedra domain~\cite{fastpoly}, the DeepPoly~\cite{deeppoly} domain, etc., have been successful in analyzing deterministic computations, but cannot reason about probabilistic computations. Despite five decades of advances in abstract interpretation, probabilistic abstract interpretation remains largely unexplored, with no existing abstract domain capable of efficiently reasoning about BNNs~\cite{probAI,probAI2,probAI3}. We identify five main challenges.

\textit{Challenge 1: Establishing suitable probabilistic concrete semantics.}
Existing abstract domains are primarily designed to represent the value domain of deterministic programs, whereas probabilistic reasoning requires tracking both the values of random variables and their associated probability measures~\cite{measure, probabilitymeasure}. This necessitates a probabilistic concrete domain. Conventional denotational semantics for probabilistic programs are based on joint probability measures that capture dependencies among all program variables~\cite{jointmeasuresemantics1, jointmeasuresemantics2,semantics1, semantics2, semantics3, semantics4, semantics5,unnormalizesemantics1,unnormalizesemantics2}, introducing substantial complexity when directly used for abstract interpretation in high-dimensional settings and also making fine-grained uncertainty analysis at individual neurons difficult. Therefore, a key challenge is to establish a concrete semantics that soundly captures probabilistic behaviors while remaining suitable for abstraction and neuron-level analysis.

\textit{Challenge 2: Expressively abstracting probability measures.}
Most existing abstract domains rely on linear representations, such as intervals or polyhedral constraints. However, probability measures arising in BNNs can exhibit complex representations, including continuous components with highly nonlinear density shapes, such as exponential decay, as well as discrete probability masses introduced by nonlinear activation functions such as ReLU. Purely linear representations often fail to capture these structures precisely, resulting in overly conservative approximations. Developing expressive abstract domains for soundly and precisely representing mixed probability distributions with nonlinear densities remains a challenge.

\textit{Challenge 3: Handling Unbounded Tails.}
BNNs with Gaussian weight distributions naturally induce probability distributions with unbounded support.
Soundly representing and propagating tail regions is challenging. Some approximate approaches~\cite{AURA} naively truncate tails to maintain computational feasibility, but such truncation may discard non-negligible probability mass and thus violate soundness guarantees. On the other hand, explicitly tracking unbounded tails often introduces significant computational overhead.

\textit{Challenge 4: Supporting compositional nonlinear probabilistic operations.}
BNNs involve operators that are fundamentally more complex than their deterministic counterparts, as even basic operations such as the addition or multiplication of random variables require reasoning about convolutions and integrations over probability measures. More advanced operations, such as affine transformations, further involve marginalization integrals that are often analytically intractable.

\textit{Challenge 5: Scaling to high-dimensional settings.}
The above challenges become even more significant in high-dimensional settings of Bayesian neural networks where the repeated composition of affine and nonlinear activations leads to increasingly complex distributions. Meanwhile, interactions among random variables induce nontrivial dependencies that must be handled carefully to preserve soundness throughout the analysis. As the network size grows, the dimensionality of the underlying probability space increases rapidly, resulting in severe scalability bottlenecks.

\textbf{This work:} To address these challenges, we first establish a novel marginal collecting semantics for probabilistic abstract interpretation, which associates each program variable with a marginal Markov kernel. We prove that this semantics is a sound abstraction of the classical joint-measure semantics. Building on this foundation, we propose \textbf{the first nonlinear} abstract domain for sound reasoning about Bayesian neural networks. Our abstract domain represents probability measures using partitions equipped with exponential-affine bounds for continuous density components and constant bounds as fallback for discrete point masses, enabling precise abstractions of complex probability distributions, including those with unbounded support. We develop sound abstract transformers for common BNN operations, including high-dimensional affine transformations and nonlinear ReLU activations, enabling compositional reasoning about uncertainty not only at the final output but also at individual intermediate neurons. The resulting analysis is highly parallelizable, enabling efficient GPU-based implementations. We further demonstrate the potential of the framework by extending it to a more general probabilistic program setting with a broader class of continuous distributions, bounded loops, and Bayesian inference, where we provide a general abstraction function and a sound Bayesian inference abstract transformer. 
By combining expressive nonlinear abstractions with formal soundness guarantees, our framework makes a step forward in developing principled and scalable probabilistic abstract interpretation. 

\textbf{Contributions:} Our work makes the following contributions:
\begin{enumerate}[leftmargin=1.5em, itemsep=0pt, topsep=0pt, parsep=0pt]
    \item We introduce a novel marginal collecting semantics that provides a tractable formal foundation for sound abstract interpretation of Bayesian neural networks.
    \item We propose the first nonlinear abstract domain with sound abstract transformers for principled fine-grained reasoning over Bayesian neural networks.
    \item We demonstrate the generality of the framework by extending it beyond BNNs to probabilistic programs supporting a broader class of distributions, Bayesian inference, and bounded loops.
    \item  We implement the abstract interpreter on GPUs and evaluate it on a broad collection of challenging Bayesian neural networks and probabilistic programs. Experimental results show that our approach captures the complete distribution while achieving significantly higher precision compared to the existing methods, demonstrating both efficiency and scalability.
\end{enumerate}

%% file: tex/02_pre.tex
\section{Preliminary}

\textit{Notation.} 
For clarity, we use bold symbols (e.g., $\mathbf{x}, \mathbf{y}$) to denote vectors, and $[n]:=\{1,2,\ldots,n\}$. 

\textit{Measure Theory.}
Given a set $S$, a measure $\mu$ assigns a non-negative value $\mu(A)\in[0,\infty]$ to a collection of subsets $A\subseteq S$, representing their size or mass. Measures satisfy $\mu(\emptyset)=0$ and are countably additive over disjoint sets. A probability measure is a measure satisfying $\mu(S)=1$. The Lebesgue measure $\lambda$ on $\mathbb{R}$ assigns $\lambda([a,b])=b-a$ for any $a\leq b$.
%
The collection of subsets on which a measure is defined is formalized as a $\sigma$-algebra.
A $\sigma$-algebra $\Sigma\subseteq2^S$ is a collection of subsets of $S$ that is closed under complement and countable unions. 
The pair $(S,\Sigma)$ is called a measurable space. If $S$ is a topological space, such as $\mathbb{R}$, the smallest $\sigma$-algebra containing all open subsets of $S$ is called the Borel $\sigma$-algebra, denoted by $\mathcal{B}(S)$.
%
Given two measurable spaces $(X,\mathcal A)$ and $(Y,\mathcal B)$, a Markov kernel $K:X\times \mathcal B \rightarrow[0,1]$ satisfies:
(i) for every $x\in X$, $K(x,\cdot)$ is a probability measure on $(Y,\mathcal B)$;
and (ii) $K(\cdot,B)$ is $\mathcal A$-measurable for every $B\in\mathcal B$.

\textit{Bayesian Neural Network.}
BNNs are probabilistic functions
$f_{\mathbf{w}}:\mathbb{R}^m\rightarrow\mathbb{R}^n$,
where the parameter vector $\mathbf{w}$ is sampled from a posterior distribution
$q(\mathbf{w})$ learned during training. In this work, we follow the standard mean-field Gaussian assumption~\cite{independentweights}, where
\(
q(\mathbf{w})
=
\prod_{i=1}^{n_w}
\mathcal{N}(w_i;\mu_i,\sigma_i^2).
\)
$n_w$ denotes the number of parameters. We consider pre-trained feed-forward BNNs with ReLU activations as our primary focus, while our methods can be extended to other architectures, such as residual architectures.
A feed-forward BNN is given by a composition of layers.
Each layer is either
(i) an affine transformation, which can be represented through neuron-wise assignments:
\(
x_i \leftarrow b_i+\sum_j w_{ij}x_j,
\)
where 
\(
w_{ij}\sim\mathcal N(\mu_{ij},\sigma_{ij}^2),
b_i\sim\mathcal N(\mu_i,\sigma_i^2);
\)
or (ii) a ReLU activation:
\(
x_i\leftarrow\max(0,x_j).
\)
We consider BNNs for regression tasks, where a fixed input $\mathbf x\in\mathbb R^m$ induces a random output $\mathbf y=f_{\mathbf w}(\mathbf x)$ through $\mathbf w\sim q(\mathbf w)$. 



\textit{Abstract Interpretation.}
Abstract interpretation computes an over-approximation of system behaviors (e.g., programs or neural networks) over potentially infinite executions~\cite{AI2}. It operates over two domains: a concrete domain and an abstract domain.
A concrete domain $(\mathcal{C}, \sqsubseteq_{\mathcal C})$ consists of a set $\mathcal{C}$ of concrete elements partially ordered by the partial order $\sqsubseteq_{\mathcal C}$ and a set of concrete transformers that model the effect of applying different program statements on concrete elements.
An abstract domain $(\mathcal{A}, \sqsubseteq_\mathcal{A})$ consists of a set $\mathcal{A}$ of abstract elements $a \in \mathcal{A}$ partially ordered by the relation $\sqsubseteq_\mathcal{A}$ and a set of abstract transformers $T^\sharp$ that model the effect of program statements on abstract elements.
The concrete and the abstract domain are connected by two functions: an abstraction function $\alpha: \mathcal{C} \rightarrow \mathcal{A}$ and a concretization function $\gamma: \mathcal{A} \rightarrow \mathcal{C}$. 
Based on these two functions, 
a given abstract transformer $T^\sharp$  is sound w.r.t its concrete transformer $T$ iff for all abstract elements $a\in\mathcal{A}$, $T(\gamma(a)) \sqsubseteq_\mathcal{C} \gamma(T^\sharp(a))$.

%% file: tex/03_overview.tex
\section{Overview}
\seclabel{overview}
In this section, we illustrate our approach using a small running example. Full formal details are provided in later sections.

\begin{figure}[t]
\centering

\begin{tikzpicture}[
scale=0.8,
transform shape,
    neuron/.style={circle,draw,minimum size=5mm},
    input/.style={circle,draw,purple,minimum size=5mm},
    relu/.style={rectangle,draw,rounded corners,minimum width=10mm,minimum height=6mm},
    output/.style={circle,draw,red,minimum size=5mm},
    >=stealth
]

\node[input] (z1) at (0,2) {$z_1$};
\node[input] (z2) at (0,0.5) {$z_2$};

\node[neuron] (x1) at (3,2) {$x_1$};
\node[neuron] (x2) at (3,0.5) {$x_2$};

\node at (3,2.7) {$b_1$};
\node at (3,-0.2) {$b_2$};

\node[neuron] (x3) at (5,2) {$x_3$};
\node[neuron] (x4) at (5,0.5) {$x_4$};

\node[neuron] (x5) at (8,2) {$x_5$};
\node[neuron] (x6) at (8,0.5) {$x_6$};

\node at (8,2.7) {$b_3$};
\node at (8,-0.2) {$b_4$};

\node[neuron] (x7) at (10,2) {$x_7$};
\node[neuron] (x8) at (10,0.5) {$x_8$};

\node[output] (y) at (12.5,1.24) {$y$};

\node at (12.5,0.54) {$b_y$};

\draw[->] (z1) -- node[above] {$W_{11}$} (x1);

\draw[->] (z1) -- node[pos=0.1,right] {$W_{12}$} (x2);

\draw[->] (z2) -- node[pos=0.1, right] {$W_{21}$} (x1);

\draw[->] (z2) -- node[below] {$W_{22}$} (x2);

\draw[->] (x1) -- node[above] {ReLU} (x3);
\draw[->] (x2) -- node[above] {ReLU} (x4);


\draw[->] (x3) -- node[above] {$W_{35}$} (x5);
\draw[->] (x3) -- node[pos=0.1,right] {$W_{36}$} (x6);

\draw[->] (x4) -- node[pos=0.1, right] {$W_{45}$}  (x5);
\draw[->] (x4) -- node[below] {$W_{46}$} (x6);


\draw[->] (x5) -- node[above] {ReLU} (x7);
\draw[->] (x6) -- node[above] {ReLU} (x8);


\draw[->] (x7) -- node[above] {$W_{7}$} (y);

\draw[->] (x8) -- node[below] {$W_{8}$} (y);


\node at (0,3.2) {Input Layer};
\node at (6,3.2) {Hidden Layers};
\node at (12.5,3.2) {Output Layer};

\end{tikzpicture}

\caption{
Example BNN with ReLU activations.
All weights \(W_{ij}\) and biases \(b_j\) are random variables drawn from posterior distributions.
}
\figlabel{fig:bnn}
\vspace{-10pt}
\end{figure}
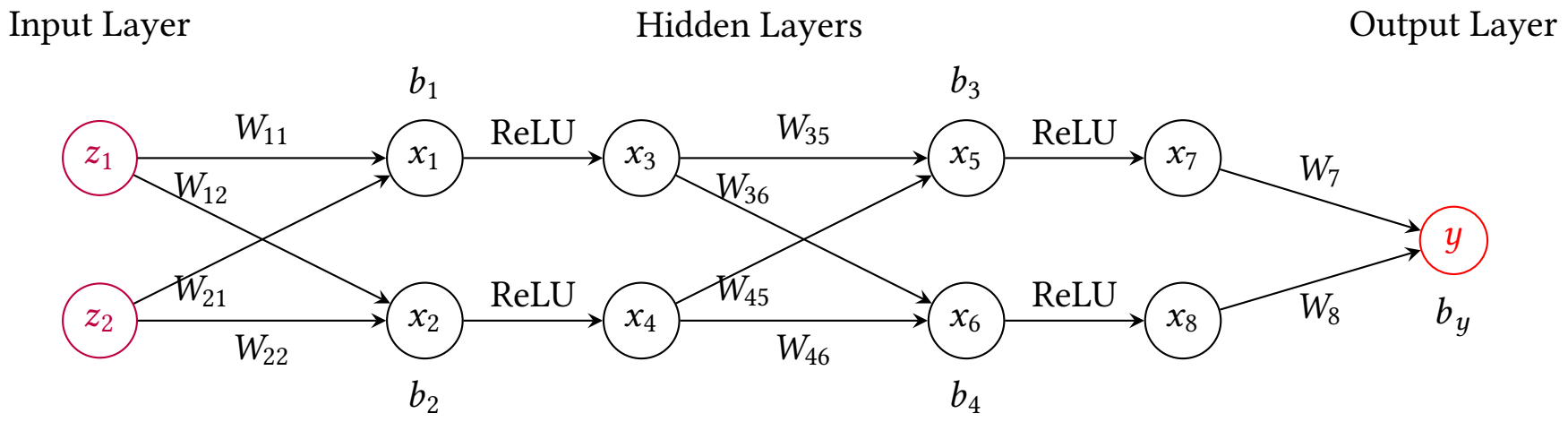

\textit{Running example on a pre-trained feedforward BNN with ReLU activations.} 
We consider the fully connected toy BNN shown in~\figref{fig:bnn}, which consists of two hidden layers, each comprising an affine transformation followed by a ReLU activation. For our analysis, we treat these operations as separate layers. All weights and biases are independently drawn from learned Gaussian posterior distributions. For our example, 
\(W_{11}, W_{22},W_{35}, W_{46}, W_7 \sim \mathcal N(1,1),
W_{12}, W_{21},W_{36},  W_{45},W_8 \sim \mathcal N(-1,1),
\)
and
\(
b_1 \sim \mathcal N(1,1), 
b_2 \sim \mathcal N(-1,1),
b_3,b_4,b_y \sim \mathcal N(0,1)
\). The network performs a regression task, taking two deterministic inputs
\(
z_1,\ z_2
\) and producing a single output $y$, 

\textit{Specification.}
We consider evaluating the network on the concrete input
\(
z_1=1,z_2=-1.
\)
Our objective is to conduct fine-grained uncertainty analysis at the level of individual neurons and ultimately compute a sound approximation of the entire output distribution, capturing its complete probability mass. 
This enables further predictive uncertainty quantification through probabilistic queries such as $\Pr(Y>\tau)$, where $Y$ is the output random variable and $\tau$ is a given threshold.

\textit{Marginal Collecting Semantics.}
To avoid explicitly tracking complex dependencies among variables while enabling fine-grained analysis of individual neurons, we define a novel marginal collecting semantics and a corresponding concrete domain tailored to our setting, 
associating each program variable with a marginal Markov kernel while soundly over-approximating dependencies. Specifically, our marginal collecting semantics is a sound abstraction of the classical joint-measure semantics, ensuring that every joint probability measure induced by the original semantics at each program point is included by our semantics.
The details are provided in~\secref{sec:semantics}.
Returning to the running example, the concrete inputs \(z_1=1,\ z_2=-1\) are first propagated through the first Bayesian affine layer. Since a linear combination of independent Gaussian random variables remains Gaussian, the outputs of this layer also follow Gaussian distributions, i.e., \(
x_1=W_{11}-W_{21}+b_1\sim\mathcal N(3,3),
x_2=W_{12}-W_{22}+b_2\sim\mathcal N(-3,3).
\) 
These Gaussian distributions serve as the starting point for our analysis.
Under our marginal collecting semantics, the distributions of $x_1$ and $x_2$ are represented by marginal Markov kernels $K_1$ and $K_2$, respectively, where
\(
K_1(\emptyset, dx) = p_1(x)dx,
K_2(\emptyset, dx) = p_2(x)dx
\) with
$p_1$ and $p_2$ representing the density function. After the ReLU assignments, the output random variable $x_3$ is obtained by applying the function
\(
x_3 := \max(0,x_1)
\)
to the probability distribution of $x_1$. 
The resulting marginal kernel $K_3$ is obtained by pushing forward the probability measure represented by the input kernel $K_1$ through the ReLU function. Formally, for any $A\in\mathcal B(\mathbb R)$, we have
$
K_3(\emptyset,A)
=
K_1(\emptyset,(-\infty,0])\,\delta_0(A)
+
K_1(\emptyset,A\cap(0,\infty)).
$
That is, the probability mass over the non-positive region is accumulated at zero, while the probability measure over the positive region remains unchanged. $K_4$ is constructed in the same way for $x_4:= \max(0,x_2)$.
The need to account for approximation of dependencies becomes evident in the next affine layer. Specifically, consider
\(
x_5 = W_{35}x_3 + W_{45}x_4 + b_3
\).
We model the affine output as a conditional Gaussian model followed by marginalization. We first consider fixing values of \(x_3\) and \(x_4\). 
The only remaining randomness therefore comes from the Gaussian weights and bias. After the linear combination, we obtain
\(
x_5 \mid (x_3,x_4)
\sim
\mathcal N\!\left(
x_3-x_4,\;
x_3^2+x_4^2+1
\right).
\)
This intermediate conditional Gaussian distribution is represented by a conditional Markov kernel $K_{\mathrm{cond}}$, given by
$
K_{\mathrm{cond}}((x_3,x_4),dx_5)
=
\frac{1}{\sqrt{2\pi(x_3^2+x_4^2+1)}}
\exp\ (
-\frac{(x_5-(x_3-x_4))^2}
{2(x_3^2+x_4^2+1)}
)dx_5.
$
By marginalizing the conditional kernel over the joint probability measure of $(x_3,x_4)$, we obtain the marginal Markov kernel associated with $x_5$:
$
K_5(\emptyset,dx_5)
=
\int_{\mathbb R^2}
K_{\mathrm{cond}}((x_3,x_4),dx_5)
\,\mu_{34}(d(x_3,x_4)),
$
where $\mu_{34}$ denotes the joint probability measure of $(x_3,x_4)$. Although the joint measure $\mu_{34}$ cannot be uniquely recovered from $K_3$ and $K_4$ due to the loss of dependency information, these marginal kernels characterize a set of possible joint measures that includes the original $\mu_{34}$. Each joint measure in this set induces a corresponding marginal kernel for $x_5$ through the above integration, thereby soundly over-approximating the distribution of $x_5$. 
Each resulting marginal kernel is then propagated through subsequent layers.

\indent \textit{Abstract Domain.} As the distributions of neurons are generally intractable, we introduce a new nonlinear abstract domain that soundly approximates the probability measures represented by marginal Markov kernels. Building on our marginal collecting semantics, the proposed domain provides sound abstract transformers for efficiently propagating distributions through BNNs. 
It abstracts both the support and probability distribution of each neuron, based on two key ideas.
The first key idea is to partition the support of each neuron and construct dedicated bounds on the probability measure over each partition. This partition-based representation offers two main advantages.
Firstly, although our analysis starts with neurons following continuous Gaussian distributions, ReLU activations inevitably introduce point masses, resulting in mixed distributions. Partitioning allows separate representation and approximation of continuous and discrete components. Secondly, for continuous components with highly nonlinear density functions, a single approximation over the entire support is often too coarse to capture their local behavior accurately. Constructing dedicated local bounds over each partition enables tighter approximations.
Formally, an abstract element
\(
a\in\mathcal A_n
\)
over \(n\) variables is represented as
\(
a=(a_1,\ldots,a_n),
\)
where
\(
a_i=
\{\langle l_i^k,u_i^k,L_i^k,U_i^k\rangle\}_{k\in[m_i]}.
\)
$m_i$ denotes the total number of partitions associated with the neuron $x_i$, and $l_i^k, u_i^k \in \mathbb{R}\cup\{-\infty, \infty\}$ \ denote the lower and upper bounds of the $k$-th partition of the support of $x_i$. $L_i^k$ and $U_i^k$  are constraints on the probability measure over that partition. Specifically, for non-degenerate intervals where \(l_i^k<u_i^k\), \(L_i^k\) and \(U_i^k\) provide lower and upper bounds on the probability density of the continuous component. For degenerate intervals where \(l_i^k=u_i^k\), \(L_i^k\) and \(U_i^k\) represent lower and upper bounds on the probability mass concentrated at that point.

While the abstract domain itself supports mixed distributions arising from subsequent propagation, the abstraction function is defined only for continuous distributions, as the analysis starts with neurons following continuous Gaussian distributions. The main challenge is therefore to construct sound bounds $L_i^k$ and $U_i^k$ on the nonlinear density functions of the initialized neurons over each partition to build the abstract inputs.
As linear abstractions often incur substantial precision loss by failing to capture the curvature of nonlinear density functions, a natural question then arises: can we design a nonlinear abstraction that can more accurately capture the shape of nonlinear densities? 
We therefore seek structural properties of density functions that can be exploited to help the derivation. Taking the Gaussian distribution as an example.
We observe that Gaussian distributions exhibit the global log-concavity property in their probability densities. That is, let \(p(x)\) denote the density function. Then \(\log p(x)\) is always concave over any given interval. This is a desirable property, since bounds for concave functions can be constructed through tangents and secants. This motivates our second key idea: reasoning about the log-density rather than the density itself. By constructing affine lower and upper bounds on the log-density and exponentiating them back, we obtain nonlinear exponential-affine lower and upper bounds on the density, i.e., $L^k_i, U^k_i$ are of the form $\exp(a x_i+b)$ where $a,b\in \mathbb{R}$.
The exponential-affine representation also facilitates key operations through its favorable closure properties under integration and multiplication, enabling the construction of tractable abstract transformers. 
By composing these transformers across layers, our analysis yields a sound approximation of the complete output distribution. 
Details are provided in~\secref{expo}.
Beyond BNNs, the same principles extend to broader classes of continuous distributions, including Laplace, Pareto, and Student's $t$ distributions, as detailed in~\appref{app:abstraction}, as well as probabilistic programs with bounded loops and Bayesian inference, as discussed in~\appref{extend}.

\textit{Abstract interpretation of the network.} 
\begin{figure}[t]
    \centering
    \includegraphics[width=\linewidth]{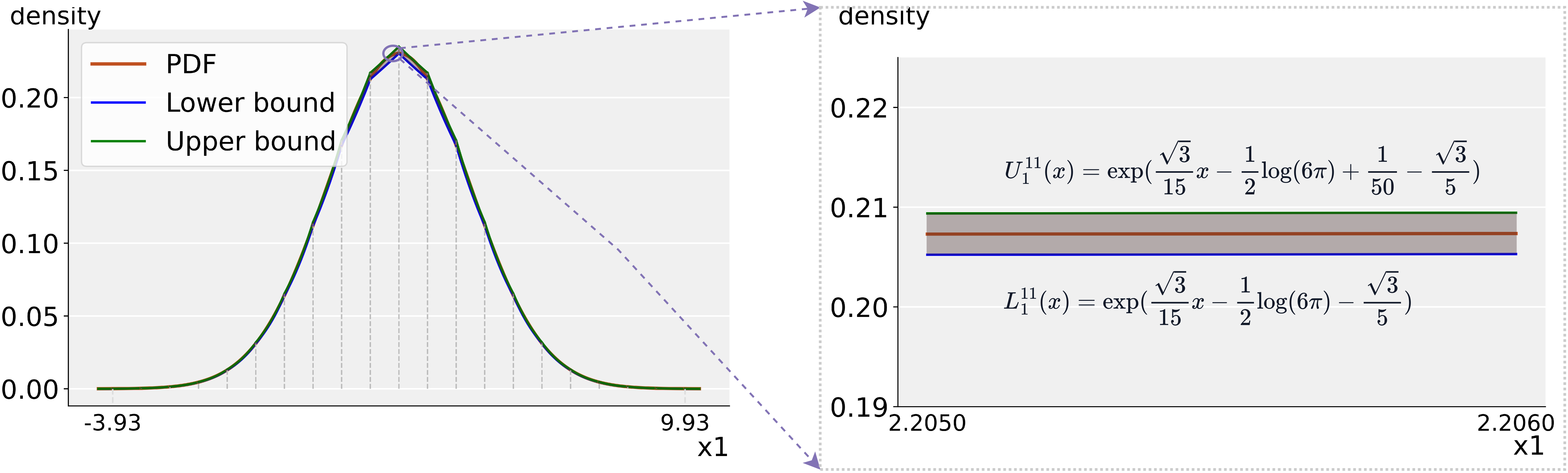}
    \caption{Visualization of the computed abstraction $a_1$ for neuron $x_1$ and comparison with the exact probability density function, as the concrete output distribution admits closed-form expressions. The right figure shows a zoomed-in view of the highlighted region in the left figure.}
    \figlabel{fig:density_zoom}
    \vspace{-16pt}
\end{figure}
We now illustrate the operation of our abstract interpreter on the example network.
We use the exact marginal distributions of \(x_1\) and \(x_2\) as the starting point of our analysis and abstract them as the abstract inputs to the subsequent transformation. 
Taking the abstraction of \(x_1\) as an example, since its associated marginal Markov kernel admits a density function, abstracting its probability measure reduces to constructing sound bounds on the corresponding density function \(p_1\).
We first partition its support $(-\infty, \infty)$.
During partitioning, instead of splitting the entire support uniformly, we allocate the approximation effort unevenly, placing more emphasis on regions with higher probability mass.
Specifically, based on the empirical rule for Gaussian distributions~\cite {empiricalrule}, we select a bounded main region
\(
[\mu-k\sigma,\ \mu+k\sigma]
\) to capture the majority of the probability mass of the Gaussian distribution $N(\mu, \sigma^2)$. 
The main region will be further partitioned into \(N\) subintervals, while the two residual tails are retained as unbounded tail intervals. The choice of $k$ and the partitioning strategy for the main region can be adjusted according to the desired trade-off between precision and efficiency.
In this example, we set $k=4$, when $[\mu-4\sigma,\ \mu+4\sigma]$ covers around 99.9936\% of the entire distribution, and we partition the main region uniformly with $N=20$ for simplicity.
This yields 22 partitions of the support of \(x_1\):
\(
\{[l_1^k,u_1^k]\}_{k=1}^{22}
=
\Big\{
(-\infty,\,3-4\sqrt3],\,
[3-4\sqrt3,\;3-\tfrac{18}{5}\sqrt3],\,\dots,
[3+\tfrac{18}{5}\sqrt3,\;3+4\sqrt3],\,
[3+4\sqrt3,\infty)
\Big\}.
\)
We then construct sound bounds for nonlinear density functions on each partition. Consider the $11$-th partition
\(
[l_1^{11}, u_1^{11}]=\left[3-\frac25\sqrt3\ ,\ 3\right]
\).
Let
$
\phi_1(x)=\log p_1(x)
=
-\frac12\log(6\pi)
-\frac{(x-3)^2}{6}.
$
We can easily prove that $\phi_1(x)$ is concave on this partition, i.e., 
$
\phi_1''(x)=-\frac13<0.
$
Therefore the tangent at any point of this partition provides a sound upper bound of $\phi_1(x)$, while the secant line over the interval provides a sound lower bound. 
We use the midpoint $m=3-\frac15\sqrt3$ as the tangent point for simplicity and efficient construction per partition, while other tangent points are also sound and can be optimized for tighter bounds at additional cost. Therefore, we obtain the affine upper and the lower bound of $\phi_1(x)$. Exponentiating the two bounds yields exponential-affine bounds $L_1^{11}(x)$ and $U_1^{11}(x)$ on the density $p_1(x)$, as shown in the right panel of~\figref{fig:density_zoom}. 
The same construction can be applied to every bounded partition, except for tail intervals. As in tail regions, a secant line cannot be constructed because one of the interval endpoints lies at infinity. Therefore, we handle tails separately by conservatively setting the lower bounds to zero, i.e., $L_1(x)=0$. 
Applying the above construction to all partitions yields the abstract representation of $x_1$:
{
\setlength{\abovedisplayskip}{2pt}
\setlength{\belowdisplayskip}{2pt}
\[
a_1
=
\Big\{
\langle l_1^k,u_1^k,L_1^k,U_1^k\rangle
\Big\}_{k\in[22]},
\
\text{where}
\
\forall k \in [22], \ \forall x \in [l^k,u^k],\  L_1^k(x) \le p_1(x) \le U_1^k(x)
\]
}
with $\bigcup_{k=1}^{22}[l^k,u^k]
=
(-\infty,\infty)$.
Therefore, the soundness of the abstraction $a_1$ for the distribution \(x_1\sim\mathcal N(3,3)\) can be expressed as
\(
\forall A \in \mathcal B(\mathbb R),
\int_L^\sharp(a_1,A)
\le
K_1(\emptyset,A)
=
\int_A p_1(x)\,dx
\le
\int_U^\sharp(a_1,A),
\)
where $\int_b^{\sharp}(a_i,A)$ is a unified abstract integration operation that handles the integration of an abstract element $a_i$ over a measurable set $A\in\mathcal B(\mathbb R)$, accounting for both integration with respect to the Lebesgue measure for continuous components and probability mass contributions for discrete components. It is defined as
$\int_b^{\#}(a_i,A)
=\sum_{k:l_i^k<u_i^k}\int_{A\cap(l_i^k,u_i^k)}b_i^k(x)\,dx
+\sum_{k:l_i^k=u_i^k}\mathbf{1}_A(l_i^k)b_i^k$,
where $b\in\{L,U\}$ denotes the lower or upper bound, respectively.  ~\figref{fig:density_zoom} visualizes $a_1$ and compares it with the exact density function, including a zoomed-in view of its bounds, showing that the abstraction achieves a high degree of precision. 
%
Similarly, we obtain $a_2$ for $x_2$, yielding the initial abstract element
\(
a=(a_1,a_2,\top,\ldots,\top),
\)
where the abstractions of the remaining seven neurons are initialized to $\top$, indicating that no constraints have yet been associated with them. This abstract element serves as the abstract input to the abstract interpreter of the network.

\begin{figure}[t]
    \centering
    \includegraphics[width=\linewidth]{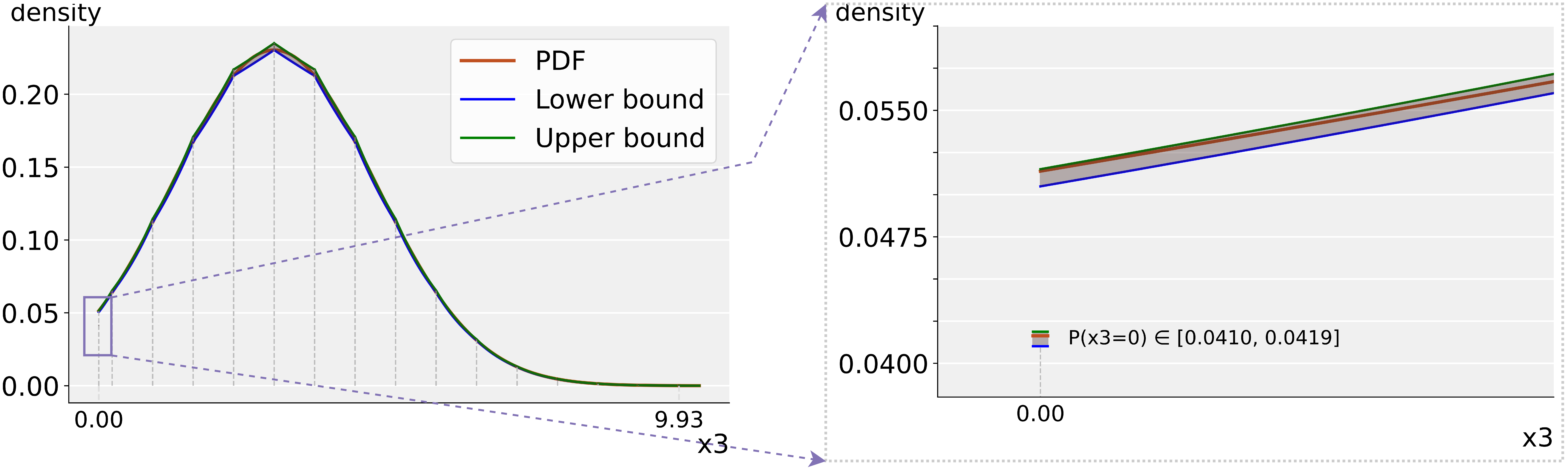}
    \caption{Visualization of the computed abstraction $a_3$ for neuron $x_3$ and comparison with the exact probability density function, as the concrete output distribution admits closed-form expressions. The right figure shows a zoomed-in view of the point mass region in the left figure.}
    \figlabel{fig:x3_density_zoom}
    \vspace{-15pt}
\end{figure}
After constructing the abstract input, we next apply the abstract transformer for the ReLU activation. 
Based on the marginal collecting semantics, since the support of $x_1$ spans both negative and positive regions, ReLU accumulates its non-positive probability mass at zero while preserving the original distribution over the positive region.
Specifically, the resulting abstraction $a_3$ consists of two parts.
The first part is a degenerate tuple
\(
\langle l_3^1,u_3^1,L_3^1,U_3^1\rangle,
\)
representing the accumulated probability mass at zero, where
$l_3^1=u_3^1=0$, and
\(
L_3^1
=
\int_L^\sharp(a_1,(-\infty,0]),
U_3^1
=
\int_U^\sharp(a_1,(-\infty,0]).
\)
The second part is a collection of tuples corresponding to the partitions of $a_1$ over the positive region, retaining their original bounds.
Combining these two parts yields the resulting abstraction $a_3$.
As the unified abstract integration operation accounts for both continuous and discrete probability components and the exponential-affine bounds admit closed-form integration, the accumulated probability mass at zero can therefore be bounded without additional integration approximation.
Consequently, our ReLU abstract transformer is optimal within our abstract domain.
\figref{fig:x3_density_zoom} visualizes $a_3$, the abstraction of $x_3$ computed by the ReLU abstract transformer. Similarly, we can get the abstraction $a_4$ corresponding to $x_4$. Then the ReLU abstract transformer outputs 
$a'=(a_1, a_2, a_3, a_4,\top,\ldots,\top )$.

Next, we apply the abstract affine transformer. Unlike the ReLU activation, propagating probability distributions through an affine layer is substantially more challenging for two reasons. First, affine transformations involve repeated additions and multiplications of random variables, requiring computationally expensive integration over probability measures. Second, dependencies among input variables must be accounted for, further complicating the computation of output distributions.
To address these challenges, our abstract affine transformer builds on three key ideas. First, we model affine transformations as conditional Gaussian distributions and further reformulate the resulting marginalization as a sum of conditional expectations weighted by probability masses over partitioned input regions, avoiding direct high-dimensional integration. Second, we adopt a novel event-based approach to soundly bound probabilities under the joint measure, thus avoiding explicit reasoning about dependencies. Third, we exploit structural properties of the result to construct sound exponential-affine bounds on the output density, ensuring closure within our abstract domain.
Specifically, consider the second affine layer in the running example. Based on marginal collecting semantics, for $x_5$ we have
$
K(\emptyset,dx_5)
=
\int_{\mathbb R^2}
K_{\mathrm{cond}}((x_3,x_4),dx_5)
\,\mu_{34}(d(x_3,x_4)).
$
$\mu_{34}$ denotes any joint probability measure of $(x_3,x_4)$ induced by the marginal kernels of $x_3$ and $x_4$.
This formulation requires integration and therefore is not convenient enough to provide an easy way to construct our abstraction. Nevertheless, it reveals an important property: since the conditional Gaussian variance $x_3^2+x_4^2+1$ is strictly positive, the resulting marginal kernel admits a density function $p_5$,
such that
\(
K(\emptyset,dx_5)=p_5(x_5)\,dx_5.
\)
We detail and prove this property in~\secref{affine}.
Therefore, constructing a sound abstraction of the marginal kernel of the output reduces to deriving sound bounds on its density function $p_5$ over each partition of the output support.
Recall that \(x_3\) and \(x_4\) are represented by the abstractions \(a_3\) and \(a_4\), respectively, each consisting of a finite partition of its support.
Accordingly, the integration domain $\mathbb R^2$ can be partitioned into $k$ Cartesian-product boxes $B_k$, formed from the intervals in the partitions of $a_3$ and $a_4$. We further reformulate the marginal density based on this partitioning as:
\(
p_5(x_5)
=
\sum_{k\in[K]}
\Pr[(x_3,x_4)\in B_k]
\mathbb{E}\left[
g(x_5,x_3,x_4)
\mid (x_3,x_4)\in B_k
\right],
\)
where $g(x_5,x_3,x_4)$ denotes the conditional Gaussian density associated with $K_{cond}$.
Detailed derivation can be found in~\secref{affine}. This reformulation involves only mathematically equivalent transformations and therefore introduces no approximation. It avoids tracking the joint distribution \(\mu_{34}\), requiring only the probability mass of each box and the conditional expectation within that box, both of which can be soundly approximated using the information maintained by our abstract domain.

Specifically, we first approximate the probability mass of each box by interpreting it as the probability of an event, which helps bypass reasoning about their dependency. We next approximate each conditional expectation term with sound exponential-affine bounds by exploring log-concavity properties of $g$ over that box. Finally, the resulting summation of exponential-affine functions is approximated by exploiting its log-convexity, yielding a sound exponential-affine abstraction of the output density. As a result, the abstract domain is closed under the affine transformer. Importantly, the derived bounds hold for every joint measure consistent with the input marginal kernels, thereby soundly over-approximating all induced output marginal kernels in the concrete semantics. More details can be found in ~\secref{affine}. We visualize the resulting abstraction of $x_5$ in~\figref{fig:x5} and compare it with the distribution estimated by Monte Carlo (MC) sampling. The results demonstrate that our method soundly captures the entire output distribution, whereas MC sampling only yields a limited empirical approximation based on finite samples.  By composing the affine and ReLU abstract transformers across layers, we eventually obtain a sound approximation of the entire output distribution of $y$, as visualized in~\figref{fig:y}. Any further arbitrary probabilistic queries on the network output can be answered directly from this abstraction without requiring additional effort.
\begin{figure}[t]
\centering
\begin{subfigure}[b]{0.48\textwidth}
    \centering
    \includegraphics[width=\linewidth]{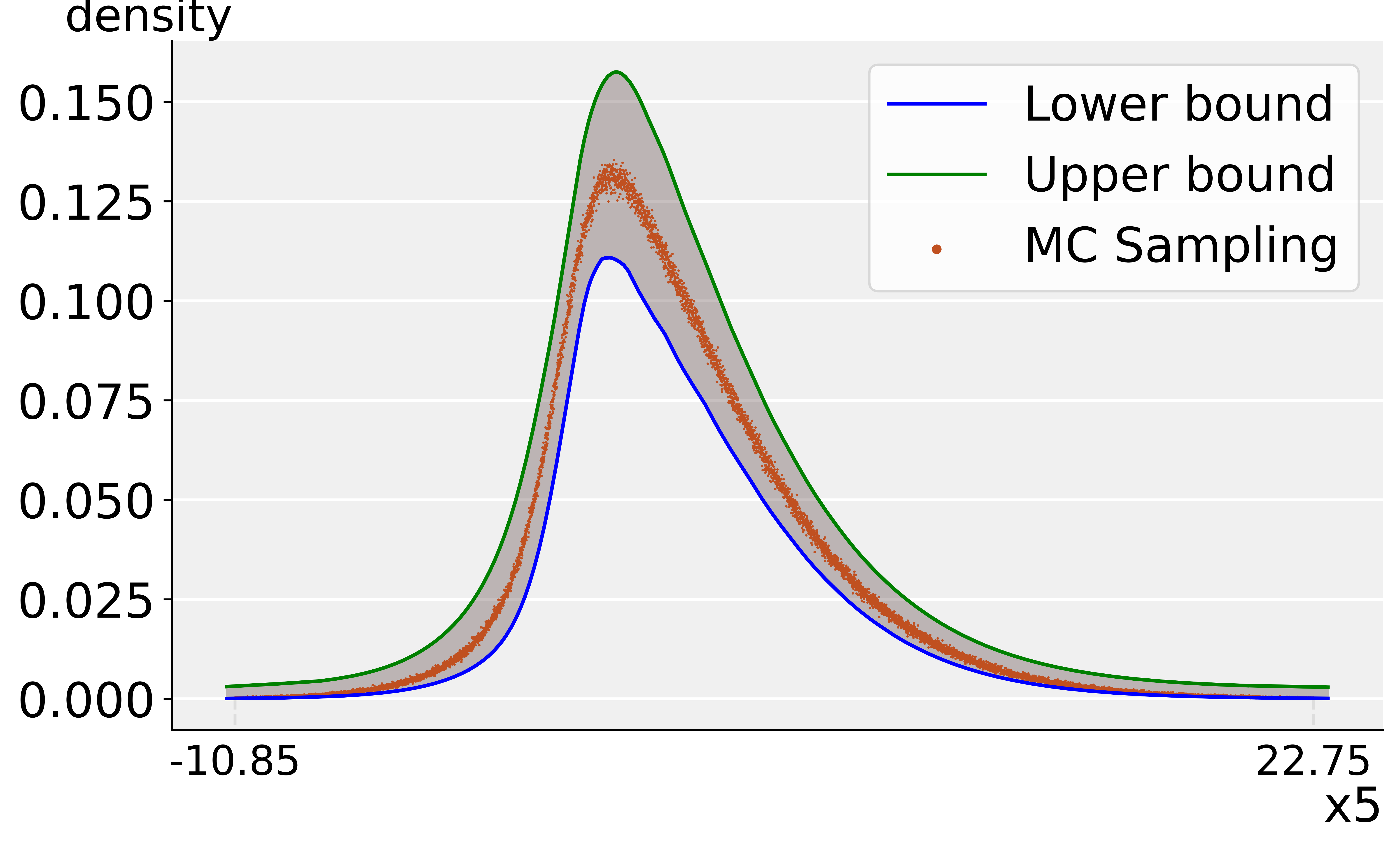}
    \caption{Abstraction of neuron $x_5$ after applying the affine abstract transformer.}
    \figlabel{fig:x5}
\end{subfigure}
\hfill
\begin{subfigure}[b]{0.48\textwidth}
    \centering
    \includegraphics[width=\linewidth]{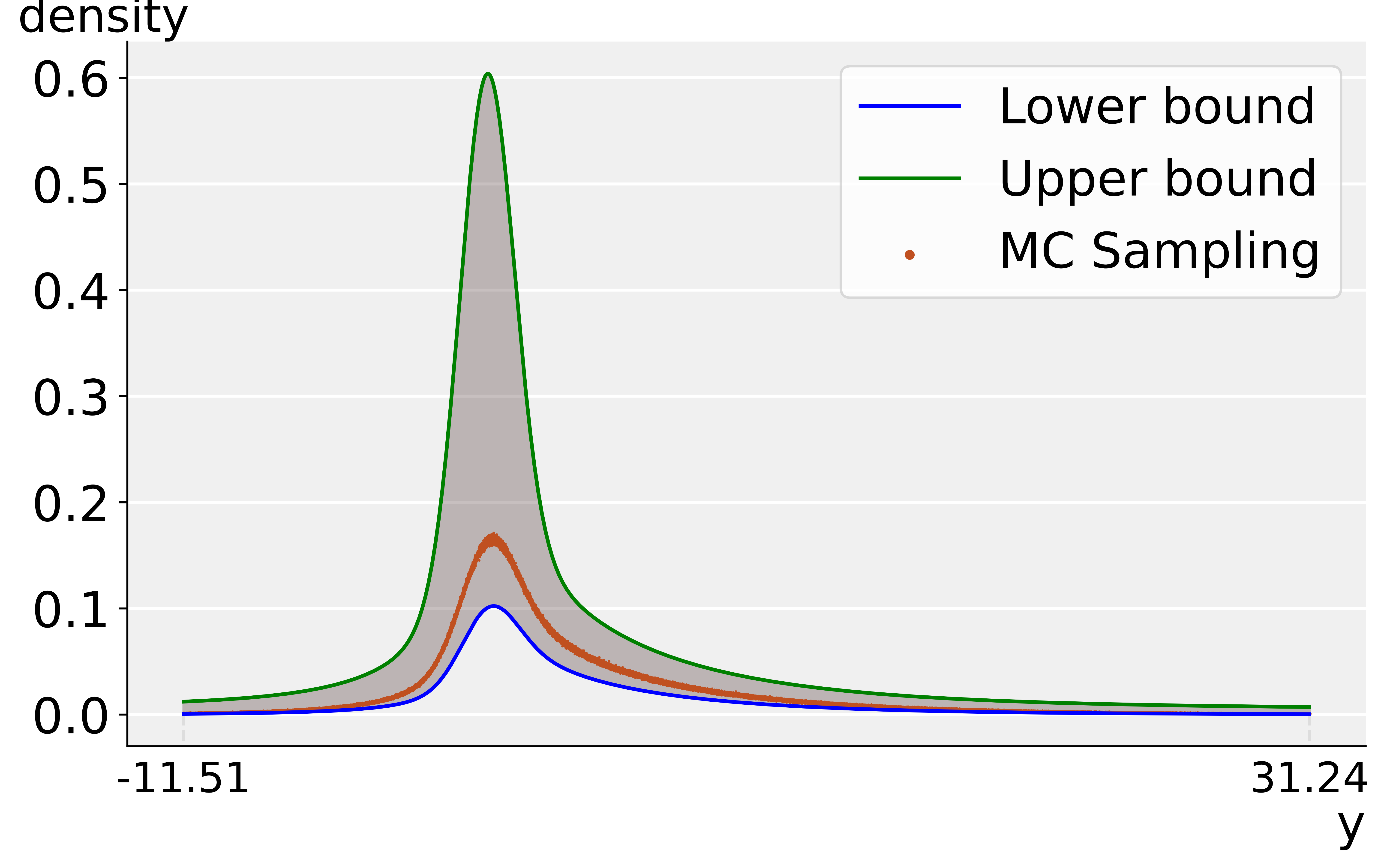}
    \caption{Abstraction of the network output $y$.}
    \figlabel{fig:y}

\end{subfigure}
\vspace{-10pt}
\caption{
~\figref{fig:x5} and~\figref{fig:y} visualize the computed abstractions for neurons $x_5$ and $y$, respectively, and compare them with histogram-based probability density estimates obtained from $10^7$ Monte Carlo samples, as their exact output distributions are intractable.
}
\figlabel{fig:overview}
\vspace{-15pt}
\end{figure}
\\
\indent \textit{Soundness Guarantee.}
The soundness of the abstract domain is therefore established in two steps. First, our marginal collecting semantics soundly over-approximates the classical joint-measure concrete semantics. Second, our abstract domain and its transformers soundly over-approximate the marginal collecting semantics. Together, these guarantees ensure that the proposed domain soundly over-approximates the output probability distribution of the original BNN.

%% file: tex/04_concrete.tex
\section{Marginal Collecting Semantics}
\seclabel{sec:semantics}
In this section, we formalize the marginal collecting semantics underlying our abstract interpretation framework based on Markov kernels. We see BNNs as straight-line probabilistic programs.
We 
first introduce the syntax of the probabilistic language and the marginal collecting semantics, then establish the soundness with respect to classical joint-measure semantics. We finally define the concrete domain based on our semantics, laying the foundation for the abstract domain.

We assume any random variable $x$ in a BNN is obtained from a finite sequence $\mathcal{X} = \{x_1, x_2 ,\ldots, x_n\}$ with $n$ variables. The syntax of the probabilistic programming language considered in our setting for expressing BNNs is shown below. Following the convention in the literature~\cite{jointmeasuresemantics2}, we explicitly index all program variables in the syntax to identify the variable being updated and the variables involved in the computation for the convenience of semantics definition.
\[
P  ::=  s \mid s;P \quad
s ::= x_i \sim \mathcal N(m,v) \mid x_i := E \quad
E  ::=\text{Affine}(\mathbf{x}_J)\mid \text{ReLU}(x_i)
\] 
The language consists of Gaussian sampling and assignment statements. 
In BNNs, given deterministic inputs, each neuron in the first affine layer computes a weighted sum of deterministic inputs. As the weights and bias are independent Gaussian random variables, their linear combination results in an output that also follows a Gaussian distribution. We therefore model this initialization of the stochastic neuron $x_i$ in the first affine layer of the network using the Gaussian sampling statement $x_i \sim \mathcal N(m,v)$, where $m,v \in\mathbb R, v>0$. 
Subsequent layers are expressed through assignments \(x_i := E\), where \(E\) is either an affine transformation $\text{Affine}(\mathbf{x}_J)$ or a ReLU activation \(\mathrm{ReLU}(x_i)\). We define $\text{Affine}(\mathbf{x}_J) = \sum_{j\in [J]}W_jx_j+b$, where \(\mathbf{x}_J, J \in [n]\) denotes a vector of $J$ input variables involved in the affine transformation. The weights \(W_j\sim\mathcal N(\mu_j,\sigma_j^2)\) and the bias \(b\sim\mathcal N(\mu_b,\sigma_b^2)\) are mutually independent Gaussian random variables, independent of the input variables \(\mathbf{x}_J\). 
This restricted syntax captures the fundamental building blocks of Bayesian neural networks.


\subsection{Marginal Collecting Semantics Rules}
\seclabel{marginalcollecting}
To capture the probabilistic information associated with each program variable while providing a unified representation of continuous, discrete, and mixed probability distributions arising in our setting, we introduce the novel definition of a probabilistic store: \textit{probstore}, which associates each random variable with a Markov kernel. A Markov kernel is defined over two measurable spaces, where the input space can be viewed as encoding dependencies between program variables, and the output space represents the possible values of a random variable. Thus, Markov kernels are commonly used to formalize conditional distributions in measure-theoretic probability theory~\cite{conditionalindependence1,conditionalindependence2, conditionalindependence3}. We refer to Markov kernels whose input valuation spaces correspond to non-empty sets of program variables as conditional Markov kernels. Since our analysis focuses on marginal distributions without tracking dependencies between program variables, we consider kernels whose input valuation space is the singleton \(\Sigma_{\emptyset}={\emptyset}\), which we refer to as marginal Markov kernels. 
Formally, let $\mathcal K$ denote the set of marginal Markov kernels
$K:\Sigma_{\emptyset}\times\mathcal B(\mathbb R)\to[0,1]$.

\begin{definition}[ProbStore]
A probabilistic store (probstore) $\sigma_p:\mathcal X\rightarrow \mathcal K$ is a total mapping from program variables
$x\in\mathcal X$ to Markov kernels in $\mathcal K$. We write $\operatorname{dom}(\sigma_p)$ to denote the domain of $\sigma_p$. For each $x\in\operatorname{dom}(\sigma_p)$, $\sigma_p(x)$ denotes the kernel associated with $x$, where $\sigma_p(x)(\emptyset,A)$ gives the probability of the event $A\in\mathcal B(\mathbb R)$ happening.
\end{definition}

\begin{figure}[h]
\centering
\small
\[
\begin{aligned}
&
\llbracket \mathrm{Affine}(\mathbf{x}_J)
\rrbracket_p^0(\sigma_p)
=
\left\{
K_\mu
\ \middle|\
\mu\in\gamma_{c_0}(\sigma_p)
\right\},
\
\text{where}
\
K_\mu(\emptyset,dy)
=
\int_{\mathbb R^J}
\frac{1}{\sqrt{2\pi v(\mathbf{x}_J)}}
\exp\left(
-\frac{(y-m(\mathbf{x}_J))^2}{2v(\mathbf{x}_J)}
\right)
dy\,\mu_J(d\mathbf{x}_J),
\\[1pt]
&
\qquad\qquad\qquad\quad
\mu_J(A)
=
\int_{\mathbb R^n}
\mathbf{1}_A(\mathbf{x}_J)\,\mu(d\mathbf{x}),
\
\forall A\in\mathcal B(\mathbb R^J).
\\[1pt]
&
\llbracket \mathrm{ReLU}(x_i)
\rrbracket_p^0(\sigma_p)
=
\{K\},
\
\text{where}
\
K(\emptyset,dy)
=
\sigma_p(x_i)(\emptyset,(-\infty,0])\,
\delta_0(dy)
+
\int_{(0,\infty)}
\delta_{y=x_i}(dy)\,
\sigma_p(x_i)(\emptyset,dx_i),
\\[1pt]
&
\llbracket x_i\sim\mathcal N(m,v)
\rrbracket_p^0(\sigma_p)
=
\{\sigma_p[x_i\mapsto K]\},
\
\text{where}
\
K(\emptyset,dx)
=
\frac{1}{\sqrt{2\pi v}}
\exp\left(
-\frac{(x-m)^2}{2v}
\right)dx,
\\[1pt]
&
\llbracket x_i:=E\rrbracket_p^0(\sigma_p)
=
\left\{
\sigma_p[x_i\mapsto K]
\ \middle|\
K\in\llbracket E\rrbracket_p^0(\sigma_p)
\right\},
\qquad
\llbracket s_1;s_2\rrbracket_p(c')
=
\llbracket s_2\rrbracket_p
\left(
\llbracket s_1\rrbracket_p(c')
\right).
\end{aligned}
\]
\vspace{-15pt}
\caption{Marginal collecting semantic rules.}
\figlabel{fig:marginalsemantics}
\vspace{-15pt}
\end{figure}


Let $\Sigma_p$ denote the set of all probstores. We denote our marginal concrete domain as $\mathcal C'= \mathcal P(\Sigma_p)$. As our concrete semantics maintains the marginal distribution of each program variable while abstracting away dependencies among variables, a probstore $\sigma_p\in\Sigma_p$ therefore cannot uniquely recover the original joint probability measure. Instead, $\sigma_p$ provides a sound approximation by representing a set of possible joint probability measures that includes the original one. 
We use $\gamma_{c_0}(\sigma_p)$ to denote the set of joint measures consistent with its marginals, formally defined in~\secref{sec:soundconcretedomain}.
We now introduce our marginal collecting semantics, denoted by $\llbracket\cdot\rrbracket_p$. For each statement $s$, let
$
\llbracket s\rrbracket_p^0:
\Sigma_p
\rightarrow
\mathcal P(\Sigma_p)
$
denote its semantics over an individual probabilistic store,  mapping each probstore to a set of possible resulting probstores due to the loss of dependency information between program variables. We then lift this semantics to the domain $\mathcal C'$ by defining
$
\llbracket s\rrbracket_p(c')
=
\bigcup_{\sigma_p\in c'}
\llbracket s\rrbracket_p^0(\sigma_p)
$ for $c' \in \mathcal C'$, 
where
$\llbracket s\rrbracket_p:\mathcal C'\to\mathcal C'$. 
Our marginal collecting semantics rules of Bayesian neural networks are summarized in~\figref{fig:marginalsemantics}. The first two rules define the semantics of expressions by constructing the kernels induced by the corresponding probabilistic operations. For the affine operation, conditioning on the input vector $\mathbf X_J=\mathbf x_J$ yields a Gaussian distribution with mean
$m(\mathbf x_J)=\sum_{j\in J}\mu_jx_j+\mu_b$
and variance
$v(\mathbf x_J)=\sum_{j\in J}\sigma_j^2x_j^2+\sigma_b^2$.
Marginalizing this conditional Gaussian over every joint input measure consistent with the probstore yields the affine rule.
The ReLU rule is the pushforward of the input marginal measure under \(\max(0,x)\), accumulating non-positive probability mass at zero while preserving the positive region.
We provide a detailed explanation and derivation for these two operations in~\appref{deriveaffinerelu}. The sampling statement initializes a random variable by associating it with a kernel represented by the probability density function of the specified Gaussian distribution. The assignment statement \(x:=E\) evaluates \(E\) and updates the probstore by mapping \(x\) to the resulting kernel, while leaving all other variables unchanged. The last rule states that, generally, our semantics supports compositional reasoning.

\subsection{Sound Abstraction of Joint-Measure Concrete Semantics}
\seclabel{sec:soundconcretedomain}
Our semantics is a sound abstraction of the classical joint-measure semantics~\cite{jointmeasuresemantics1,jointmeasuresemantics2}. Formally, let $\mathcal U(\mathbb R^n)$ denote the set of joint probability measures on the measurable space $(\mathbb R^n,\mathcal B(\mathbb R^n))$ over $n$ random variables. We denote the classical joint-measure concrete domain as $\mathcal C=\mathcal P (\mathcal U(\mathbb R^n))$. We establish soundness in two steps. First, we define the abstraction function  $\alpha_c:\mathcal C\to\mathcal C'$ and the concretization function $\gamma_c:\mathcal C'\to\mathcal C$ between $\mathcal C$ and $\mathcal C'$ and prove that they form a Galois connection. Second, we prove that, for every statement $s$, our marginal collecting semantics soundly approximates the classical joint-measure semantics, i.e., $\bigl(
\llbracket s\rrbracket_\mu(\gamma_c(c'))
\bigr)
\subseteq
\gamma_c(\llbracket s\rrbracket_p(c'))$ for any $c' \in \mathcal C'$, where $\llbracket s\rrbracket_\mu:\mathcal C\to\mathcal C$ denotes the classical joint-measure semantics of each statement.



\begin{definition}[Abstraction and Concretization]
Let
$\pi:\mathcal U(\mathbb R^n)\to\Sigma_p$
denote the marginalization function that maps a
joint probability measure to a probstore containing
the marginal distribution of each program variable.
Formally, $\pi(\mu)=\sigma_p$, where
$
\sigma_p(x_i)(\emptyset,A)
=
\int_{\mathbb R^n}
\mathbf 1_A(x_i)\,\mu(d\mathbf{x}),
\forall i\in[n],\
\forall A\in\mathcal B(\mathbb R).
$
For each variable $x_i$, $\pi(\mu)(x_i)$ denotes the marginal probability measure of $\mu\in\mathcal U(\mathbb R^n)$ with respect to $x_i$.
The abstraction function $\alpha_c:\mathcal C\to\mathcal C'$ is defined as 
$\alpha_c (c)=\{\sigma_p | \sigma_p = \pi(\mu), \mu \in c\}$ for $c \in \mathcal \mathcal C$.
Let $\gamma_{c_0}:\Sigma_p\to\mathcal C$ denotes the mapping from a single probabilistic store to a set of joint measures consistent with marginal kernels, where
$
\gamma_{c_0}(\sigma_p)
=
\{
\mu\in\mathcal U(\mathbb R^n)
\;|\;
\forall i\in[1,n],\
\pi(\mu)(x_i)
=
\sigma_p(x_i)(\emptyset,\cdot)\}
$ for $\sigma_p \in \Sigma_p$. 
The concretization function
$\gamma_c:\mathcal C'\to\mathcal C$ is defined as
$
\gamma_c(c')
=
\bigcup_{\sigma_p\in c'}
\gamma_{c_0}(\sigma_p)
$
 for $c' \in \mathcal C'$. 
\end{definition}

\begin{theorem}[Galois Connection]
The pair $(\alpha_c,\gamma_c)$ forms a Galois
connection between $(\mathcal C,\subseteq)$
and $(\mathcal C',\subseteq)$.
Formally, for every $c\in\mathcal C$ and
$c'\in\mathcal C'$,
$
\alpha_c(c)\subseteq c'
\iff
c\subseteq\gamma_c(c').
$
\theoremlabel{galois}
\end{theorem}

%

\begin{theorem}[Soundness of Marginal Collecting Semantics]
\theoremlabel{marginal-soundness}
For every statement $s$ and every $c'\in\mathcal C'$, we have
$
\llbracket s\rrbracket_\mu(\gamma_c(c'))
\subseteq
\gamma_c\bigl(\llbracket s\rrbracket_p(c')\bigr).
$
\end{theorem}
The complete proofs of the above two theorems are provided in~\appref{app:proofofgalois} and ~\appref{app:proofofmarginal-soundness}.

\subsection{Concrete Domain}
We next define concrete elements and concrete transformers 
based on marginal collecting semantics.

\begin{definition}[Concrete Element]
A concrete element $c' \in \mathcal{C'},\mathcal{C'}=\mathcal{P}(\Sigma_p) $ is a set of probstores.
\deflabel{def:concrete_element}
\end{definition}


The set of all possible concrete elements forms a lattice $(\mathcal{C'}, \sqsubseteq_{\mathcal C'}, \cup, \cap, \top, \bot)$. 
Here, the least upper bound and greatest lower bound of two concrete elements are computed via set union $\cup$ and intersection $\cap$, respectively. $\top$ is the top element in the lattice containing all possible probstores for all random variables in $\mathcal{X}$. $\bot$ is the bottom element representing an empty set of probstores. 
\begin{definition}[Concrete Transformer]
A concrete transformer $T: \mathcal{C'} \rightarrow \mathcal{C'}$ corresponding to a probabilistic program statement (e.g., assignment, observe) maps a concrete element $c_1' \in \mathcal{C'}$ to another concrete element $c_2' \in \mathcal{C'}$ by updating the probstores in the concrete element according to the marginal collecting semantics of the statement.
\end{definition}

%% file: tex/05_expohedron.tex
\section{Nonlinear Abstract Domain}
\seclabel{expo}
We now formally introduce our nonlinear abstract domain for sound probabilistic reasoning over BNNs. 
%
%
Formally, an abstract element $a\in\mathcal{A}_n$ is a tuple
\(
a=(a_1,\dots,a_n)
\)
over $n$ program variables,
where each component $a_i$ abstracts the probability measure associated with the $i$-th program variable $x_i$ and is represented as a collection of tuples with four components:
\(
a_i
=
\{
\langle l_i^k,u_i^k,L_i^k,U_i^k\rangle
\}_{k\in[m_i]}.
\)
Here, $m_i\in\mathbb N_{>0}$ denotes the number of partitions for $x_i$, and $l_i^k,u_i^k\in\mathbb{R}\cup\{-\infty,+\infty\}$ define the $k$-th interval of the partition.
%
For each non-degenerate interval, i.e., $l_i^k<u_i^k$, the functions $L_i^k$ and $U_i^k$ provide lower and upper bounds on the probability density of the continuous component, taking either exponential-affine or fallback constant forms whenever exponential-affine bounds cannot be constructed:
\(
L_i^k,U_i^k
\in
\left\{
x_i\mapsto\exp(\alpha x_i+\beta)
\;\middle|\;
\alpha,\beta\in\mathbb R
\right\}
\cup
\left\{
x_i\mapsto c
\;\middle|\;
c\in[0,\infty)
\right\}.
\)
For each degenerate interval, i.e., $l_i^k=u_i^k$, the bounds $L_i^k, U_i^k$ represent lower and upper bounds on the probability mass at the corresponding point of the form
\(
L_i^k,U_i^k
\in
\left\{
x_i\mapsto c
\;\middle|\;
c\in[0,\infty)
\right\}
\).
The concretization function
$\gamma_n$
is defined as
\[
\gamma_n:
\mathcal A_n\rightarrow
\mathcal P(\mathcal K^n),
\quad
\gamma_n(a)
=
\left\{
\mathbf{k}\in \mathcal K^n
\;\middle|\;
\forall i\in[n],\
\forall A\in\mathcal B(\mathbb R),
\int_{L}^{\#}(a_i,A)
\le
k_i(\emptyset,A)
\le
\int_{U}^{\#}(a_i,A)
\right\},
\]
where $\int^{\#}(a_i,A)$ is the integration operation of an abstract element $a_i$ over a measurable set $A\in\mathcal B(\mathbb R)$,
defined as:
\[
\int_{b}^{\#}(a_i,A)
=
\sum_{k\in[m_i]}
\begin{cases}
\displaystyle
\int_{A\cap(l_i^k,u_i^k)}
b_i^k(x)\,dx,
& l_i^k<u_i^k
\\[8pt]
\mathbf{1}_A(l_i^k)\,
b_i^k,
& l_i^k=u_i^k
\end{cases},
\quad b\in\{L,U\}.
\]
For convenience, we also use
$
\gamma:\mathcal{A}\rightarrow\mathcal P(\mathcal K)
$
to denote the operation over each abstraction, where
$
\forall i \in [n], a_i=\{(l_i^k,u_i^k,L_i^k,U_i^k)\}_{k\in[m_i]},
$
$
\gamma(a_i)
=
\{
k_i\in\mathcal K \mid
\forall A\in\mathcal B(\mathbb R),
\int_{L}^{\#}(a_i,A)
\le
k_i(\emptyset,A)
\le
\int_{U}^{\#}(a_i,A)
\}.
$
Below we first introduce the abstraction function and then abstract transformers. Our abstract transformers $T_f^\sharp: \mathcal{A}_n \rightarrow \mathcal{A}_n$ for a probabilistic function $f$
satisfy the following soundness property:
\(
T_f(\gamma_n(a))
\subseteq
\gamma_n\ (T_f^\sharp(a))
\)
for all $a\in\mathcal{A}_n$, where $T_f:\mathcal{C'} \rightarrow \mathcal{C'}$ is the corresponding concrete transformer induced by the concrete semantics of $f$ (defined in~\secref{sec:semantics}), given by
\(
T_f(c')
=
\bigcup_{\sigma_p\in c'} f(\sigma_p)
\)
where $c' \in \mathcal C'$ and $f:\Sigma_p\rightarrow\mathcal{P}(\Sigma_p)$
maps each probstore to the set of possible resulting probstores.

\subsection{Abstraction Function}
\seclabel{sec:abstraction}

As the BNN analysis is initialized from continuous Gaussian distributions, we define the abstraction function only for these initial concrete elements.
Formally, let $\mathcal K_{\mathrm{con}}\subseteq\mathcal K$ denote the set of kernels whose associated probability measures are absolutely continuous with respect to the Lebesgue measure. We establish the definition of $\mathcal K_{\mathrm{con}}$ through Radon–Nikodym theorem~\cite{rntheorem}:
\(
\mathcal K_{\mathrm{con}}
=
\{
K\in\mathcal K
| \
\exists f:\mathbb R\to[0,\infty), s.t.,
\forall B\in\mathcal B([0,\infty)),
f^{-1}(B)\in\mathcal B(\mathbb R),
\forall A\in\mathcal B(\mathbb R),
K(\emptyset,A)
=
\int_A f(x)\,\lambda(dx)
\}.
\)
where $\lambda$ denotes the Lebesgue measure on $\mathbb R$,
and $f$ is the probability density function associated with
the probability measure $K(\emptyset,\cdot)$.
Let
\(
\mathcal C^{'\mathrm{init}}
=
\mathcal P(\mathcal K_{\mathrm{con}}^n).
\)
The abstraction function
\(
\alpha:
\mathcal C^{'\mathrm{init}}
\rightarrow
\mathcal A_n
\)
maps an initial concrete element
$c\in\mathcal C^{'\mathrm{init}}$
to an abstract element $a=(a_1,\ldots,a_n)$.
Let
\(
\Pi_i(c')=
\{\frac{d\sigma_p(x_i)(\emptyset,\cdot)}{d\lambda}
|
\sigma_p\in c', c\in \mathcal C^{'\mathrm{init}}
\}
\)
denotes the set of all probability density functions associated with the $i$-th program variable in the concrete element $c'$. The high-level idea of the abstraction function is to construct each $a_i$ by first abstracting every density function $f_i^{j}\in\Pi_i$ individually, supposing $|\Pi_i|=t$ and $j\in [t]$. Then aggregate the resulting individual abstractions by selecting a common lower and upper bound that enclose all individual abstractions within each partition. The resulting $\ a_i=\{\langle l_i^k,u_i^k,L_i^k,U_i^k\rangle\}_{k\in[m_i]}$ therefore soundly encloses every function in $\Pi_i$.
We first introduce the abstraction procedure for an individual density function and then the aggregation strategy.

\textit{Partitioning.}
For each probability density function $f_i^{j}$, we partition its support into a finite set of intervals
$\{[l_i^{j,k},u_i^{j,k}]\}_{k\in[m_i^{j}]}, $ where $m_i^j \in\mathbb N_{>0}$ denotes the number of partitions and $l_i^{j,k},u_i^{j,k}\in\mathbb{R}\cup\{-\infty,+\infty\}$.
For Gaussian distributions $N(\mu, \sigma^2)$ in our setting, we use a non-uniform partitioning strategy that allocates more intervals around the mean, where most probability mass is concentrated, and fewer intervals in the tails. Specifically, we first identify the main region
\(
[\mu-n\sigma,\;\mu+n\sigma], n \in \mathbb{R},
\)
and partition this region into multiple bounded intervals. The remaining probability mass is represented by two unbounded tail intervals,
\(
(-\infty,\mu-n\sigma]
\)
and
\(
[\mu+n\sigma,+\infty).
\)
This partitioning strategy yields tighter abstractions under a fixed partition budget, while the number of partitions and the choice of $n$ determine the trade-off between precision and efficiency.

\textit{Constructing Exponential-Affine Bounds.} 
We first construct sound bounds on the bounded intervals and then discuss the treatment of the unbounded tail intervals.
Given a function $f_i^{j}\in\Pi_i$ within each bounded interval \([l_i^{j,k},u_i^{j,k}]\), we construct lower and upper bounds on it in the form of exponential-affine functions:
$
L_i^{j,k}(x_i)
=
\exp(a^{\le}x_i+b^{\le}),
U_i^{j,k}(x_i)
=
\exp(a^{\ge}x_i+b^{\ge}),
$
where $a^{\le}, b^{\le},a^{\ge},b^{\ge}\in \mathbb{R}$, such that
$
L_i^{j,k}(x_i)
\le
f_i^{j}(x_i)
\le
U_i^{j,k}(x_i),
\forall x_i\in[l_i^{j,k},u_i^{j,k}].
$
This reduces the problem to constructing affine bounds on the log-density
$
a^{\le}x_i+b^{\le}
\le
\phi_i^{j}(x_i)
\le
a^{\ge}x_i+b^{\ge},
\forall x_i\in[l_i^{j,k},u_i^{j,k}],
$
where $\phi_i^{j}(x_i) = \log f_i^{j}(x_i)$.
The construction of the coefficients $(a^{\le},b^{\le},a^{\ge},b^{\ge})$ depends on the geometric properties of $f_i^{j}$ over the interval. 
As in our setting, the initialized continuous probability distributions are Gaussian. Accordingly, we restrict the presentation below to the abstraction of Gaussian distributions. A complete abstraction procedure for more general classes of density functions is provided in~\appref{app:abstraction}.
For clarity, we simplify the notation here. We denote the Gaussian density function as $f$. Let
\(
\phi(x)=\log f(x), x\in[l,u]
\).
We can easily prove that $\phi$ is concave for any $x\in[l,u]$, i.e. 
\(
\phi\bigl(\lambda a+(1-\lambda)b\bigr)
\;\ge\;
\lambda\phi(a)+(1-\lambda)\phi(b),
\forall a,b\in[l,u],\;
\forall\lambda\in[0,1].
\)
In this case, the lower bound of $\phi$ is given by the secant line over $[l, u]$, and the upper bound is given by a tangent line at some $x_0 \in [l, u]$. Proof is provided in ~\appref{app: proof}. Therefore we have
\(
a^{\le} = \frac{\phi(u)-\phi(l)}{u-l},  b^{\le} = \frac{u\,\phi(l)-l\,\phi(u)}{u-l}, 
a^{\ge}=g,
b^{\ge}=\phi(x_0)-g\,x_0.
\)
where $g\in\partial \phi(x_0)$ is a subgradient of \(\phi\) at \(x_0\).
By exponentiating these affine bounds, we obtain exponential-affine bounds on the density.

We next describe the abstraction of tail intervals. Unlike bounded intervals, a tail interval has one endpoint at infinity, and therefore a secant line cannot be constructed for the log-density. Consequently, we construct the upper bound in the same way as for bounded intervals, using the tangent line to the log-density at the finite endpoint of the tail interval. For the lower bound, we employ a conservative fallback by setting it to the constant $0$. Formally,
\(
L(x)=0,
U(x)=\exp\bigl(\phi(t)+\phi'(t)(x-t)\bigr),
\)
where $t=l$ for a right tail $[l,+\infty)$ and $t=u$ for a left tail $(-\infty,u]$.

\textit{Constructing a Unified Abstraction.}
After constructing an abstraction for each individual function $f_i^{j}\in\Pi_i$, we construct a unified abstraction $a_i$ that soundly encloses every function in \(\Pi_i\) by finding a common upper and lower bound on each partition. We prove the existence of such a unified abstraction in~\appref{app:aggre} and establish the theoretical feasibility of the abstraction function. In our setting, since each initialized variable has a single Gaussian density (i.e., each $\Pi_i$ contains only one function), the unified abstraction construction step is trivial and is omitted.

\subsection{ReLU Abstract Transformer}
Let
$
f:\mathcal K\rightarrow\mathcal K
$
be the function that executes the assignment
$
x_i \leftarrow \max(0,x_j)
$
on the random variables, for some $j<i$.
The corresponding abstract transformer is defined as
\(
T_f^\sharp(a) = a^\prime\), where $a, a^\prime \in \mathcal{A}_n, a = (a_1,\ldots, a_n), a^\prime = (a_1^\prime, \ldots, a_n^\prime)$. For all $t<i$ and $t>i$, $a_t^\prime= \{(l_t^{\prime k},u_t^{\prime k},L_t^{\prime k},U_t^{\prime k})\}_{k\in[m^\prime _t]}=a_t = \{(l_t^{k},u_t^{k},L_t^{k},U_t^{k})\}_{k\in[m_t]}$. For the new component $a_i^\prime$, there are three cases.
If multiple cases apply, only the first applicable case is selected. Let  $a_j = \{(l_j^{k},u_j^{k},L_j^{k},U_j^{k})\}_{k\in[m_j]}$,
    \begin{itemize}[leftmargin=*]
        \item \textbf{Case 1: \(u_j^{m_j}\le0\).} $m_j$ denotes the last partition of $a_j$, and $u_j^{m_j}$ is the upper endpoint of the support, which is non-positive. Therefore, the entire support is mapped to the single point \(0\). Since the resulting probability mass is concentrated on a zero-width interval, it cannot be represented by a probability density function. We therefore record the accumulated probability mass at \(0\), i.e., $m_i^\prime=1$ and
\(
a_i^\prime=\{(l_i^{\prime k},u_i^{\prime k},L_i^{\prime k},U_i^{\prime k})\}_{k\in[m_i']}
=
\left\{
\left(
0,\,
0,\,
\int_L^\#(a_j,(-\infty,0]),\,
    \int_U^\#(a_j,(-\infty,0])
\right)
\right\}.
\)
        \item \textbf{Case 2: \(l_j^1\ge0\).} Therefore the entire support is non-negative. ReLU acts as the identity mapping. Therefore $m_i^\prime=m_j$ and 
        \(
a_i^\prime = \{(l_i^{\prime k},u_i^{\prime k},L_i^{\prime k},U_i^{\prime k})\}_{k\in[m_i']}
=
\{(l_j^{k},u_j^{k},L_j^{k},U_j^{k})\}_{k\in[m_j]}
\)
    \item \textbf{Case 3: \(l_j^1<0<u_j^{m_j}\).}
In this case, the support spans both sides of 0. The density over the non-positive support is transformed into a probability mass concentrated at \(0\), while the density over the positive support remains unchanged. Therefore
\(
a_i^\prime 
= \{(l_i^{\prime k},u_i^{\prime k},L_i^{\prime k},U_i^{\prime k})\}_{k\in[m_i']}
=
\{
(
0,\,
0,\,
\int_L^\#(a_j,(-\infty,0]),\,
\int_U^\#(a_j,(-\infty,0])
)
\} \bigcup
\{
(
\max(0,l_j^{k}),\,
u_j^{k},\,
L_j^{k},\,
U_j^{k}
)
\}_{k\in[m_j],\ u_j^{k}>0}.
\)
    \end{itemize}

\begin{theorem}
The ReLU abstract transformer is sound.
\end{theorem}

We provide the complete soundness proof in~\appref{soundrelu}.
The proof also implies its optimality.

\subsection{Affine Abstract Transformer}
\seclabel{affine}
Let
\(
f:\mathcal K^{J}\rightarrow \mathcal P (\mathcal K)
\)
be the function that executes the affine transformation
\(
x_i\leftarrow \sum_{j\in[J]} W_j\cdot x_j+b,
\)
where $W_j, j\in[J], J \in [i-1]$ and $b$ are Gaussian random variables.
The corresponding abstract affine transformer is
\(
T_f^\sharp(a) = a^\prime\), where $a, a^\prime \in \mathcal{A}_n, a = (a_1,\ldots, a_n), a^\prime = (a_1^\prime, \ldots, a_n^\prime)$. For all $t<i$ and $t>i$, $a_t^\prime= \{(l_t^{\prime k},u_t^{\prime k},L_t^{\prime k},U_t^{\prime k})\}_{k\in[m^\prime _t]}=a_t = \{(l_t^{k},u_t^{k},L_t^{k},U_t^{k})\}_{k\in[m_t]}$. 
We construct the new component $a_i^\prime = \{(l_i^{\prime k},u_i^{\prime k},L_i^{\prime k},U_i^{\prime k})\}_{k\in[m^\prime _i]}$ based on the concrete semantics we derive in~\secref{sec:semantics}.

To facilitate the subsequent development of our partition-based abstract domain, we first further decompose~\eqnref{eq:py1} in the concrete semantics over
a finite partition of the vector space $\mathbf X_J$, and reinterpret the resulting marginalization as a conditional expectation, providing a basis for tractable approximation in the abstract domain.
Specifically, 
as each input variable $x_j$ is represented by an abstract element with a finite partition of its support $
\{I_j^{k_j}\}_{k_j \in [m_j]} = \{[l_j^{k_j}, u_j^{k_j}]\}_{k_j \in [m_j]},
$
where $m_j$ denotes the number of partitions of the $j$-th variable $x_j$.
The joint input space is therefore partitioned into a collection of multi-dimensional boxes, where each box is defined as the Cartesian product of intervals across all dimensions:
\(
B_k = I_1^{k_1} \times I_2^{k_2} \times \cdots \times I_J^{k_J}.
\)
Here $k_j\in[m_j]$ for each $j\in[J]$, $k \in [\prod_{j=1}^{J} m_j]$.
For each input region $B_k$, the integral
can be further equivalently expressed using conditional expectation. 
The output kernel of the affine transformation is finally rewritten as
\(
K(\emptyset,dy)
=
\left(
\sum_k
\left\{
\mathbb{E}\left[
\left.
\frac{1}{\sqrt{2\pi v(\mathbf{X}_J)}}
\exp\left(
-\frac{(y-m(\mathbf{X}_J))^2}{2v(\mathbf{X}_J)}
\right)
\,\right|\,
\mathbf{X}_J\in B_k
\right]
\cdot \Pr[\mathbf{X}_J\in B_k]
\right\}
\right)dy.
\)
where  $\Pr[\mathbf X_J\in B_k]=
\mu_{J}(B_k),
\mu_J(A)
=
\int_{\mathbb R^n}
\mathbf{1}_A(\mathbf{x}_J)\,\mu(d\mathbf{x}),
\
\forall A\in\mathcal B(\mathbb R^J), 
\mu\in \gamma_{c0}(\sigma_p)
$, with $\sigma_p$
denoting the input probabilistic store, as defined in ~\secref{marginalcollecting}. We provide a detailed derivation in~\appref{app:derivationcond}. 
The equality holds without any independence assumptions, and it provides a useful decomposition of the integral into two components: the probability mass within the region $B_k$, and the conditional expectation of the output kernel over that region.
We prove that, when the Affine bias $b$ has strictly positive variance $\sigma_b^2>0$, as in our setting, for any joint input probability measure $\mu_J$, the output probability measure $K(\emptyset,dy)$ is absolutely continuous with
respect to the Lebesgue measure $\lambda$.
Specifically, the function 
\(
f_Y(y)=
\sum_k
\{
\mathbb{E}[
\frac{1}{\sqrt{2\pi v(\mathbf{X}_J)}}
\exp(
-\frac{(y-m(\mathbf{X}_J))^2}{2v(\mathbf{X}_J)}
)
\,|\,
\mathbf{X}_J\in B_k
]
\cdot \Pr[\mathbf{X}_J\in B_k]
\}
\)
is a probability density
function of the output distribution.
We provide the complete proof in~\appref{affine-density}.
Therefore, our affine abstract transformer only needs
to construct lower and upper bounds on this density function.
We complete this goal in two steps.
We first soundly approximate each term in $f_Y(y)$, and then approximate the summation. The goal is to represent the measure of $y$ as an abstract element by deriving exponential-affine upper and lower bounds on each partition of its support.

\textit{Event-based bounding of $\ \Pr[\mathbf{X}_J \in B_k\rbrack$.}
First, we consider bounding $\Pr[\mathbf{X}_J\in B_k]$ in $f_Y(y)$. The insight here is to take an event-based perspective and reason about the probability of $\mathbf{X}_J$ lying in the box $B_k$ by decomposing it into per-dimension events. The advantage of this method is that it allows us to derive sound bounds of $\Pr[\mathbf{X}_J\in B_k]$ without requiring any assumptions on the underlying joint distribution of $\mathbf{X}_J$, such as independence across dimensions. 
We first rewrite the event
$
\{\mathbf{X}_J \in B_k\}
$
in terms of per-dimension events. Specifically, recalling $B_k = I_1^{k_1} \times I_2^{k_2} \times \cdots \times I_J^{k_J}$, for each dimension $h$, we further define the event
$
A_h^{k_h} = \{X_h \in I_h^{k_h}\}.
$
Then the event that the input lies in the box $B_k$ is equivalent to the intersection of these events:
$
\{\mathbf{X}_J \in B_k\} = \bigcap_{h=1}^J A_h^{k_h}.
$
Although the joint distribution of $\mathbf{X}_J$ is unknown,
we can still derive sound bounds on the box probability.
For the upper bound, observe that
$
\bigcap_{h=1}^J A_h^{k_h}\subseteq A_j^{k_j}
$
for any $j\in[J]$, i.e., satisfying all conditions
implies satisfying any single one.
For the lower bound, we consider the complementary event
and apply the union bound:
\(
\Pr[\mathbf{X}_J \in B_k]
=\Pr(\bigcap_{h=1}^J A_h^{k_h})= 1 - \Pr\ (\bigcup_{h=1}^{J} (A_h^{k_h})^c)
\ge
1 - \sum_{h=1}^J (1 - \Pr(A_h^{k_h})).
\)
Taking probabilities of the event inclusion
and combining these results, we can finally obtain
\(
\max\ (
0,\sum_{h=1}^J
\Pr(X_h\in I_h^{k_h})-(J-1)
)
\le
\Pr[\mathbf{X}_J\in B_k]
\le
\min_{h\in[J]}\Pr(X_h\in I_h^{k_h}).
\)
We provide a detailed explanation in~\appref{boxbounds}.
We use $\underline{P}_k$ and $\overline{P}_k$ to represent the lower bound and upper bound of $\Pr[\mathbf{X}_J \in B_k]$, respectively, in the following paper for simplicity.

\textit{Bounding the conditional expectation over $B_k$.} Considering bounding $\mathbb{E}[g(y,\mathbf{X}_J) \mid \mathbf{X}_J \in B_k]$ in $f_Y(y)$. The key idea here is to eliminate the expectation operator by deriving bounds that hold uniformly over all $\mathbf{X}_J \in B_k$. 
Let
\(
g(y,\mathbf{x}_J)
=
\mathcal{N}(y; m(\mathbf{x}_J), v(\mathbf{x}_J))
=
\frac{1}{\sqrt{2\pi v(\mathbf{x}_J)}}
\exp\ \left(
-\frac{(y - m(\mathbf{x})_J)^2}{2 v(\mathbf{x}_J)}
\right),
\)
where
\(
m(\mathbf{x}_J) = \sum_{j=1}^J \mu_{j} x_j + \mu_{b},
v(\mathbf{x}) = \sum_{j=1}^J \sigma_{j}^2 x_j^2 + \sigma_{b}^2.
\)
Since for $\mathbf{x}_J \in B_k$, each dimension satisfies
$
x_h \in [l_h^{k_h}, u_h^{k_h}],
$ which implies that the mean and variance are bounded as
$m(\mathbf{x}_J) \in [\underline{m}_{k}, \overline{m}_{k}],
v(\mathbf{x}_J) \in [\underline{v}_{k}, \overline{v}_{k}],
$
where the bounds can be easily obtained via interval arithmetic.
Note that some input intervals may be unbounded, i.e., $l_h^{k_h}=-\infty$ or $u_h^{k_h}=+\infty$, which may lead to unbounded interval estimates of the conditional mean and variance. To avoid complicating the following derivation, we handle such tail regions separately using
a conservative approximation strategy, as detailed in~\appref{tailregion}. Therefore, in the following derivation, we only consider bounded input boxes $B_k$, for which the bounds
$\underline{m}_k$, $\overline{m}_k$, $\underline{v}_k$, and $\overline{v}_k$ are all finite.
For all $\mathbf{x}_J \in B_k$, we derive $\underline{g}_{k}(y)\le g(y,\mathbf{x}_J) \le \overline{g}_{k}(y),$
where
\(
\overline{g}_{k}(y)
=
\sup_{\substack{
m \in [\underline{m}_{k}, \overline{m}_{k}]\\
v \in [\underline{v}_{k}, \overline{v}_{k}]
}}
\mathcal{N}(y; m, v),
\underline{g}_{k}(y)
=
\inf_{\substack{
m \in [\underline{m}_{k}, \overline{m}_{k}]\\
v \in [\underline{v}_{k}, \overline{v}_{k}]
}}
\mathcal{N}(y; m, v).
\)
Since the bounds hold for every realization \(\mathbf{x}_J\in B_k\), they also hold for the random input vector \(\mathbf{X}_J\) on the event \(\mathbf{X}_J\in B_k\), i.e.,
\(
\underline{g}_{k}(y)
\;\le\;
g(y,\mathbf{X}_J)
\;\le\;
\overline{g}_{k}(y),
\)
for $\mathbf{X}_J\in B_k.$
Taking conditional expectation on both sides, we obtain
\(
\mathbb{E}[\underline{g}_{k}(y) \mid \mathbf{X}_J \in B_k]
\;\le\;
\mathbb{E}[g(y,\mathbf{X}_J) \mid \mathbf{X}_J \in B_k]
\;\le\;
\mathbb{E}[\overline{g}_{k}(y) \mid \mathbf{X}_J \in B_k].
\)
Since $\underline{g}_{k}(y)$ and $\overline{g}_{k}(y)$ do not depend on $\mathbf{X}_J$, we have
\(
\underline{g}_{k}(y)
\;\le\;
\mathbb{E}[g(y,\mathbf{X}_J) \mid \mathbf{X}_J \in B_k]
\;\le\;
\overline{g}_{k}(y).
\)
We detail the derivation of $\underline{g}_{k}(y)$ and $\overline{g}_{k}(y)$ in~\appref{proof}. Furthermore, we show that the derived bounds exhibit desired properties. Specifically, we can prove that $\underline g(y)$ is log-concave, while
$\overline g(y)$ is either log-concave or log-convex in $y$.
The proof is also provided in~\appref{proof}.

\textit{Approximating the summation.} So far, we have obtained sound upper and lower bounds for each term in $f_Y(y)$. Specifically, for each box $B_k$, we have 
\(
\underline{g}_{k}(y)
\le
\mathbb{E}[g(y,\mathbf{X}_J) \mid \mathbf{X}_J \in B_k]
\le
\overline{g}_{k}(y),
\)
and
\(
\underline{P}_k
\le
\Pr[\mathbf{X}_J \in B_k]
\le
\overline{P}_k.
\)
As all bounds are positive,
summing over all boxes yields pointwise upper and lower bounds on the output density: 
\(
\sum_k \underline{P}_k \, \underline{g}_{k}(y)
\le
f_{Y}(y)
\le
\sum_k \overline{P}_k \, \overline{g}_{k}(y).
\)
The next step is therefore to approximate this summation to obtain the abstraction over output.
We complete this goal in two steps.
First, we partition the output domain into intervals and exploit the log-concavity or log-convexity of $\overline{g}_{k}(y)$ and $\underline{g}_{k}(y)$ to derive exponential-affine bounds for each term in the summation.
Second, we exploit the convexity of the log-sum-exp function to obtain exponential-affine bounds on the entire summation.
For unbounded intervals, we adopt conservative bounds to ensure soundness over the entire support.
The detailed construction is provided in~\appref{boundsummation}.
Finally, on each interval $J^r$ of the output $y$, we will obtain exponential-affine bounds of the form
\(
\exp(a_r^L y + b_r^L)
\le
f_{Y}(y)
\le
\exp(a_r^U y + b_r^U),
\)
Therefore
the output abstract element $a_i'$ corresponding to $y$ (which is $x_i$) is defined as: 
\(
a_i'
=
\left\{
\left\langle
l^r, u^r,
\exp(a_{r}^L y+b_{r}^L),
\exp(a_{r}^U y+b_{r}^U)
\right\rangle
\right\}_{r},
\)
which is aligned with our abstract domain.

\begin{theorem}
The Affine abstract transformer is sound.
\end{theorem}

We provide the formal proof in~\appref{affinesound}.

%% file: tex/08_eval.tex
\section{Evaluation}
Beyond BNNs, we extend our framework to general probabilistic programs with bounded loops and Bayesian inference, which is detailed in~\appref{extend}.
We then evaluate our approach to answer the following research questions:
\begin{itemize}[topsep=0pt, leftmargin=9pt]
    \item \textbf{RQ1 (Precision)}: How precisely does it approximate probability distributions compared to existing interval-based abstract domains? (\secref{sec: eval1},~\secref{sec: eval2})
    \item \textbf{RQ2 (Coverage)}: How effectively does it capture the entire output probability distribution compared to sampling-based approaches? (\secref{sec: eval1})
    \item \textbf{RQ3 (Scalability)}: How well does it scale with increasing network complexity and loop iterations? (\secref{sec: eval1},~\secref{sec: eval2})
    \item \textbf{RQ4 (Generality)}: How effectively does it generalize to probabilistic programs with Bayesian inference and bounded loops? (\secref{sec: eval2})
\end{itemize}

\textit{Key Observations.}
To the best of our knowledge, there is currently no approximation inference work capable of soundly reasoning the entire output distributions of BNNs. Therefore we implement an interval-based abstract domain as a baseline for comparison for RQ1, which is the most common approach for probabilistic abstract interpretation~\cite{intervalprobAI}. 
For RQ2, we compare our framework against Monte Carlo sampling. 
For RQ3, we evaluate it on a challenging benchmark consisting of BNNs with ReLU activations, trained for regression tasks across diverse datasets, architectures, and input dimensions. For RQ4, we compare it against the interval-based domain, AURA~\cite{AURA}, a state-of-the-art sound approximate inference framework with limitations, and PSI~\cite{psi}, an exact symbolic inference framework.
We find that our framework substantially improves precision over existing methods without sacrificing scalability (RQ1, RQ3).
Also, unlike sampling-based approaches, our approach can soundly capture the entire output probability distribution of BNNs, especially in low-probability tail regions  (RQ2). Moreover, the results demonstrate that our framework generalizes effectively to probabilistic programs, consistently outperforming the interval domain and AURA in both precision and runtime on all comparable benchmarks, while supporting complex probabilistic models that AURA and PSI cannot fully handle (RQ4).

\textit{Experimental Setup.}
Our abstract interpreter is implemented in Python. BNN experiments run on an NVIDIA H100 GPU with 
230~GB RAM, and probabilistic program experiments use four NVIDIA A100 GPUs with 
256~GB RAM.
All methods in each comparison use the same hardware, partition budgets, and applicable GPU optimizations to ensure fairness.

\textit{Optimization for instantiation.}
As the direct implementation of abstract transformers suffers from the curse of dimensionality and rapid partition growth. To address these practical challenges, we incorporate several optimization strategies that preserve soundness while substantially improving scalability.
The complete algorithms of these optimizations are provided in~\appref{optimization}.

\subsection{Evaluation on Bayesian Neural Networks}
\seclabel{sec: eval1}
\textit{Benchmarks.}
Targeted at the uncertainty quantification task, we train a set of Bayesian neural networks using Bayes by Backprop~\cite{independentweights,VI} with diagonal Gaussian variational posteriors on three UCI regression datasets separately: Concrete Compressive Strength~\cite{uci_concrete} (with 8 features), Energy Efficiency~\cite{uci_energy} (with 8 features), and Communities and Crime~\cite{uci_crime} (with 99 features). We consider architectures with hidden widths ranging from 4 to 128, including both single-hidden-layer and two-hidden-layer BNNs with high predictive accuracy. These model architectures are commonly used in previous works and are suitable for the benchmark tasks~\cite{sample1, sample2}.

\noindent \textit{Baselines.}
We implement the interval domain by replacing the exponential-affine constraints with scalar constraints, keeping all other design and optimization techniques the same as our framework. 
Since constant bounds on unbounded tails are non-integrable, the interval domain
cannot support abstract transformers that require integration over the full support. Therefore, for experimental evaluation, we restrict the setting to bounded supports. We denote the interval-based domain and our abstract domain under this setting as \textit{$\text{Interval}^\star$} and \textit{$\text{Ours}^\star$}, respectively, and refer to the complete version of our domain as \textit{$\text{Ours}$}. We also compare our approach against Monte Carlo sampling. 

\subsubsection{Illustrative Evaluation on Small BNNs}
\begin{figure}[t]
\centering
\begin{subfigure}[b]{0.48\textwidth}
    \centering
    \includegraphics[width=\linewidth]{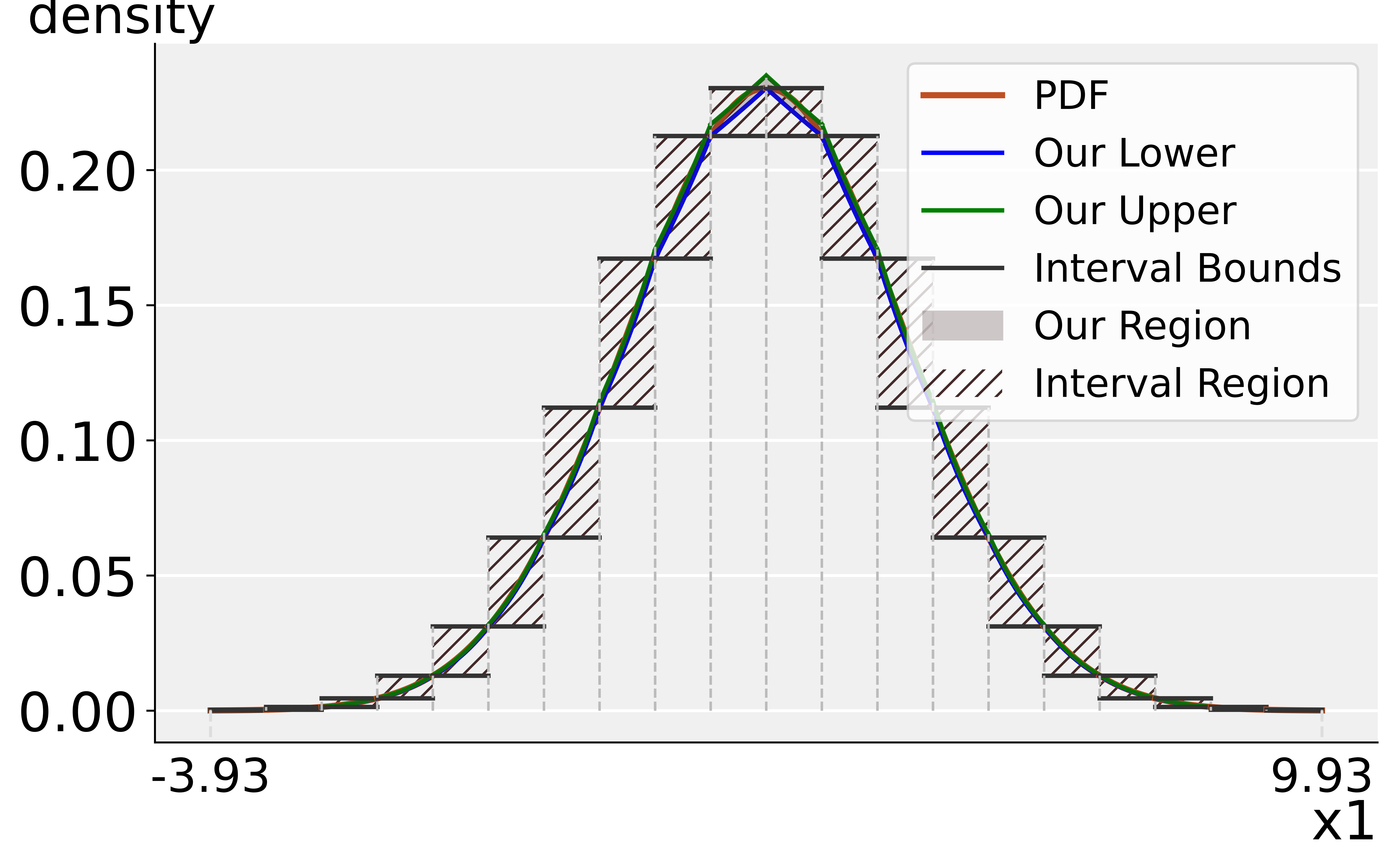}
    \caption{Comparison of abstractions of neuron $x_1$ with different domains.}
    \figlabel{fig:sub1}
\end{subfigure}
\hfill
\begin{subfigure}[b]{0.48\textwidth}
    \centering
    \includegraphics[width=\linewidth]{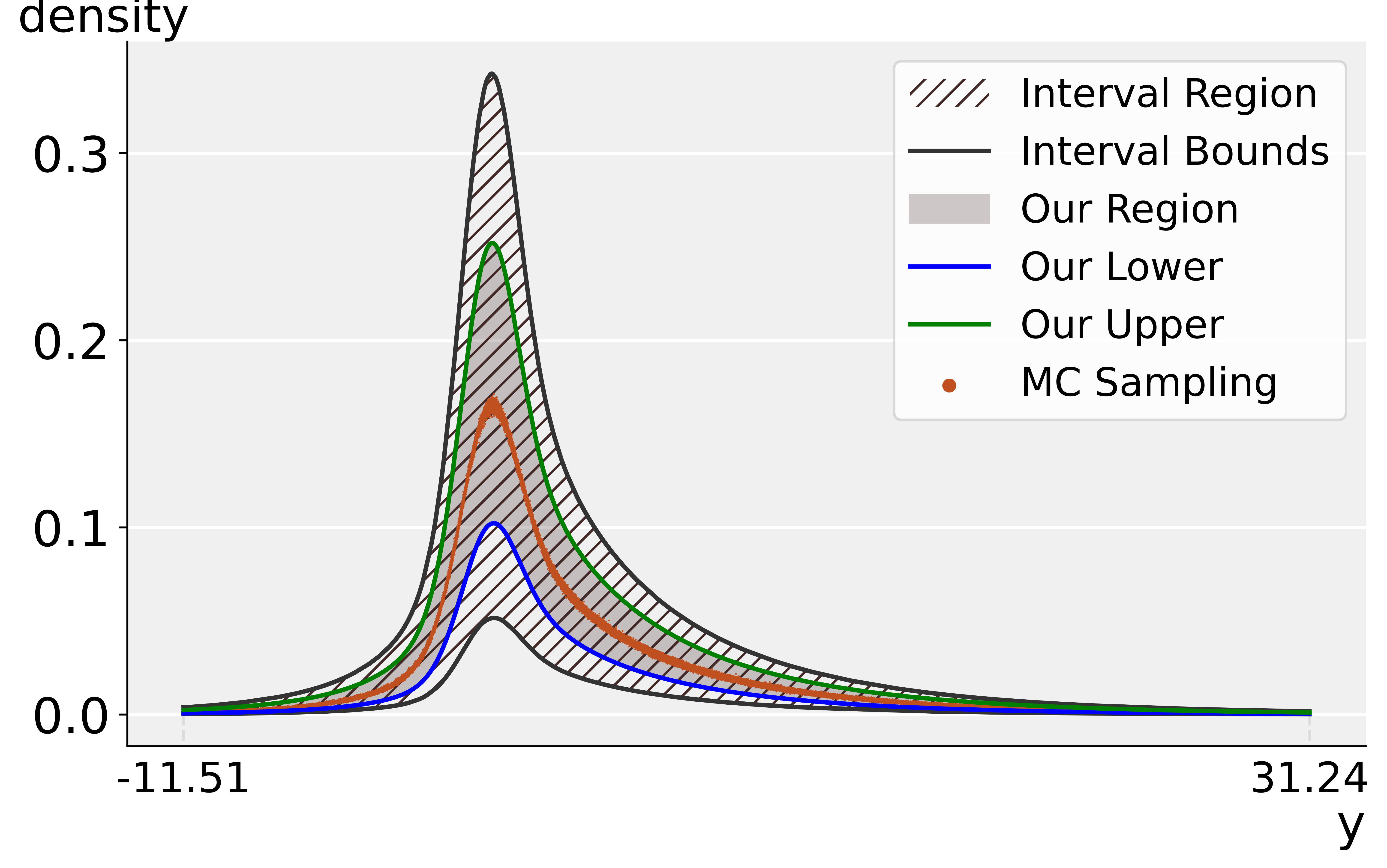}
    \caption{Comparison of abstractions of neuron $y$ with different domains.}
    \figlabel{fig:sub4}
\end{subfigure}
\vspace{-5pt}
\caption{
Comparison between our domain and the interval-based domain on the toy BNN~\figref{fig:bnn} introduced in~\secref{overview}. Full results are provided in~\appref{fulltoymodel1}.
}
\figlabel{fig:toymodel1}
\vspace{-10pt}
\end{figure}
\begin{figure}[t]
    \centering
\includegraphics[width=\linewidth]{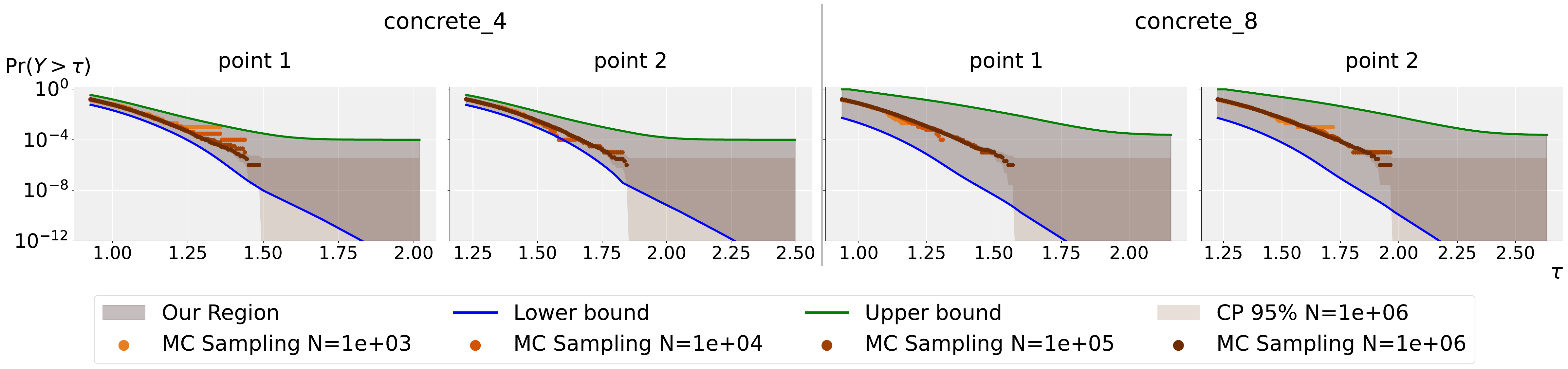}
    \caption{
    Quantifying the probability mass of tail regions for two BNNs (left: 4 hidden units; right: 8 hidden units). Both models are evaluated on the same two test inputs (point 1 and point 2). The figure shows the right-tail probability $\Pr(Y>\tau)$, while the complete results including left-tail probability are shown in~\appref{fulltoymodel1}.
  }
    \figlabel{fig:toymodel2}
    \vspace{-15pt}
\end{figure}
We begin with using toy BNNs to visualize the precision and coverage advantages of our approach.
We first compare \textit{$\text{Ours}^\star$} with \textit{$\text{Interval}^\star$} using the toy BNN introduced in~\secref{overview}. As shown in~\figref{fig:toymodel1}, the exponential-affine bounds of \textit{$\text{Ours}^\star$} capture the shape of the probability distributions more precisely than the constant bounds of \textit{$\text{Interval}^\star$}, resulting in tighter abstractions throughout the network. Full results are provided in~\appref{fulltoymodel1}.

We then apply our approach to tail probability estimation, an important uncertainty quantification task for BNNs. Unlike deterministic DNNs with fixed parameters, BNNs capture parameter uncertainty arising from limited training data, enabling the estimation of probabilities of rare but potentially harmful outcomes, such as $\Pr(Y>\tau)$, where $\tau$ is a predefined risk threshold. Such probabilities are crucial for risk-sensitive decision-making~\cite{uncertaintyapplication1,cancer}, and sound bounds on these probabilities enable reliable estimation without underestimating any true risk. We use two one-hidden-layer ReLU BNNs with 4 and 8 hidden units, trained on the UCI Concrete Compressive Strength dataset~\cite{uci_concrete}. For each network, we evaluate the same two test inputs and compute sound bounds on the right- and left-tail probabilities, $\Pr(Y>\tau)$ and $\Pr(Y\le\tau)$, across different thresholds $\tau$. We compare the bounds produced by our abstract domain with Monte Carlo estimates obtained using $10^3$, $10^4$, $10^5$, and $10^6$ samples. We additionally add a comparison with Clopper--Pearson intervals~\cite{cp}, which provide probability bounds with statistical confidence based on independent sampling, without soundness guarantees. We use a 95\% confidence level with $10^6$ samples. As shown in~\figref{fig:toymodel2}, our domain consistently provides sound probability bounds across different thresholds, with precision decreasing as the network size increases. In contrast, Monte Carlo sampling becomes increasingly ineffective (with zero probability) in extreme tail regions, as rare events can hardly be observed with a finite number of samples. And the Clopper--Pearson confidence intervals yield a zero lower bound when no events are observed. 
These results highlight the advantage of our approach for quantifying low-probability events.

\subsubsection{Evaluation on Larger BNNs}
We then scale up our evaluation to larger BNNs to assess the precision, coverage, and efficiency of our approach through statistics.

\textit{Metrics.}
We use three metrics: precision, coverage, and efficiency. For precision and efficiency, we compare \textit{$\text{Ours}^\star$} with \textit{$\text{Interval}^\star$}. Precision is measured using the standard Total Variation Distances (TVDs) metrics~\cite{AURA}. Given lower and upper density bounds on an approximation with \(k\) partitions, TVD is given by:
\(
\mathrm{TVD}
=
\frac12
\sum_{i=1}^{k}
\int_{l_i}^{u_i}
(U_i(x)-L_i(x))\,dx.
\) Smaller TVD values indicate tighter bounds. For each benchmark, we report the average TVD over 5 test inputs.
We use different partition budgets for different benchmarks to balance precision and computational cost. We additionally report the ratio of the TVDs, which is defined as
\(
\mathrm{Ratio}
=
\frac{\mathrm{TVD}(\text{Interval}^\star)}
{\mathrm{TVD}(\text{Ours}^\star)},
\)
where a larger ratio indicates a more precise abstraction produced by our domain. We measure the runtime of both methods as the end-to-end execution time of the abstract interpreters. For coverage, we compare our approach against MC sampling. 
We follow a histogram-based estimator~\cite{histogram}, measuring the reference probability mass covered by histogram bins visited by sampled outputs. We use $10^6$ bins, an independent reference pool of $8\times10^6$ samples, and Monte Carlo sample sizes of $10^5$ and $10^6$.

\textit{Observations.} 
As shown in~\tabref{tab:prec_bnn}, our framework consistently produces more precise abstractions than the interval domain across all benchmarks, achieving up to a $950\times$ improvement with only modest computational overhead. In terms of coverage, our framework captures the entire output probability mass by construction. In contrast, Monte Carlo sampling cannot capture the complete distribution, with several benchmarks achieving less than 90\% coverage even with $10^6$ samples. 

{
\small
\vspace{-5pt}
\setlength{\tabcolsep}{4pt}
\centering
\begin{longtable}{@{}llrrrrrrrr@{}}
\caption{Precision, probability mass coverage, and runtime comparison across different BNN benchmarks.}
\vspace{-5pt}
\tablabel{tab:prec_bnn}\\

\toprule
\multirow{2}{*}{\textbf{Dataset}}
& \multirow{2}{*}{\textbf{Arch.}}
& \multicolumn{3}{c}{\textbf{Precision}}
& \multicolumn{3}{c}{\textbf{Coverage}}
& \multicolumn{2}{c}{\textbf{Time (s)}} \\
\cmidrule(lr){3-5}
\cmidrule(lr){6-8}
\cmidrule(lr){9-10}
&
&
\textbf{Interval*}
&
\textbf{Ours*}
&
\textbf{Ratio}
&
\textbf{MC($10^5$)}
&
\textbf{MC($10^6$)}
&
\textbf{Ours}
&
\textbf{Interval*}
&
\textbf{Ours*}
\\
\midrule

\multirow{3}{*}{Concrete}
& $1\times32$  & 2874.05 & 15.10 & \textbf{190.32} & 29.05\% & 90.61\% & \textbf{100\%} & 38.71 & 48.30 \\
& $2\times32$  & 1397.41 & 622.29 & 2.25  & 35.36\% & 92.44\% & \textbf{100\%} & 282.82 & 309.46 \\
& $2\times64$  & 757.12  & 573.32 & 	1.32  & 31.43\% & 91.42\% & \textbf{100\%} & 52.02 & 96.39 \\

\midrule

\multirow{6}{*}{Energy}
& $1\times32$  & 3665.16  & 20.45   & \textbf{179.24} & 30.30\% & 91.00\% & \textbf{100\%} & 252.28 & 303.03 \\
& $1\times64$  & 21826.74 & 36.46   & \textbf{598.62} & 29.31\% & 90.77\% & \textbf{100\%} & 126.56 & 233.12 \\
& $1\times128$ & 46514.66 & 48.94   & \textbf{950.43} & 27.10\% & 89.68\% & \textbf{100\%} & 57.87 & 50.07 \\
& $2\times32$  & 298206.33 & 16858.80 & \textbf{17.69} & 37.06\% & 92.78\% & \textbf{100\%} & 394.32 & 459.20 \\
& $2\times64$  & 598.68   & 570.65  & 1.05   & 33.31\% & 92.11\% & \textbf{100\%} & 51.65 & 46.19 \\
& $2\times128$ & 1651.88  & 1541.85 & 1.07   & 32.74\% & 91.93\% & \textbf{100\%}& 32.37 & 36.70 \\

\midrule

\multirow{6}{*}{\shortstack{Communities\\\& Crime}}
& $1\times32$  & 6621.06  & 33.89   & \textbf{195.36} & 27.94\% & 89.68\% & \textbf{100\%} & 104.23 & 140.71 \\
& $1\times64$  & 38996.32 & 305.72  & \textbf{127.56} & 27.67\% & 89.74\% & \textbf{100\%} & 134.95 & 231.36 \\
& $1\times128$ & 1530.69  & 55.30   & \textbf{27.68}  & 27.36\% & 89.79\% & \textbf{100\%} & 28.58 & 17.15 \\
& $2\times32$  & 51223.49 & 38683.77 & 1.32  & 38.10\% & 92.68\% & \textbf{100\%} & 4.95 & 6.49 \\
& $2\times64$  & 2760.06  & 2346.15 & 1.18   & 49.73\% & 95.40\% & \textbf{100\%} & 6.98 & 8.28 \\
& $2\times128$ & 3193.27  & 2942.39 & 1.09   & 58.62\% & 96.01\% & \textbf{100\%} & 21.68 & 24.41 \\

\bottomrule
\end{longtable}
\vspace{-10pt}
}

\subsection{Evaluation on Programs with Bayesian Inference}
\seclabel{sec: eval2}
\textit{Benchmarks.}
We further evaluate our approach on the Bayesian inference benchmarks to demonstrate its generality. We derive our benchmark suite by filtering and simplifying models from prior studies~\cite{InferNET,AURA,psi} as well as representative probabilistic models from other application domains. ~\tabref{tab:benchmarks} in~\appref{benchmarksforbayes} details the resulting benchmarks. Despite simplifications, the benchmarks preserve the core probabilistic inference tasks and remain practically meaningful.

\textit{Baselines.}
We compare our abstract domain against the interval domain and PSI~\cite{psi}. We additionally report the results of our full framework. AURA~\cite{AURA} is excluded from the main comparison as it does not support compositional abstract transformers and its soundness guarantees exclude tail regions.
Nevertheless, we provide a comparison with AURA in~\appref{aura} under a setting that favors AURA, demonstrating that our approach still outperforms it in both precision and runtime.

\textit{Metrics.}
We evaluate both precision and efficiency. For the approximated inference framework, precision is still measured using TVD.
We set the maximum number of partitions to $10^6$.
For the exact inference framework, following the evaluation metrics of PSI~\cite{psi},  we report the simplification status of the resulting symbolic posterior.
Efficiency is measured by inference time.

\textit{Observations.}
~\tabref{tab:bayes_prec} shows that our framework generalizes well to a broad range of Bayesian inference tasks with diverse combinations of prior and likelihood distributions, including heavy-tailed distributions, mixture priors, truncated distributions, and non-conjugate likelihoods, while consistently producing significantly tighter approximations
compared with the interval domain, even when using the full version of our domain. In particular, the interval abstraction becomes excessively conservative for complex posteriors and even fails on the \textit{RegLogistic} benchmark as the partition cost becomes too large to handle, whereas our approach completes the analysis successfully. The comparison with PSI highlights the generality of our framework, as PSI relies on exact symbolic inference and therefore either leaves symbolic integrals unsimplified or fails to support complex probabilistic models. 

{
\small
\vspace{-3pt}
\setlength{\tabcolsep}{2.5pt}
\centering
\begin{longtable}{@{}lllrrrrrrrr@{}}
\caption{
Precision and computation time comparison across benchmarks.
Symbols:
$\mathcal{N}$: Normal;
$\mathcal{N}_t$: truncated Normal;
$\mathcal N_{\mathrm{lin}}$:
Normal with a linear combination of latent variables in the mean;
$\mathcal N_{\mathrm{prod}}$:
Normal whose mean contains products of latent variables;
$\mathcal{S}$: Logistic;
$\mathcal{G}$: Gamma;
$\mathcal{P}$: Pareto;
$\mathcal{T}$: Student-t;
$+$ denotes a mixture distribution;
$^\alpha$ indicates the number of loops;
\cmark\: fully simplified;
\pcmark\: partially simplified;
\pcmark$^{\int}_n$: $n$ remaining symbolic integrals;
\xmark\: unsupported;
$\triangle$: partition explosion.
}
\vspace{-10pt}
\tablabel{tab:bayes_prec}\\[4pt]

\toprule
\multicolumn{3}{c}{\textbf{Benchmark}} &
\multicolumn{4}{c}{\textbf{Precision}} &
\multicolumn{4}{c}{\textbf{Time (s)}} \\

\cmidrule(lr){1-3}
\cmidrule(lr){4-7}
\cmidrule(lr){8-11}

\textbf{Program} &
\textbf{Prior} &
\textbf{Lik.} &
\textbf{Interval*} &
\textbf{Ours*} &
\textbf{Ours} &
\textbf{PSI} &
\textbf{Interval*} &
\textbf{Ours*} &
\textbf{Ours} &
\textbf{PSI} \\
\midrule

Basic
& $\mathcal{N}$ & $\mathcal{N}$
& 0.030377 & \textbf{0.000182} & \textbf{0.001200} & \cmark
& 0.99 & 0.98 & 0.99 & 0.08 \\

WealthEstimation
& $\mathcal{P}$ & $\mathcal{N}$
& 0.003712 & \textbf{0.000006} & \textbf{0.004760} & \pcmark$^{\int}_3$
& 1.00 & 1.14 & 1.71 & 0.06 \\

RobustBayesian
& $\mathcal{G}$ & $\mathcal{T}$
& 0.048138 & \textbf{0.000901} & \textbf{0.000932} & \pcmark$^{\int}_3$
& 0.14 & 0.21 & 0.24 & 0.08 \\

LearningGaussian
& $\mathcal{N}$ & $\mathcal{N}^{20}$
& $1.05\times 10^{105}$ & \textbf{0.011856} & \textbf{0.014730} & \cmark
& 2.27 & 2.51 & 2.90 & 0.52 \\

ClinicalTrial
& $\mathcal{N}+\mathcal{N}$ & $\mathcal{N}^{10}$
& 0.183803 & \textbf{0.000011} & \textbf{0.000037} & \cmark
& 1.56 & 1.67 & 1.34 & 0.36 \\

HumanHeight
& $\mathcal{N}_{t}+\mathcal{N}_{t}$ & $\mathcal{N}^{3}$
& 0.640451 & \textbf{0.005743} & \textbf{0.005743} & \pcmark
& 0.34 & 0.35 & 0.36 & 0.20 \\

RegLogistic
& $\mathcal{N}_{t}$ & $\mathcal{S}^{30}$
& $\triangle$ & \textbf{0.040490} & \textbf{0.040490} & \xmark
& $\triangle$ & 36.58 & 36.15 & \xmark \\

Privacy
& $\mathcal{N}_{\mathrm{lin}}$ & $\mathcal{N}$
& 0.139192 & \textbf{0.000286} & \textbf{0.000325} & \cmark
& 0.13 & 0.17 & 0.17 & 0.07 \\

ProbabilisticPCA
& $\mathcal{N}_{\mathrm{prod}}$ & $\mathcal{N}$
& 0.010338 & \textbf{0.002904} & \textbf{0.002919} & \pcmark$^{\int}_3$
& 28.26 & 15.99 & 13.00 & 0.11 \\

\bottomrule
\end{longtable}
\vspace{-10pt}
}

%% file: tex/09_related.tex
\section{Related Work}
Reasoning about probabilistic programs has been extensively studied using symbolic and approximate inference techniques~\cite{Wonyeol,Holtzen,Atkinson,zaiser2}. PSI~\cite{psi} performs exact symbolic Bayesian inference, while $\lambda$PSI~\cite{lambdapsi} extends it to higher-order and recursive probabilistic programs. However, symbolic inference can become intractable under nonlinear transformations or in high-dimensional settings. In contrast, GuBPI~\cite{gubpi} computes guaranteed posterior bounds for probabilistic programs using interval-based trace semantics and a weight-aware type system. AURA~\cite{AURA} provides sound posterior bounds through gradient-based optimization, supporting a broader class of probabilistic programs.
Both cannot scale to perform probabilistic reasoning on BNNs.
For BNNs, early work~\cite{sample2} studies probabilistic robustness by reasoning over intervals in the posterior weight space and computing certified lower bounds on the probability that a sampled network satisfies a robustness specification.
Subsequent work~\cite{sample1} extends this framework to certify both probabilistic robustness and Bayesian decision robustness through unified reasoning over the posterior weight distribution.
More recently, RAC-BNN~\cite{sample3} introduces risk-aware certification by incorporating Conditional Value-at-Risk (CVaR) into BNN certification, enabling certified robustness analysis under worst-case risk scenarios.
All of these approaches reason about BNNs through sampling.

%% file: tex/10_conclusion.tex
\section{Conclusion}

We present the first nonlinear abstract domain for sound, expressive BNN reasoning. 
Our domain is built upon a novel marginal collecting semantics that represents probability distributions with dependencies abstracted away. 
We further extend the framework to general probabilistic programs. Experimental results demonstrate that our approach consistently achieves substantially higher precision while maintaining decent scalability. 
We believe it provides a promising foundation for probabilistic abstract interpretation.

%% file: tex/appendix.tex
\appendix

\section{Derivation of Marginal Collecting Semantics of Affine and ReLU}
\applabel{deriveaffinerelu}
\paragraph{Derivation of concrete semantics for Affine operation.} Consider the affine transformation $\mathrm{Affine}(\mathbf{x}_J)$, where $\mathbf{x}_J=(x_1,\ldots,x_J), J \in [n]$ denotes the $J$ input program variables. Let $\mathbf{X}_J=(X_1,\ldots,X_J)$ denote their corresponding random variables induced by the probabilistic state $\sigma_p$. The output random variable is given by $Y=\sum_{j=1}^{J}W_jX_j+b$.
The high-level idea behind the concrete semantics is to model the output as a conditional Gaussian distribution followed by marginalization. 
We first fix the input vector
$\mathbf X_J=\mathbf{x}_J=(x_1,\ldots,x_J)\in\mathbb R^J$. Conditioned on $\mathbf X_J=\mathbf x_J$, the only remaining randomness comes from the Gaussian weights and bias. 
Since a linear combination of independent Gaussian random variables is still Gaussian, we have
\(
Y \mid (\mathbf{X}_J = \mathbf{x}_J) \sim \mathcal{N}(m(\mathbf{x}_J), v(\mathbf{x}_J)),
\)
where the conditional mean is
$m(\mathbf{x}_J) =
\mathbb{E}[Y \mid \mathbf{X}_J = \mathbf{x}_J]
=
\sum_{j=1}^{J} \mu_{j} x_j + \mu_{b},
$
and the conditional variance is
$
v(\mathbf{x}_J)
=
\mathrm{Var}(Y \mid \mathbf{X}_J = \mathbf{x}_J)
=
\sum_{j=1}^{J} \sigma_{j}^2 x_j^2 + \sigma_{b}^2.
$
This gives rise to an intermediate conditional Markov kernel
$K_{cond}$:
\begin{equation}
K_{\mathrm{cond}}(\mathbf x_J,dy)
=
\mathcal N(m(\mathbf x_J),v(\mathbf x_J))dy
=
\frac{1}{\sqrt{2\pi v(\mathbf{x}_J)}}
\exp\left(
-\frac{(y-m(\mathbf{x}_J))^2}{2v(\mathbf{x}_J)}
\right)dy.
\eqnlabel{eq:gj}
\end{equation}
Let $\mu_J$ denote the joint probability measure of the random vector $\mathbf{X}_J$. By marginalizing the conditional kernel over the input distribution, we can obtain the output marginalized kernel
\begin{equation}
K(\emptyset,dy)
=
\int_{\mathbb R^J}
K_{\mathrm{cond}}(\mathbf x_J,dy)
\,\mu_J(d\mathbf x_J)
=
\left(
\int_{\mathbb R^J}
\frac{1}{\sqrt{2\pi v(\mathbf x_J)}}
\exp\left(
-\frac{(y-m(\mathbf x_J))^2}{2v(\mathbf x_J)}
\right)
\mu_J(d\mathbf x_J)
\right)dy.
\eqnlabel{eq:py1}
\end{equation}
where
\(
\mu_J(A)
=
\int_{\mathbb R^n}
\mathbf{1}_A(\mathbf{x}_J)\,\mu(d\mathbf{x}),
\
\forall A\in\mathcal B(\mathbb R^J), 
\mu\in \gamma_{c_0}(\sigma_p).
\)

\paragraph{Derivation of concrete semantics for ReLU operation.} Consider the ReLU activation. Let \(Y=\mathrm{ReLU}(X_i)=\max(0,X_i)\), where \(X_i\) is the random variable corresponding to the input program variable \(x_i\). The output kernel is obtained by pushing forward the input probability measure through the ReLU function. Specifically, we derive the output kernel by partitioning the input domain into non-positive and positive regions, where ReLU maps all non-positive values to zero and acts as the identity on positive values:
\begin{equation*}
\begin{aligned}
K(\emptyset,dy)
&=
\int_{\mathbb R}
\delta_{y=\mathrm{ReLU}(x_i)}(dy)\,
\sigma_p(x_i)(\emptyset,dx_i)
\\
&=
\int_{(-\infty, 0]}
\delta_{0}(dy)\,
\sigma_p(x_i)(\emptyset,dx_i)
+
\int_{(0,\infty)}
\delta_{y=x_i}(dy)\,
\sigma_p(x_i)(\emptyset,dx_i)
\\
&=
\sigma_p(x_i)(\emptyset,(-\infty,0])\,
\delta_0(dy)
+
\int_{(0,\infty)}
\delta_{y=x_i}(dy)\,
\sigma_p(x_i)(\emptyset,dx_i)
\end{aligned}
\end{equation*}
where $\delta_{y=v}$ denotes the Dirac measure concentrated at $v \in \mathbb R$, defined by
\(
\delta_{y=v}(A)
=
\begin{cases}
1, & v\in A,\\
0, & v\notin A,
\end{cases}
\ A\in\mathcal B(\mathbb R).
\)

\section{Proof of~\theoremref{galois}}
\applabel{app:proofofgalois}
\begin{proof}
By the definitions of $\alpha_c$ and $\gamma_c$,
\[
\begin{aligned}
\alpha_c(c)\subseteq c'
&\iff
\forall\mu\in c,\ \pi(\mu)\in c'
\iff
\forall\mu\in c, \exists \sigma_p \in c', \pi(\mu) = \sigma_p\\
&\iff
\forall\mu\in c, \exists \sigma_p \in c', \forall i \in [n], \pi(\mu)(x_i) = \sigma_p(x_i)(\emptyset, \cdot)\\
&\iff
\forall\mu\in c, \mu\in
\cup_{\sigma_p\in c'}
\gamma_{c_0}(\sigma_p)
\iff
\forall\mu\in c,
\mu\in\gamma_c(c')
\iff
c\subseteq\gamma_c(c').
\end{aligned}
\]
\end{proof}

\section{Proof of~\theoremref{marginal-soundness}}
\applabel{app:proofofmarginal-soundness}

To establish the soundness theorem that shows that our marginal collecting semantics soundly approximates the joint-measure concrete semantics, we first briefly recall the classical joint-measure semantics~\cite{jointmeasuresemantics1,jointmeasuresemantics2}.
For semantic convenience, we assume that every program variable has an initial probability distribution, allowing us to reason over a fixed $n$-dimensional space. These initial distributions are overwritten by sampling or assignment statements before the corresponding variables are used.

Let $\llbracket\cdot\rrbracket_\mu$ denote the classical joint-measure based concrete semantics of each statement. Similarly, for each statement $s$, we use
$
\llbracket s\rrbracket_\mu^0:
\mathcal U(\mathbb R^n)
\rightarrow
\mathcal U(\mathbb R^n)
$
to denote its semantics over individual joint probability measures. We then lift this semantics to the domain $\mathcal C$ by defining
$
\llbracket s\rrbracket_\mu(c)
=
\left\{
\llbracket s\rrbracket_\mu^0(\mu)
\mid \mu\in c
\right\}
$ for $c \in \mathcal C$,
where
$\llbracket s\rrbracket_\mu:\mathcal C\to\mathcal C$.
The semantics is shown in~\figref{fig:jointsemantics}. For each expression $E$, its semantics is defined as a conditional Markov kernel that maps an $n$-dimensional program valuation to a probability distribution over $\mathbb R$. Formally, we write $\llbracket E\rrbracket_K\in\mathcal K_{\mathrm{cond}}(\mathbb R^n,\mathbb R)$.
Specifically, affine expressions are interpreted as Gaussian kernels with mean
$m(\mathbf{x}_J)=\sum_{j\in J}\mu_jx_j+\mu_b$ and variance $v(\mathbf{x}_J)=\sum_{j\in J}\sigma_j^2x_j^2+\sigma_b^2$, where $\mu_j$ and $\sigma_j^2$ denote the mean and
variance of the weight, and $\mu_b$ and $\sigma_b^2$ denote those of the bias, respectively. ReLU expressions are interpreted as deterministic kernels given by Dirac measures concentrated at $\max(0,x_j)$. 
The Gaussian sampling statement updates the joint probability measure by sampling a new value $y$ for $x_i$ independently from the specified Gaussian distribution $\mathcal N(m,v)$. Here, $(x_1,\ldots,x_{i-1},y,x_{i+1},\ldots,x_n)$ denotes the updated valuation with all other variables remaining unchanged. The resulting joint measure is obtained by integrating over the original joint measure $\mu$ and the Gaussian density.
Similarly, the assignment statement updates the joint probability measure by integrating the conditional Markov kernel $\llbracket E\rrbracket_K$ over the original joint measure $\mu$.
The joint-measure semantics also supports sequential composition by applying the semantics of statements in execution order.
%
\begin{figure}[h]
\centering
\small
\vspace{-15pt}
\[
\begin{aligned}
&\llbracket \mathrm{Affine}(\mathbf{x}_J)
\rrbracket_K
= K,
\text{where}
\
K((x_1, x_2, \cdots, x_n), dy)=
\frac{1}{\sqrt{2\pi v(\mathbf{x}_J)}}
\exp\left(
-\frac{(y-m(\mathbf{x}_J))^2}{2v(\mathbf{x}_J)}
\right)dy,
\\[6pt]
&
\llbracket \mathrm{ReLU}(x_j)
\rrbracket_K
= K,
\text{where}
\
K((x_1, x_2, \cdots, x_n), dy)=
\delta_{\max(0,x_j)}(dy),
\\[6pt]
&
\llbracket x_i\sim\mathcal N(m,v)
\rrbracket_\mu^0(\mu)
=
\mu',
\ \text{where} \ \forall A\in\mathcal B(\mathbb R^n),
\mu'(A)=
\\
&
\quad
\int_{\mathbb R^n}
\int_{\mathbb R}
\mathbf{1}_A(x_1,\cdots, x_{i-1}, y, x_{i+1}, \cdots, x_n)
\frac{1}{\sqrt{2\pi v}}
\exp\left(
-\frac{(y-m)^2}{2v}
\right)dy\,
\mu(d(x_1, \cdots, x_n)),
\\[6pt]
&
\llbracket x_i:=E\rrbracket_\mu^0(\mu)
=
\mu',
\
\text{where} \ \forall A\in\mathcal B(\mathbb R^n), 
\mu'(A)=
\\
&
\quad
\int_{\mathbb R^{n}}
\int_{\mathbb R}
\mathbf{1}_A(x_1, \cdots, x_{i-1}, y, x_{i+1}, \cdots, x_n)
\llbracket E\rrbracket_K
((x_1, \cdots, x_n),dy)\,
\mu(d(x_1,\cdots,x_n)),
\\[6pt]
&
\llbracket s_1;s_2\rrbracket_\mu(c)
=
\llbracket s_2\rrbracket_\mu
\left(
\llbracket s_1\rrbracket_\mu(c)
\right).
\end{aligned}
\]
\vspace{-10pt}
\caption{Joint-measure concrete semantic rules.}
\figlabel{fig:jointsemantics}
\end{figure}

Based on this knowledge, we now prove~\theoremref{marginal-soundness}.

\begin{proof}
Our goal is to prove that $\forall c' \in \mathcal C'$,
\[
\llbracket s\rrbracket_\mu(\gamma_c(c'))
\subseteq
\gamma_c\bigl(\llbracket s\rrbracket_p(c')\bigr).
\]
We prove the theorem by considering each statement in our probabilistic programming language separately.
There are three base statements, including the Gaussian sampling statement, the affine assignment statement, and the ReLU assignment statement, together with sequential composition.
For the three base statements, we prove the theorem in two steps. We first prove that, for any probabilistic store
$\sigma_p\in\Sigma_p$ and any joint probability measure $\mu\in\gamma_{c_0}(\sigma_p)$,
\begin{equation}
\alpha_0\bigl(
\llbracket s\rrbracket_\mu^0(\mu)
\bigr)
\in
\llbracket s\rrbracket_p^0(\sigma_p).
\eqnlabel{eq:single-store-soundness}
\end{equation}
Specifically, we start from an arbitrary joint probability measure, apply the joint-measure semantics, and show that its marginalization is included in the result of our marginal collecting semantics. We then lift the result to $\mathcal C$ and $\mathcal C'$ to establish the desired soundness relation for the three base statements.
Finally, we prove the theorem for sequential composition using induction.

\begin{itemize}[leftmargin=*]
    \item \textbf{Case 1: Gaussian sampling statement}.
Let $s\equiv x_i\sim\mathcal N(m,v)$ and
$\mu'=\llbracket s\rrbracket_\mu^0(\mu)$.
By the joint-measure semantic rule, for any
$A\in\mathcal B(\mathbb R^n)$, we have
\[
\mu'(A)
=
\int_{\mathbb R^n}
\int_{\mathbb R}
\mathbf{1}_A(x_1,\ldots,x_{i-1},y,x_{i+1},\ldots,x_n)
\cdot
\frac{1}{\sqrt{2\pi v}}
\exp\left(-\frac{(y-m)^2}{2v}\right)
dy\,\mu(d\mathbf{x}).
\]
Marginalizing $\mu'$ with respect to $x_i$ gives
\begin{equation}
\pi(\mu')(x_i)=K_i,
\ \text{where} \
K_i(\emptyset,A)
=
\int_A
\frac{1}{\sqrt{2\pi v}}
\exp\left(
-\frac{(y-m)^2}{2v}
\right)dy.
\eqnlabel{case1.1}
\end{equation}
Since the sampling rule preserves the joint distribution defined over all variables except $x_i$, for every $j\neq i$, marginalizing $\mu'$ with respect to $x_j$ gives
\[
\pi(\mu')(x_j)=\pi(\mu)(x_j).
\]
Since $\mu\in\gamma_{c_0}(\sigma_p)$, the marginal distributions of $\mu$ agree with the kernels stored in $\sigma_p$.
Therefore, we have
\begin{equation}
\pi(\mu')(x_j)=\pi(\mu)(x_j)
=\sigma_p(x_j),
\quad \forall j\neq i.
\eqnlabel{case1.2}
\end{equation}
Combining ~\eqnref{case1.1} and \eqnref{case1.2}, we have
\[
\pi(\mu') = \sigma_p', \text{where} \
\sigma_p'(x_j)=
\begin{cases}
K_i, K_i(\emptyset, dx_i)=\frac{1}{\sqrt{2\pi v}}
\exp\left(
-\frac{(x_i-m)^2}{2v}
\right)dx_i& j=i,\\
\sigma_p(x_j), & j\neq i,
\end{cases}
\]
While by the marginal collecting semantic rule, we know
$\{\sigma_p'\}=\llbracket s\rrbracket_p^0(\sigma_p)$. Therefore we have
\[
\pi(\mu')
=\sigma_p'
\in
\{\sigma_p'\}
=
\llbracket s\rrbracket_p^0(\sigma_p).
\]
Thus,~\eqnref{eq:single-store-soundness} holds.

\item \textbf{Case 2: Affine Assignment statement}.
Let $s\equiv x_i:=E \equiv x_i := \mathrm{Affine}(\mathbf{x}_J)$ and
$\mu'=\llbracket s\rrbracket_\mu^0(\mu)$.
By the joint-measure semantic rule, for any
$A\in\mathcal B(\mathbb R^n)$, we have
\[
\begin{aligned}
&\mu'(A)
=
\int_{\mathbb R^n}
\int_{\mathbb R}
\mathbf{1}_A(x_1,\ldots,x_{i-1},y,x_{i+1},\ldots,x_n)\cdot
\llbracket E\rrbracket_K(\mathbf{x},dy)
\,\mu(d\mathbf{x}),
\\
&\llbracket E\rrbracket_K(\mathbf{x},dy)
=
\frac{1}{\sqrt{2\pi v(\mathbf{x}_J)}}
\exp\left(
-\frac{(y-m(\mathbf{x}_J))^2}
{2v(\mathbf{x}_J)}
\right)dy.
\end{aligned}
\]
Marginalizing $\mu'$ with respect to $x_i$ gives
\begin{equation}
\begin{aligned}
\pi(\mu')(x_i)= K_i,
\quad
K_i(\emptyset, A)
&=
\int_{\mathbb R^n}
\llbracket E\rrbracket_K
(\mathbf{x},A)
\,\mu(d\mathbf{x})\\
&=
\int_{\mathbb R^n}
\int_A
\frac{1}{\sqrt{2\pi v(\mathbf{x}_J)}}
\exp\left(
-\frac{(y-m(\mathbf{x}_J))^2}
{2v(\mathbf{x}_J)}
\right)
dy\,\mu(d\mathbf{x})\\
&=
\int_{\mathbb R^{|J|}}
\int_A
\frac{1}{\sqrt{2\pi v(\mathbf{x}_J)}}
\exp\left(
-\frac{(y-m(\mathbf{x}_J))^2}
{2v(\mathbf{x}_J)}
\right)
dy\,\mu_J(d\mathbf{x}_J).
\end{aligned}
\eqnlabel{case2.1}
\end{equation}
for every $A\in\mathcal B(\mathbb R)$, where $\mu_J$ is the joint measure of the input variables $\mathbf{x}_J$.
Since the assignment rule preserves the joint distribution defined over all variables except $x_i$, for every $j\neq i$, marginalizing $\mu'$ with respect to $x_j$ gives
\[
\pi(\mu')(x_j)=\pi(\mu)(x_j).
\]
Since $\mu\in\gamma_{c_0}(\sigma_p)$, the marginal distributions of $\mu$ agree with the kernels stored in $\sigma_p$.
Therefore, we have
\begin{equation}
\pi(\mu')(x_j)=\pi(\mu)(x_j)
=\sigma_p(x_j),
\quad \forall j\neq i.
\eqnlabel{case2.2}
\end{equation}
Combining ~\eqnref{case2.1} and \eqnref{case2.2}, we have
\[
\pi(\mu') = \sigma_p', \text{where} \
\sigma_p'(x_j)=
\begin{cases}
K_i, K_i(\emptyset, dx_i)
=
\displaystyle\int_{\mathbb R^{|J|}}
\frac{1}{\sqrt{2\pi v(\mathbf{x}_J)}}
\exp\left(
-\frac{(x_i-m(\mathbf{x}_J))^2}
{2v(\mathbf{x}_J)}
\right)
\mu_J(d\mathbf{x}_J)\,dx_i
& j=i,\\
\sigma_p(x_j), & j\neq i,
\end{cases}
\]
While by the marginal collecting semantic rule, we know
$\sigma_p'\in\llbracket s\rrbracket_p^0(\sigma_p)$. Therefore we have
\[
\pi(\mu')
= \sigma_p'
\in
\llbracket s\rrbracket_p^0(\sigma_p).
\]
Thus,~\eqnref{eq:single-store-soundness} holds.

\item \textbf{Case 3: ReLU Assignment Statement.}
Let $s\equiv x_i:=E \equiv x_i:=\mathrm{ReLU}(x_j)$ and $\mu'=\llbracket s\rrbracket_\mu^0(\mu)$.
By the joint-measure semantic rule, for any
$A\in\mathcal B(\mathbb R^n)$, we have
\[
\begin{aligned}
&\mu'(A)=
\int_{\mathbb R^n}
\int_{\mathbb R}
\mathbf{1}_A(x_1,\ldots,x_{i-1},y,x_{i+1},\ldots,x_n)\cdot
\llbracket E\rrbracket_K(\mathbf{x},dy)
\,\mu(d\mathbf{x}),
\\
&
\llbracket E\rrbracket_K(\mathbf{x},dy)
=
\delta_{\max(0,x_j)}(dy).
\end{aligned}
\]
Marginalizing $\mu'$ with respect to $x_i$ gives
\begin{equation}
\begin{aligned}
\pi(\mu')(x_i)
=K_i,
\text{where} \
K_i(\emptyset,A)
&=
\int_{\mathbb R^n}
\llbracket E\rrbracket_K(\mathbf{x},A)
\,\mu(d\mathbf{x})\\
&=
\int_{\mathbb R^n}
\delta_{\max(0,x_j)}(A)
\,\mu(d\mathbf{x})\\
&=
\int_{\mathbb R}
\delta_{\max(0,x_j)}(A)
\,\pi(\mu)(x_j)(\emptyset,dx_j)\\
&=
\int_{\mathbb R}
\delta_{\max(0,x_j)}(A)
\,\sigma_p(x_j)(\emptyset,dx_j)\\
&=
\sigma_p(x_j)(\emptyset,(-\infty,0])
\delta_0(A)+
\int_{(0,\infty)}
\delta_{x_j}(A)\,
\sigma_p(x_j)(\emptyset,dx_j).
\end{aligned}
\eqnlabel{case3.2}
\end{equation}
$\forall A\in\mathcal B(\mathbb R).$
Since the assignment rule preserves the joint distribution defined over all variables except $x_i$, for every $k\neq i$, marginalizing $\mu'$ with respect to $x_k$ gives
\[
\pi(\mu')(x_k)
=
\pi(\mu)(x_k).
\]
Since $\mu\in\gamma_{c_0}(\sigma_p)$, the marginal distributions of $\mu$ agree with the kernels stored in $\sigma_p$.
Therefore, we have
\begin{equation}
\pi(\mu')(x_k)
=
\pi(\mu)(x_k)
=
\sigma_p(x_k),
\quad \forall k\neq i.
\eqnlabel{case3.1}
\end{equation}

Combining~\eqnref{case3.1} and~\eqnref{case3.2}, we have
\[
\pi(\mu')=\sigma_p',
\]
\[
\sigma_p'(x_k)=
\begin{cases}
K_i, K_i(\emptyset,dx_i)
=
\sigma_p(x_j)(\emptyset,(-\infty,0])
\delta_0(dx_i)+
\displaystyle\int_{(0,\infty)}
\delta_{x_j}(dx_i)\,
\sigma_p(x_j)(\emptyset,dx_j),
& k=i,\\
\sigma_p(x_k), & k\neq i,
\end{cases}
\]
By the marginal collecting semantic rule, we know $\{\sigma_p'\}=\llbracket s\rrbracket_p^0(\sigma_p)$.
Therefore, we have
\[
\pi(\mu')
=
\sigma_p'
\in
\{\sigma_p'\}
=
\llbracket s\rrbracket_p^0(\sigma_p).
\]
Thus,~\eqnref{eq:single-store-soundness} holds.

\end{itemize}
We now lift the result to $\mathcal C'$ for the three base cases. Let $c'\in\mathcal C'$. We prove $\llbracket s\rrbracket_\mu(\gamma_c(c'))
\subseteq
\gamma_c\bigl(\llbracket s\rrbracket_p(c')\bigr)$ by proving each element
$\mu'\in \llbracket s\rrbracket_\mu(\gamma_c(c'))$ also belongs to $\gamma_c\bigl(\llbracket s\rrbracket_p(c')\bigr)$.
For any $\mu'\in \llbracket s\rrbracket_\mu(\gamma_c(c'))$, by the definition of collecting semantics, there exists $\mu\in\gamma_c(c')$ such that
\[
\mu'=\llbracket s\rrbracket_\mu^0(\mu).
\]
By the definition of $\gamma_c$, there exists
$\sigma_p\in c'$ satisfying
$\mu\in\gamma_{c_0}(\sigma_p)$.
Therefore by~\eqnref{eq:single-store-soundness},
\[
\alpha_0(\mu')
=\alpha_0(\llbracket s\rrbracket_\mu^0(\mu))
\in
\llbracket s\rrbracket_p^0(\sigma_p)
\subseteq
\llbracket s\rrbracket_p(c').
\]
By the Galois connection between $\gamma_c$ and $\alpha_c$, this implies
\[
\mu'
\in
\gamma_c\bigl(\llbracket s\rrbracket_p(c')\bigr).
\]
Therefore,
\[
\llbracket s\rrbracket_\mu(\gamma_c(c'))
\subseteq
\gamma_c\bigl(\llbracket s\rrbracket_p(c')\bigr).
\]
Thus, the theorem holds for the three base cases.

\begin{itemize}[leftmargin=*]
\item \textbf{Case 4: Sequential Composition.}
Let $s\equiv s_1;s_2$ and $c'\in\mathcal C'$.
By the joint-measure semantic rule, we have
\begin{equation}
\llbracket s_1;s_2\rrbracket_\mu
(\gamma_c(c'))
=
\llbracket s_2\rrbracket_\mu
\left(
\llbracket s_1\rrbracket_\mu(\gamma_c(c'))
\right).
\eqnlabel{case4.1}
\end{equation}

Since the theorem holds for the three base cases, by induction
we have
\begin{equation*}
\llbracket s_1\rrbracket_\mu(\gamma_c(c'))
\subseteq
\gamma_c\left(
\llbracket s_1\rrbracket_p(c')
\right).
\end{equation*}
\begin{equation}
\llbracket s_2\rrbracket_\mu
\left(
\gamma_c\left(
\llbracket s_1\rrbracket_p(c')
\right)
\right)
\subseteq
\gamma_c\left(
\llbracket s_2\rrbracket_p
\left(
\llbracket s_1\rrbracket_p(c')
\right)
\right).
\eqnlabel{case4.2}
\end{equation}

Combining~\eqnref{case4.1} and~\eqnref{case4.2}, we have
\[
\begin{aligned}
&
\llbracket s_1;s_2\rrbracket_\mu
(\gamma_c(c'))=
\llbracket s_2\rrbracket_\mu
\left(
\llbracket s_1\rrbracket_\mu
(\gamma_c(c'))
\right)\subseteq
\llbracket s_2\rrbracket_\mu
\left(
\gamma_c\left(
\llbracket s_1\rrbracket_p(c')
\right)
\right)\subseteq
\gamma_c\left(
\llbracket s_2\rrbracket_p
\left(
\llbracket s_1\rrbracket_p(c')
\right)
\right)\\
&=
\gamma_c\left(
\llbracket s_1;s_2\rrbracket_p(c')
\right).
\end{aligned}
\]
Thus, the theorem holds for sequential composition.
\end{itemize}
\end{proof}

\section{Complete Abstraction Function and Performance Evaluation}
\applabel{app:abstraction}
In this section, we introduce the complete abstraction function work for general continuous distributions in our framework.
We first show the derivation, and then demonstrate the performance of the abstraction function by using the interval-based abstract domain as a baseline. For clarity, we simplify the notation here and fix an arbitrary density function $f$ and let $\phi(x) = \log f(x)$, $x\in [l,u]$.

\noindent\textbf{(1) Case 1: Log-Concave.}
If $\mu$ is log-concave on the interval $[l,u]$, i.e., 
\[
\phi\bigl(\lambda a+(1-\lambda)b\bigr)
\;\ge\;
\lambda\phi(a)+(1-\lambda)\phi(b),
\quad \forall a,b\in [l, u], \;\forall \lambda \in [0,1].
\]
In this case, the lower bound of $\phi$ is given by the secant line over $[l, u]$, and the upper bound is given by a tangent line at some $x_0 \in [l, u]$. Proof is provided in ~\appref{app: proof}. Therefore we have
\[
a^{\le} = \frac{\phi(u)-\phi(l)}{u-l}, \quad b^{\le} = \frac{u\,\phi(l)-l\,\phi(u)}{u-l}, \quad
a^{\ge}=g,
\qquad
b^{\ge}=\phi(x_0)-g\,x_0.
\]
where $g\in\partial \phi(x_0)$.

\noindent\textbf{(2) Case 2: Log-Convex.}
If $\mu$ is log-convex on the interval $[l,u]$, i.e., 
\[
\phi\bigl(\lambda a+(1-\lambda)b\bigr)
\;\le\;
\lambda\phi(a)+(1-\lambda)\phi(b),
\quad \forall a,b\in [l, u], \;\forall \lambda \in [0,1].
\]
In this case, the lower bound is given by a tangent line at some $x_0 \in [l, u]$, and the upper bound of $\phi$ is given by the secant line over $[l, u]$. Proof can be found in ~\appref{app: proof}. Therefore,
\[
a^{\le}=g,
\quad
b^{\le}=\phi(x_0)-g\,x_0, \quad
a^{\ge} = \frac{\phi(u)-\phi(l)}{u-l}, \quad b^{\ge} = \frac{u\,\phi(l)-l\,\phi(u)}{u-l}.
\]
where $g\in\partial \phi(x_0)$.

\noindent\textbf{(3) Case 3: General Case.}
If $\mu$ violates both log-concavity and log-convexity on the interval $[l,u]$, in such cases, we adopt a fallback strategy that constructs constant bounds using extremal values of the density, which can be regarded as a special case of exponential-affine functions with zero slope. Formally, for $x \in [l,u]$,
\begin{equation*}
a^{\le}= 0,
\quad
b^{\le} = log\big( \inf_{x\in[l,u]}\phi(x) \big),
\quad
a^{\ge}= 0,
\quad
b^{\le} = log\big( \sup_{x\in[l,u]}\phi(x) \big),
\end{equation*}


We next describe the abstraction of tail regions. Unlike bounded intervals, unbounded intervals require special treatment, since a secant line cannot be constructed over an interval with an infinite endpoint. Therefore, we adopt constant bounds as a fallback when a secant line is required. Specifically, for the upper bound, we use the maximum density value over the tail interval. 
For the lower bound, we conservatively use zero.


We demonstrate the precision of our abstraction function by applying it to three representative classes of probability distributions, including log-concave distributions, log-convex distributions, and distributions that are neither globally log-concave nor log-convex, and comparing the resulting abstractions with those of the Interval domain, as shown in \figref{fig:comparison}, which shows that our approach produces substantially tighter abstractions while remaining sound.

\begin{figure}[h]
    \centering

    \begin{subfigure}{0.32\linewidth}
        \centering
        \includegraphics[width=\linewidth]{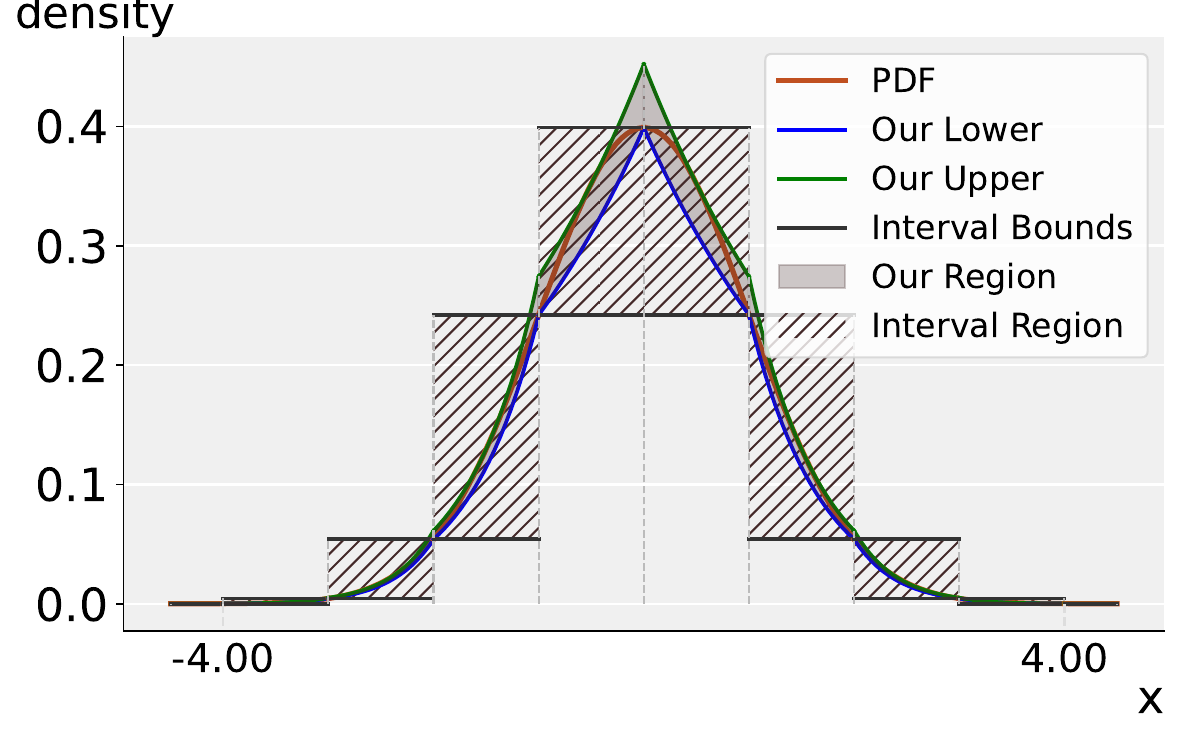}
        \caption{Abstraction of $\mathrm{Normal}(0,1)$ on $[-4,4]$.}
        \figlabel{fig:comp_normal}
    \end{subfigure}
    \hfill
    \begin{subfigure}{0.32\linewidth}
        \centering
        \includegraphics[width=\linewidth]{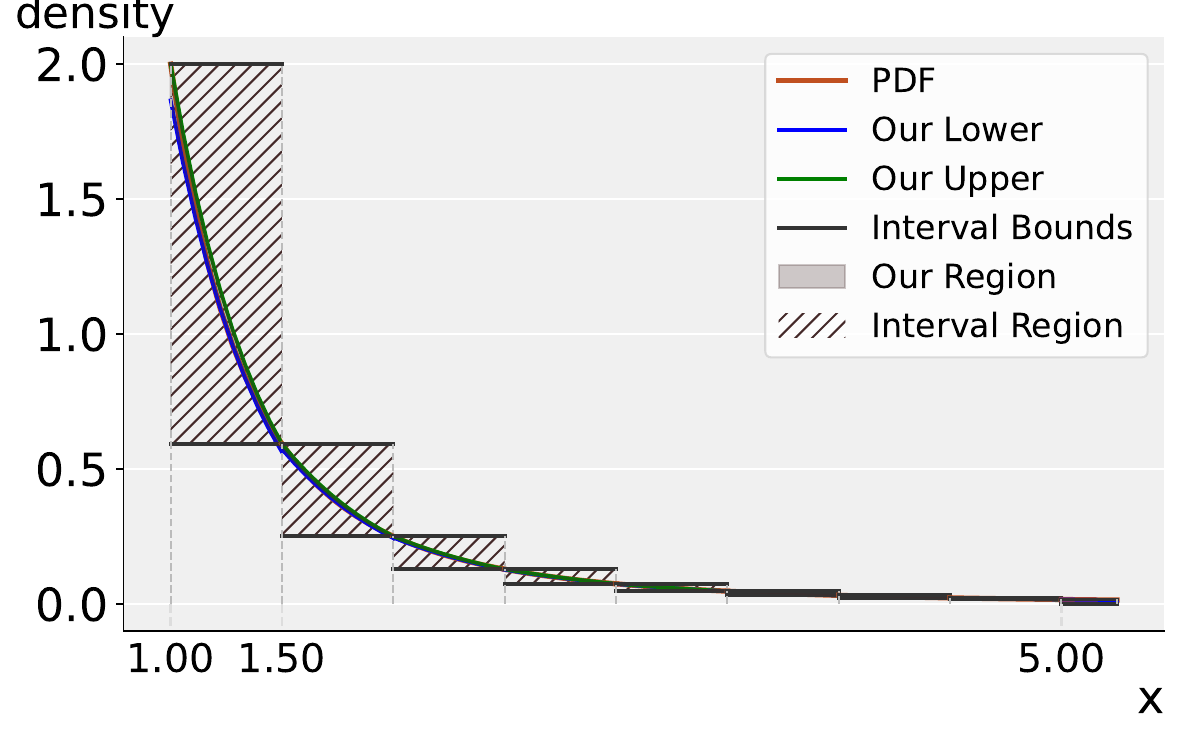}
        \caption{Abstraction of $\mathrm{Pareto}(x_m=1,\alpha=2)$ on $[1,5]$.}
        \figlabel{fig:comp_pareto}
    \end{subfigure}
    \hfill
    \begin{subfigure}{0.32\linewidth}
        \centering
        \includegraphics[width=\linewidth]{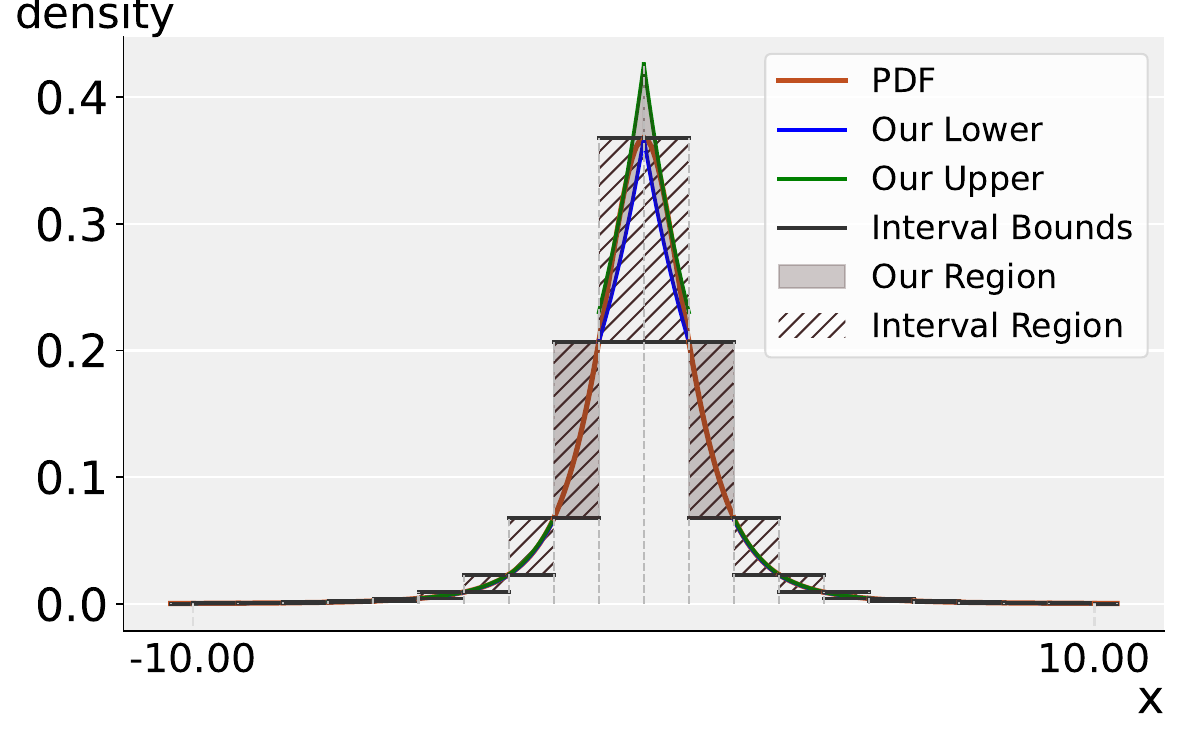}
        \caption{Abstraction of $\mathrm{Student}\text{-}t(v=3)$ on $[-10,10]$.}
        \figlabel{fig:comp_studentt}
    \end{subfigure}

    \caption{Comparison of abstraction based on our domain and the Interval domain, separately, on the same partition of the support. The three examples represent a log-concave distribution (Normal), a log-convex distribution (Pareto), and a distribution that is neither globally log-concave nor log-convex (Student's $t$). 
    }
    \figlabel{fig:comparison}
\end{figure}

\section{Proving Soundness of Exponential-Affine Bounds}
\applabel{app: proof}
In this section, we provide proofs for the sound construction of exponential-affine bounds. In particular, we show how sound affine lower and upper bounds on $\phi=\log \mu$ can be established under different geometric properties of $\phi$.

\noindent\textbf{(1) Case 1: Log-Concave.}
\begin{proposition}[Secant lower bound for concave $\phi$]
\label{prop:secant-lower}
Let $\phi$ be concave on $[l,u]$ with $l<u$. Define the secant line $a^{\le}x+b^{\le}$, where
\[
a^{\le} = \frac{\phi(u)-\phi(l)}{u-l}, \quad b^{\le} = \frac{u\,\phi(l)-l\,\phi(u)}{u-l}.
\]
Then for all $x\in[l,u]$, $
\phi(x)\;\ge\;a^{\le}x+b^{\le}.
$
\end{proposition}

\begin{proof}
Take any $x\in[l,u]$, let $\lambda = \frac{u-x}{u-l}\in[0,1]$, then we have $x = \lambda l + (1-\lambda)u$.
By concavity of $\phi$, we have
$
\phi(x) \;\ge\; \lambda\phi(l)+(1-\lambda)\phi(u).
$
The right-hand side is exactly the value of the secant line through
$(l,\phi(l))$ and $(u,\phi(u))$ evaluated at $x$, i.e.,
$\lambda\phi(l)+(1-\lambda)\phi(u)=a^{\le}x+b^{\le}$.
\end{proof}

\begin{lemma}[Subgradient inequality]
\label{lem:subgrad}
Let $\phi$ be concave on an interval and let $x_0$ be any point in its interior.
For any subgradient $g\in\partial \phi(x_0)$, we have for all $x$,
\[
\phi(x)\;\le\;\phi(x_0)+g(x-x_0).
\]
\end{lemma}

\begin{proof}
This is the standard supporting-hyperplane property of concave functions:
a subgradient at $x_0$ defines a global affine upper bound.
\end{proof}

\begin{proposition}[Tangent-line upper bound for concave $\phi$]
\label{prop:tangent-upper}
Let $\phi$ be concave on $[l,u]$ with $l<u$. $x_0 \in [l,u].$
Choose any $g\in\partial\phi(x_0)$ and define $a^{\ge}x+b^{\ge}$, where
\[
a^{\ge}=g,
\qquad
b^{\ge}=\phi(x_0)-g\,x_0.
\]
Then for all $x\in[l,u]$, $
\phi(x)\;\le\;a^{\ge}x+b^{\ge}.
$
\end{proposition}

\begin{proof}
Apply Lemma~\ref{lem:subgrad} and rearrange:
$\phi(x)\le \phi(x_0)+g(x-x_0)=g x +(\phi(x_0)-g x_0)$.
\end{proof}

Specifically, for the upper bound,
if $\phi$ is differentiable at $x_0$, then $\partial\phi(x_0)=\{\phi'(x_0)\}$
and we can take $g=\phi'(x_0)$, i.e.,
\[
a^{\ge}=\phi'(x_0),
\qquad
b^{\ge}=\phi(x_0)-\phi'(x_0)x_0,
\]
where we define $x_0=\tfrac{l+u}{2}$ (midpoint).

\noindent\textbf{(2) Case 2: Log-Convex.}
\begin{proposition}[Secant upper bound for convex $\phi$]
\label{prop:secant-upper}
Let $\phi$ be convex on $[l,u]$ with $l<u$. Define the secant line $a^{\ge}x+b^{\ge}$, where
\[
a^{\ge} = \frac{\phi(u)-\phi(l)}{u-l}, \quad b^{\ge} = \frac{u\,\phi(l)-l\,\phi(u)}{u-l}.
\]
Then for all $x\in[l,u]$, $
\phi(x)\;\le\;a^{\ge}x+b^{\ge}.
$
\end{proposition}

\begin{proof}
Take any $x\in[l,u]$, let $\lambda = \frac{u-x}{u-l}\in[0,1]$, then we have $x = \lambda l + (1-\lambda)u$.
By convexity of $\phi$, we have
$
\phi(x) \;\le\; \lambda\phi(l)+(1-\lambda)\phi(u).
$
The right-hand side is exactly the value of the secant line through
$(l,\phi(l))$ and $(u,\phi(u))$ evaluated at $x$, i.e.,
$\lambda\phi(l)+(1-\lambda)\phi(u)=a^{\ge}x+b^{\ge}$.
\end{proof}

\begin{lemma}[Subgradient inequality]
\label{lem:subgrad-convex}
Let $\phi$ be convex on an interval and let $x_0$ be any point in its interior.
For any subgradient $g\in\partial \phi(x_0)$, we have for all $x$,
\[
\phi(x)\;\ge\;\phi(x_0)+g(x-x_0).
\]
\end{lemma}

\begin{proof}
This is the standard supporting-hyperplane property of convex functions:
a subgradient at $x_0$ defines a global affine lower bound.
\end{proof}

\begin{proposition}[Tangent-line lower bound for convex $\phi$]
\label{prop:tangent-lower}
Let $\phi$ be convex on $[l,u]$ with $l<u$. $x_0 \in [l,u].$
Choose any $g\in\partial\phi(x_0)$ and define $a^{\le}x+b^{\le}$, where
\[
a^{\le}=g,
\qquad
b^{\le}=\phi(x_0)-g\,x_0.
\]
Then for all $x\in[l,u]$, $
\phi(x)\;\ge\;a^{\le}x+b^{\le}.
$
\end{proposition}

\begin{proof}
Apply Lemma~\ref{lem:subgrad-convex} and rearrange:
$\phi(x)\ge \phi(x_0)+g(x-x_0)=g x +(\phi(x_0)-g x_0)$.
\end{proof}

Specifically, for the lower bound,
if $\phi$ is differentiable at $x_0$, then $\partial\phi(x_0)=\{\phi'(x_0)\}$
and we can take $g=\phi'(x_0)$, i.e.,
\[
a^{\le}=\phi'(x_0),
\qquad
b^{\le}=\phi(x_0)-\phi'(x_0)x_0,
\]
where we define $x_0=\tfrac{l+u}{2}$ (midpoint).

\section{Aggregation Strategy of Abstraction Function}
\applabel{app:aggre}
The notation of this section follows that in \secref{sec:abstraction}.

After constructing an abstraction for each individual function $f_i^{j}\in\Pi_i$, we construct a unified abstraction $a_i$ that soundly encloses every function in \(\Pi_i\) by finding an upper bound for all individual upper bounds and a lower bound for all individual lower bounds on each partition.
The process consists of two steps. First, the interval partitions of the individual abstractions are refined into a common partition $\{[l_i^k,u_i^k]\}_{k\in[m_i]}$.
This refinement follows the requirement that, on every resulting refined interval, there always exists one individual lower bound that attains the minimum among all lower bounds, and one individual upper bound that attains the maximum among all upper bounds.
Second, the final lower and upper bounds \(L_i^k(x)\) and \(U_i^k(x)\) on each refined subinterval are constructed such that, for every refined interval
\(
[l_i^k,u_i^k],
\)
\[
L_i^k(x)
\le
f_i^{j}(x)
\le
U_i^k(x),
\quad \forall f_i^{j}\in\Pi_i
\quad
\forall x\in[l_i^k,u_i^k].
\]
We now prove that such a unified abstraction can always be constructed. 

\begin{theorem}[Closure under Aggregation]
Let
\(
a_1,\ldots,a_t
\)
be abstractions of
\(
\mu_1,\ldots,\mu_t.
\)
There exists a finite refinement of their interval partitions such that, on every refined subinterval, one lower bound is minimal among all lower bounds, and one upper bound is maximal among all upper bounds. 
Consequently, the aggregated abstraction is again an element of the
our domain.
\end{theorem}
\begin{proof}
We first consider bounded intervals.
Each lower or upper bound is an exponential-affine function of the form
\(
e^{ax+b}.
\)
For any two exponential-affine bounds
\(
e^{a_1x+b_1}
\)
and
\(
e^{a_2x+b_2},
\)
their intersection satisfies
$
a_1x+b_1=a_2x+b_2,
$
and therefore they intersect at most once. Since only finitely many exponential-affine bounds are involved, the number of pairwise intersections is finite. Refining the partition using all such intersection points yields
finitely many subintervals. Within each refined subinterval, the ordering of all lower and upper
bounds is fixed. 
Hence, on each refined subinterval, one lower bound is minimal among all lower bounds, and one upper bound is maximal among all upper bounds.
Both remain exponential-affine, implying that the aggregated abstraction over bounded intervals belongs to the our abstract domain.

For tail intervals, every individual upper bound is either an exponential-affine function or a positive constant, where the latter can be viewed as a special case of an exponential-affine function with zero slope. Hence, the same refinement argument as for bounded intervals applies to the upper bounds.
For the lower bounds, the only special case is the constant function \(0\). If at least one individual abstraction has a lower bound \(L(x)\equiv0\), then the aggregated lower bound is also \(0\), since no other lower bound can be smaller. Otherwise, all lower bounds are positive exponential-affine functions, and the same refinement argument as above applies.
\end{proof}

\section{Soundness Proof of ReLU Abstract Transformer}
\applabel{soundrelu}
\begin{proof}
$\forall a \in \mathcal{A}_n$ we have
\[
\begin{aligned}
&T_f(\gamma_n(a))
=
\left\{
T_f(\mathbf K)
\mid
\mathbf K \in\gamma_n(a)
\right\}
\\
&=
\left\{
T_f(\mathbf K)
\;\middle|\;
\mathbf K=(K_1,\ldots,K_n)\in\mathcal K^n,
\forall t\in[n],\,
\forall A\in\mathcal B(\mathbb R),
\int_L^\#(a_t,A)
\leq K_t(\emptyset,A)
\leq \int_U^\#(a_t,A)
\right\}\\
&=\left\{
T_f(K_1, K_2,\ldots, K_n) \mid
\forall t\in[n],\,
\forall A\in\mathcal B(\mathbb R),
\int_L^\#(a_t,A)
\leq K_t(\emptyset,A)
\leq \int_U^\#(a_t,A)
\right\}\\
&=\{
(K_1, K_2, \dots, K_{i-1}) \land (K_i = \text{ReLU}(K_j))\land (K_{i+1},\ldots, K_n)
\mid 
\forall t \in [n], t\neq i, 
\forall A\in\mathcal B(\mathbb R),
\\
&
\int_L^\#(a_t,A)
\leq K_t(\emptyset,A)
\leq \int_U^\#(a_t,A)
\}
\end{aligned}
\]
That is, $T_f$ updates only the $i$-th variable through the concrete semantics of the ReLU transformation, while
all other variables remain unchanged. 

And we have
\[
\begin{aligned}
&\gamma_n(T_f^\sharp(a))= \gamma_n(a')\\
&=(\gamma(a_1^\prime), \ldots,\gamma(a_{i-1}^\prime),\gamma(a_{i}^\prime),\gamma(a_{i+1}^\prime),\ldots, \gamma(a_{n}^\prime))\\
&=(\gamma(a_1), \ldots,\gamma(a_{i-1}),\gamma(a_{i}^\prime),\gamma(a_{i+1}),\ldots, \gamma(a_{n}))\\
&=\{(K_1,\ldots,K_{i-1})\land \gamma(a_i^\prime)\land(K_{i+1},\ldots, K_n)
\mid 
\forall t \in [n], t\neq i, 
\forall A\in\mathcal B(\mathbb R),
\int_L^\#(a_t,A)
\leq K_t(\emptyset,A)
\leq \int_U^\#(a_t,A)
\}\\
\end{aligned}
\]
Therefore, if we want to prove that $\forall a \in \mathcal{A}_n,T_f(\gamma_n(a))= \gamma_n(T_f^\sharp(a))$, we only need to prove that \[
\{K_i = \text{ReLU}(K_j) \mid \forall A\in\mathcal B(\mathbb R),
\int_L^\#(a_j,A) \leq K_j(\emptyset,A)
\leq \int_U^\#(a_j,A)\} \sqsubseteq_C  \gamma(a_i^\prime), \forall a \in \mathcal{A}_n.
\]
Let $LHS$ and $RHS$ denote the left-hand side and right-hand side of this inclusion, respectively.
As $RHS=\gamma(a_i^\prime) = \{K_i^\prime \mid \forall A\in\mathcal B(\mathbb R), \int_L^\#(a_i',A) \leq K_i'(\emptyset,A) \leq \int_U^\#(a_i',A)\}$.
We consider the three cases of $a_i^\prime$, respectively.

\noindent\textbf{Case 1: \(u_j^{m_j}\le0\).}
We have 
\[
a_i^\prime=\{(l_i^{\prime k},u_i^{\prime k},L_i^{\prime k},U_i^{\prime k})\}_{k\in[m_i']}
=
\left\{
\left(
0,\,
0,\,
\int_L^\#(a_j,(-\infty,0]),\,
    \int_U^\#(a_j,(-\infty,0])
\right)
\right\}.
\]
Therefore
\[
\begin{aligned}
RHS
&=\gamma(a_i^\prime)\\
&=
\left\{
K_i^\prime\in\mathcal K
\;\middle|\;
\forall A\in\mathcal B(\mathbb R),
\int_L^\#(a_j,A)
\leq K_i^\prime(\emptyset,A)
\leq \int_U^\#(a_j,A)
\right\}\\
&=
\{
K_i^\prime\in\mathcal K
\;|\;
\forall A\in\mathcal B(\mathbb R),\
0\in A,
\displaystyle
\int_L^\#(a_j,(-\infty,0])
\leq K_i^\prime(\emptyset,A)
\leq \int_U^\#(a_j,(-\infty,0])
\\
&
\land 
\forall A\in\mathcal B(\mathbb R),\
0\notin A,
K_i^\prime(\emptyset,A)=0
\}.
\end{aligned}
\]
And we have
\[
\begin{aligned}
LHS
&=
\left\{
\operatorname{ReLU}(K_j)
\;\middle|\;
K_j\in\mathcal K,\
\forall A\in\mathcal B(\mathbb R),
\displaystyle
\int_L^\#(a_j,A)
\leq K_j(\emptyset,A)
\leq \int_U^\#(a_j,A), u_j^{m_j}\leq 0,
\right\}\\
&=
\{
K_i' \in\mathcal K
\;|\;
\forall A\in\mathcal B(\mathbb R),
K_i'(\emptyset,A)
=
K_j(\emptyset,\operatorname{ReLU}^{-1}(A)), \\
&
\text{where}
\int_L^\#(a_j,A)
\leq K_j(\emptyset,A)
\leq \int_U^\#(a_j,A), 
u_j^{m_j}\leq 0
\}.
\end{aligned}
\]
Since $u_j^{m_j}\leq 0$, the support of $K_j(\emptyset,\cdot)$ is contained in $(-\infty,0]$. Hence, we have for every measurable set $A$ with
$0 \in A$, $
K_i(\emptyset,A)
=
K_i(\emptyset,\{0\})
=
K_j(\emptyset,(-\infty,0])
$
and for every measurable set $A$ with
$0\notin A$,
$
K_i(\emptyset,A)=0,
$ as the preimage under ReLU of all points except for 0 in $A$ has zero probability mass under the input distribution.
Therefore, we have 
\[
\begin{aligned}
LHS
&=
\{
K_i^\prime\in\mathcal K
\;|\;
\forall A\in\mathcal B(\mathbb R),\
 0\in A,
\displaystyle
\int_L^\#(a_j,(-\infty,0])
\leq K_i^\prime(\emptyset,A)
\leq \int_U^\#(a_j,(-\infty,0])
\\
&
\land
\forall A\in\mathcal B(\mathbb R),\
 0\notin A,
K_i^\prime(\emptyset,A)=0
\}.
\end{aligned}
\]
Therefore $LHS = RHS$.

\noindent\textbf{Case 2: \(l_j^1\geq 0\).}
We have
\[
a_i^\prime
=
\{(l_i^{\prime k},u_i^{\prime k},
L_i^{\prime k},U_i^{\prime k})\}_{k\in[m_i']}
=
\{(l_j^k,u_j^k,L_j^k,U_j^k)\}_{k\in[m_j]}.
\]
Therefore,
\[
RHS
=\gamma(a_i^\prime) =\gamma(a_j)
=
\left\{
K_i^\prime\in\mathcal K
\;\middle|\;
\forall A\in\mathcal B(\mathbb R),
\int_L^\#(a_j,A)
\leq K_i^\prime(\emptyset,A)
\leq \int_U^\#(a_j,A)
\right\}.
\]
And we have
\[
\begin{aligned}
LHS
&=
\left\{
\operatorname{ReLU}(K_j)
\;\middle|\;
K_j\in\mathcal K, l_j^1\geq0,
\forall A\in\mathcal B(\mathbb R),
\displaystyle
\int_L^\#(a_j,A)
\leq K_j(\emptyset,A)
\leq \int_U^\#(a_j,A)
\right\}\\
&=
\{
K_i^\prime\in\mathcal K
\;|\;
\forall A\in\mathcal B(\mathbb R),
K_i^\prime(\emptyset,A)
=
K_j(\emptyset,\operatorname{ReLU}^{-1}(A)),
\\
&\text{where}
\displaystyle
\int_L^\#(a_j,A)
\leq K_j(\emptyset,A)
\leq \int_U^\#(a_j,A),
l_j^1\geq0
\}.
\end{aligned}
\]
Since $l_j^1\geq0$, the support of
$K_j(\emptyset,\cdot)$ is contained in
$[0,\infty)$. As ReLU acts as the identity
on this support, for every
$A\in\mathcal B(\mathbb R)$, we have
\[
K_i^\prime(\emptyset,A)
=
K_j(\emptyset,\operatorname{ReLU}^{-1}(A))\\
=
K_j(\emptyset,A).
\]
Therefore,
\[
LHS
=
\left\{
K_i^\prime\in\mathcal K
\;\middle|\;
\forall A\in\mathcal B(\mathbb R),
\int_L^\#(a_j,A)
\leq K_i^\prime(\emptyset,A)
\leq \int_U^\#(a_j,A)
\right\}
\]
Hence, $LHS=RHS$.

\noindent\textbf{Case 3: \(l_j^1<0<u_j^{m_j}\).}
We have
\[
\begin{aligned}
a_i^\prime
&=
\{(l_i^{\prime k},u_i^{\prime k},
L_i^{\prime k},U_i^{\prime k})\}_{k\in[m_i']}\\
&=
\left\{
\left(
0,\,
0,\,
\int_L^\#(a_j,(-\infty,0]),\,
\int_U^\#(a_j,(-\infty,0])
\right)
\right\}
\cup
\left\{
\left(
\max(0,l_j^k),\,
u_j^k,\,
L_j^k,\,
U_j^k
\right)
\right\}_{k\in[m_j],\,u_j^k>0}.
\end{aligned}
\]

Therefore,
\[
\begin{aligned}
RHS
&=\gamma(a_i^\prime)\\
&
=
\left\{
K_i^\prime\in\mathcal K
\;\middle|\;
\forall A\in\mathcal B(\mathbb R),
\int_L^\#(a_j,A)
\leq K_i^\prime(\emptyset,A)
\leq \int_U^\#(a_j,A)
\right\}\\
&=
\{
K_i^\prime\in\mathcal K
\;|\;
\forall A\in\mathcal B(\mathbb R),\
\mathbf 1_A(0)\int_L^\#(a_j,(-\infty,0])
+\int_L^\#(a_j,A\cap(0,\infty))
\leq K_i^\prime(\emptyset,A)
\\
&
\leq
\mathbf 1_A(0)\int_U^\#(a_j,(-\infty,0])
+\int_U^\#(a_j,A\cap(0,\infty))
\}.
\end{aligned}
\]
And we have
\[
\begin{aligned}
LHS
&=
\left\{
\operatorname{ReLU}(K_j)
\;\middle|\;
K_j\in\mathcal K,
l_j^1<0<u_j^{m_j},
\forall A\in\mathcal B(\mathbb R),
\displaystyle
\int_L^\#(a_j,A)
\leq K_j(\emptyset,A)
\leq \int_U^\#(a_j,A)
\right\}\\
&=
\{
K_i^\prime\in\mathcal K
\;|\;
\forall A\in\mathcal B(\mathbb R),
K_i^\prime(\emptyset,A)
=
K_j(\emptyset,\operatorname{ReLU}^{-1}(A)),\\
&\text{where}
\int_L^\#(a_j,A)
\leq K_j(\emptyset,A)
\leq \int_U^\#(a_j,A),
l_j^1<0<u_j^{m_j}
\}.
\end{aligned}
\]

We expand $LHS$ by considering two cases,
depending on whether the measurable set $A$ contains $0$.
We first consider $0\in A$.
By the concrete semantics of ReLU,
\[
K_i^\prime(\emptyset,A)
=
K_j(\emptyset,\operatorname{ReLU}^{-1}(A))
=
K_j(\emptyset,(-\infty,0])
+
K_j(\emptyset,A\cap(0,\infty)).
\]
Since $K_j\in\gamma(a_j)$, by the
definition of the abstract integration
operator, we have
\[
\begin{aligned}
\int_L^\#(a_j,(-\infty,0])
&\leq K_j(\emptyset,(-\infty,0])
\leq \int_U^\#(a_j,(-\infty,0]),\\
\int_L^\#(a_j,A\cap(0,\infty))
&\leq K_j(\emptyset,A\cap(0,\infty))
\leq \int_U^\#(a_j,A\cap(0,\infty)).
\end{aligned}
\]
Consequently,
\[
\int_L^\#(a_j,(-\infty,0])
+\int_L^\#(a_j,A\cap(0,\infty)) \leq K_i^\prime(\emptyset,A)\leq
\int_U^\#(a_j,(-\infty,0])
+\int_U^\#(a_j,A\cap(0,\infty)).
\]
Next, consider any measurable set $A\in\mathcal B(\mathbb{R})$ where $0\notin A$.
Since ReLU preserves positive values,
$
\operatorname{ReLU}^{-1}(A)=A\cap(0,\infty).
$
Consequently,
\[
K_i^\prime(\emptyset,A)
=
K_j(\emptyset,\operatorname{ReLU}^{-1}(A))
=
K_j(\emptyset,A\cap(0,\infty)).
\]
Therefore
\[
\int_L^\#(a_j,A\cap(0,\infty))
\leq K_i^\prime(\emptyset,A)
\leq \int_U^\#(a_j,A\cap(0,\infty)).
\]
Combining the two cases, we have
\[
\begin{aligned}
LHS
&=
\{
K_i^\prime\in\mathcal K
\;|\;
\forall A\in\mathcal B(\mathbb R),\
\mathbf 1_A(0)\int_L^\#(a_j,(-\infty,0])
+\int_L^\#(a_j,A\cap(0,\infty))
\leq K_i^\prime(\emptyset,A)
\\
&
\leq
\mathbf 1_A(0)\int_U^\#(a_j,(-\infty,0])
+\int_U^\#(a_j,A\cap(0,\infty))
\}
\end{aligned}
\]

Therefore, $LHS= RHS$.

In summary, we actually prove that $T_f(\gamma_n(a))= \gamma_n(T_f^\sharp(a))$, meaning that the ReLU abstract transformer doesn't lose any precision.
\end{proof}

\section{Derivation of $K(\emptyset,dy)$}
\applabel{app:derivationcond}
In this section, we provide a detailed derivation of $K(\emptyset,dy)$. Specifically, we explain the rewriting step by step:
\begin{equation*}
\begin{aligned}
K(\emptyset,dy)
&=\int_{\mathbb R^J}
K_{\mathrm{cond}}(\mathbf x_J,dy)
\,\mu_J(d\mathbf x_J) \\
&=
\sum_k\int_{B_k} K_{\mathrm{cond}}(\mathbf{x}_J,dy)\,
\mu_{J}(d\mathbf{x}_J)\\
&=\sum_k\mathbb{E}[K_{\mathrm{cond}}(\mathbf{X}_J,dy) \mid \mathbf{X}_J \in B_k] \cdot \Pr[\mathbf{X}_J \in B_k]\\
&=
\left(
\sum_k
\left\{
\mathbb{E}\left[
\left.
\frac{1}{\sqrt{2\pi v(\mathbf{X}_J)}}
\exp\left(
-\frac{(y-m(\mathbf{X}_J))^2}{2v(\mathbf{X}_J)}
\right)
\,\right|\,
\mathbf{X}_J\in B_k
\right]
\cdot \Pr[\mathbf{X}_J\in B_k]
\right\}
\right)dy.
\end{aligned}
\end{equation*}
where  $\Pr[\mathbf X_J\in B_k]=
\mu_{J}(B_k),
\mu_J(A)
=
\int_{\mathbb R^n}
\mathbf{1}_A(\mathbf{x}_J)\,\mu(d\mathbf{x}),
\
\forall A\in\mathcal B(\mathbb R^J), 
\mu\in \gamma_{c0}(\sigma_p)
$, with $\sigma_p$
denoting the input probabilistic store.
The first equality follows from the marginalization
in~\eqnref{eq:py1}, while the second follows by
partitioning the input space $\mathbb R^J$ into
disjoint boxes $\{B_k\}_k$.
We now derive the third equality using the
definition of conditional expectation.
Specifically, we prove that
\begin{equation*}
\int_{B_k} K_{\mathrm{cond}}(\mathbf{x}_J,dy)\,
\mu_{J}(d\mathbf{x}_J)
=\mathbb{E}[K_{\mathrm{cond}}(\mathbf{X}_J,dy) \mid \mathbf{X}_J \in B_k] \cdot \Pr[\mathbf{X}_J \in B_k],
\eqnlabel{cond}
\end{equation*}
\begin{proof}

\begin{itemize}[leftmargin=*]
    \item \textbf{Case 1:} $\Pr[\mathbf{X}_J \in B_k]>0$.
\begin{align*}
\int_{B_k}
K_{\mathrm{cond}}(\mathbf{x}_J,dy)\,
\mu_{J}(d\mathbf{x}_J)
&=
\frac{
\int_{B_k}
K_{\mathrm{cond}}(\mathbf{x}_J,dy)\,
\mu_{J}(d\mathbf{x}_J)
}{
\Pr[\mathbf X_J\in B_k]
}
\cdot\Pr[\mathbf X_J\in B_k]
\\
&=
\frac{
\int_{\mathbb R^J}
K_{\mathrm{cond}}(\mathbf{x}_J,dy)\,
\mathbf{1}_{B_k}(\mathbf{x}_J)\,
\mu_{J}(d\mathbf{x}_J)
}{
\Pr[\mathbf X_J\in B_k]
}
\cdot\Pr[\mathbf X_J\in B_k]
\\
&=
\int_{\mathbb R^J}
K_{\mathrm{cond}}(\mathbf{x}_J,dy)\,
\frac{
\mathbf{1}_{B_k}(\mathbf{x}_J)
\mu_{J}(d\mathbf{x}_J)
}{
\Pr[\mathbf X_J\in B_k]
}
\cdot\Pr[\mathbf X_J\in B_k]
\\
&=
\int_{\mathbb R^J}
K_{\mathrm{cond}}(\mathbf{x}_J,dy)\,
\mu_{J}
(d\mathbf{x}_J\mid\mathbf X_J\in B_k)
\cdot\Pr[\mathbf X_J\in B_k]
\\
&=
\mathbb E[
K_{\mathrm{cond}}(\mathbf X_J,dy)
\mid\mathbf X_J\in B_k
]
\cdot\Pr[\mathbf X_J\in B_k].
\end{align*}
Here, the expectation of the conditional kernel
is understood as a probability measure that assigns the probability to each measurable set $A\in\mathcal B(\mathbb R)$, i.e.,
$
\left(
\mathbb E\left[
K_{\mathrm{cond}}(\mathbf X_J,\cdot)
\mid\mathbf X_J\in B_k
\right]
\right)(A)
=
\mathbb E\left[
K_{\mathrm{cond}}(\mathbf X_J,A)
\mid\mathbf X_J\in B_k
\right].
$
\item \textbf{Case 2:} $\Pr[\mathbf{X}_J \in B_k]=0$.

Since $K_{\mathrm{cond}}(\mathbf{x}_J,\cdot)$ is a probability measure, for any $A\in\mathcal B(\mathbb R)$, we have
\begin{align*}
0
\le
\int_{B_k}
K_{\mathrm{cond}}(\mathbf{x}_J,A)\,
\mu_J(d\mathbf{x}_J)
\le
\int_{B_k}1\,\mu_J(d\mathbf{x}_J)
=
\Pr[\mathbf X_J\in B_k]
=0.
\end{align*}
Therefore,
we have
$
\int_{B_k}
K_{\mathrm{cond}}(\mathbf{x}_J,dy)\,
\mu_J(d\mathbf{x}_J)=0.
$
For this case, we define the conditional expectation
$\mathbb E[K_{\mathrm{cond}}(\mathbf X_J,\cdot)
\mid\mathbf X_J\in B_k]$
as an arbitrary probability measure, since its contribution will be multiplied by zero anyway. Hence,
\[
\int_{B_k}
K_{\mathrm{cond}}(\mathbf{x}_J,dy)\,
\mu_J(d\mathbf{x}_J)
=
\mathbb E[
K_{\mathrm{cond}}(\mathbf X_J,dy)
\mid\mathbf X_J\in B_k
]
\cdot\Pr[\mathbf X_J\in B_k]
=0.
\]

\end{itemize}

Therefore, $\int_{B_k}
K_{\mathrm{cond}}(\mathbf{x}_J,dy)\,
\mu_J(d\mathbf{x}_J)
=
\mathbb E[
K_{\mathrm{cond}}(\mathbf X_J,dy)
\mid\mathbf X_J\in B_k$ holds.
\end{proof}
The final equality follows by expanding the conditional Gaussian density.

\section{Proof of Absolute Continuity of Affine Output}
\applabel{affine-density}
In this section, we prove that for any joint input probability measure $\mu_J$, the output probability measure $K(\emptyset,\cdot)$ is absolutely continuous with respect to the Lebesgue measure $\lambda$, supposing that the Gaussian bias has positive variance $\sigma_b^2>0$.
Moreover, \(
f_Y(y)=
\sum_k
\{
\mathbb{E}[
\frac{1}{\sqrt{2\pi v(\mathbf{X}_J)}}
\exp(
-\frac{(y-m(\mathbf{X}_J))^2}{2v(\mathbf{X}_J)}
)
\,|\,
\mathbf{X}_J\in B_k
]
\cdot \Pr[\mathbf{X}_J\in B_k]
\}
\)
is a probability density function of the output distribution.

\begin{proof}
We first prove that $f_Y$
is a probability density function of the output distribution,
i.e.,
$(1)\ f_Y(y)\geq 0; \ (2) \
\int_{\mathbb R} f_Y(y)\,dy=1; \ (3) \ K(\emptyset,A)=\int_A f_Y(y)\,\lambda(dy),
$
for all $y\in\mathbb R$ and $A\in\mathcal B(\mathbb R)$.
We then prove that the output
probability measure is continuous with respect to the Lebesgue measure,
i.e.,
$
\forall A\in\mathcal B(\mathbb R),
\lambda(A)=0
\implies
K(\emptyset,A)=0,
$ as it can be induced from the former proof.
Let
\[
g(y\mid\mathbf{X}_J)
=
\frac{1}{\sqrt{2\pi v(\mathbf{X}_J)}}
\exp\left(
-\frac{(y-m(\mathbf{X}_J))^2}
{2v(\mathbf{X}_J)}
\right)
\]
denote the conditional Gaussian density evaluated at the random input vector \(\mathbf X_J\), where
$
v(\mathbf{X}_J)
=
\sum_{j=1}^{J}\sigma_j^2 X_j^2+\sigma_b^2
\geq\sigma_b^2>0.
$

We first show that $f_Y$ is non-negative.
Since the conditional Gaussian density is
non-negative, its conditional expectation is
also non-negative. Thus,
$
f_Y(y)\geq0, \forall y\in\mathbb R.
$

Next, we show that $f_Y$ integrates to one:
\[
\begin{aligned}
\int_{\mathbb R}f_Y(y)\,dy
&=
\int_{\mathbb R}
\sum_k
\left\{
\mathbb E\left[
g(y\mid\mathbf X_J)
\mid \mathbf X_J\in B_k
\right]
\cdot \Pr[\mathbf X_J\in B_k]
\right\}dy
\\
&=
\sum_k
\Pr[\mathbf X_J\in B_k]
\int_{\mathbb R}
\mathbb E\left[
g(y\mid\mathbf X_J)
\mid \mathbf X_J\in B_k
\right]dy
\\
&=
\sum_k
\Pr[\mathbf X_J\in B_k]
\left(
\int_{\mathbb R}
\int_{B_k}
g(y\mid\mathbf x_J)\,
\mu_k(d\mathbf x_J)\,dy
\right)
\\
&=
\sum_k
\Pr[\mathbf X_J\in B_k]
\left(
\int_{B_k}
\int_{\mathbb R}
g(y\mid\mathbf x_J)\,dy \
\mu_k(d\mathbf x_J)\,
\right)
\\
&=
\sum_k
\Pr[\mathbf X_J\in B_k]
\mathbb E\left[
\left.
\int_{\mathbb R}
g(y\mid\mathbf X_J)\,dy
\,\right|\,
\mathbf X_J\in B_k
\right]
\\
&=
\sum_k
\Pr[\mathbf X_J\in B_k]
\\
&=1.
\end{aligned}
\]
The last equality follows because
$\{B_k\}_k$ forms a finite measurable
partition of the input space $\mathbb R^J$.

Finally, we show that $f_Y$ is the density
of the output probability measure.
For any $A\in\mathcal B(\mathbb R)$,
by the concrete affine semantics,
\[
\begin{aligned}
K(\emptyset,A)
&=
\int_{\mathbb R^J}
K_{\mathrm{cond}}(\mathbf x_J,A)
\,\mu_J(d\mathbf x_J)\\
&=
\int_{\mathbb R^J}
\left(
\int_A
g(y\mid\mathbf x_J)\,\lambda(dy)
\right)
\mu_J(d\mathbf x_J)\\
&=
\int_A
\left(
\int_{\mathbb R^J}
g(y\mid\mathbf x_J)\,
\mu_J(d\mathbf x_J)
\right)\lambda(dy)\\
&=
\int_A
\left(
\sum_k
\int_{B_k}
g(y\mid\mathbf x_J)\,
\mu_J(d\mathbf x_J)
\right)\lambda(dy)\\
&=
\int_A f_Y(y)\,\lambda(dy),
\end{aligned}
\]
where the last equality follows from
the conditional expectation decomposition
in~\eqnref{cond}.
Consequently, for any measurable set $A$
with $\lambda(A)=0$, we have
\[
K(\emptyset,A)
=
\int_A f_Y(y)\,\lambda(dy)
=0.
\]
Hence, the output probability measure $K(\emptyset,\cdot)$ is absolutely continuous with respect to the Lebesgue measure $\lambda$, and $f_Y$ is a probability density function
of the output distribution.
\end{proof}

\section{Detailed Derivation of Box Probability Bounds}
\applabel{boxbounds}
In this section, we provide a detailed explanation of obtaining bounds of $\Pr[\mathbf{X}_J\in B_k]$. 

Recalling that we have $B_k = I_1^{k_1} \times I_2^{k_2} \times \cdots \times I_J^{k_J}$, for each dimension $h$. We further define the event
$
A_h^{k_h} = \{X_h \in I_h^{k_h}\}.
$
Then the event that the input lies in the box $B_k$ is equivalent to the intersection of these events:
$
\{\mathbf{X}_J \in B_k\} = \bigcap_{h=1}^J A_h^{k_h}.
$
Although the joint distribution of $\mathbf{X}_J$ is unknown, we can still derive a sound upper bound, as for any $j \in [J]$, we have
\(
\bigcap_{h=1}^m A_h^{k_h} \subseteq A_j^{k_j},
\)
i.e., satisfying all conditions implies satisfying any single one. Taking probabilities on both sides gives
\(
\Pr[\mathbf{X}_J\in B_k] =\Pr\left(\bigcap_{h=1}^J A_h^{k_h}\right)
\le
\Pr(A_j^{k_j}),\forall j \in[J].
\)
Therefore we obtain
\begin{equation}
\Pr[\mathbf{X}_J \in B_k]
\le
\min_{1 \le h \le J} \Pr(X_h \in I_h^{k_h}), \quad \forall \Pr[\mathbf{X}_J \in B_k]=\mu_J(B_k)
\eqnlabel{box_u}
\end{equation}
A similar reasoning can be used to derive a lower bound. Starting from
$
\{\mathbf{X} \in B_k\} = \bigcap_{h=1}^J A_h^{k_h},
$
we consider its complement:
\(
\Pr\!\left(\bigcap_{h=1}^{J} A_h^{k_h}\right)
= 1 - \Pr\!\left(\left(\bigcap_{h=1}^{J} A_h^{k_h}\right)^c\right) = 1 - \Pr\!\left(\bigcup_{h=1}^{J} \left(A_h^{k_h}\right)^c\right).
\)
Applying the union bound, we obtain
\(
\Pr\left(\bigcup_{h=1}^J (A_h^{k_h})^c\right)
\le
\sum_{h=1}^J \Pr\left((A_h^{k_h})^c\right)
=
\sum_{h=1}^J (1 - \Pr(A_h^{k_h})).
\)
Therefore we have
\begin{align*}
\Pr[\mathbf{X}_J \in B_k]
&=\Pr\left(\bigcap_{h=1}^J A_h^{k_h}\right)= 1 - \Pr\!\left(\bigcup_{h=1}^{J} \left(A_h^{k_h}\right)^c\right)
\ge
1 - \sum_{h=1}^J (1 - \Pr(A_h^{k_h}))=
\sum_{h=1}^J \Pr(A_h^{k_h}) - (J-1).
\end{align*}
Since probabilities are nonnegative, we further obtain
\begin{equation}
\Pr[\mathbf{X}_J \in B_k]
\ge
\max\left(0,\,
\sum_{h=1}^J \Pr(X_h \in I_h^{k_h}) - (J-1)
\right),
 \quad \forall \Pr[\mathbf{X}_J \in B_k]=\mu_J(B_k).
\eqnlabel{box_l}
\end{equation}
Since each $X_h$ is abstracted by a collection of partitions, we bound the probability mass associated with each partition $I_h^{k_h}$ using its corresponding tuple $\langle l_h^{k_h},u_h^{k_h},L_h^{k_h},U_h^{k_h}\rangle$. For a non-degenerate partition, the bounds are obtained by integrating $L_h^{k_h}$ and $U_h^{k_h}$ over the interval. For a degenerate partition, the bounds are directly given by $L_h^{k_h}$ and $U_h^{k_h}$, which represent the probability mass at the corresponding point. Here, non-degenerate partitions exclude any atomic points represented separately by degenerate partitions, ensuring that the probability mass is counted only once. We denote the resulting lower and upper bounds by
$l_h$ and $u_h$, respectively, with $ l_h \in \mathbb{R}, u_h \in \mathbb{R}\cup \{\infty\}$. 
Therefore we have
\(
l_h \le \Pr[X_h \in I_h^{k_h}] \le u_h,
\)
 Consequently, combining~\eqnref{box_u} and ~\eqnref{box_l}, the box probability can be bounded by two deterministic constants
\(
\max\left(0, \sum_{h=1}^J l_h - (J-1)\right) \le \Pr[\mathbf{X}_J \in B_k]
\le \min_{1 \le h \le J} u_h.
\)
To be more precise, to ensure that the bounds lie within the valid probability range, we can further clamp the resulting bounds to $[0,1]$, yielding
\begin{equation}
\max\!\left(0,\; \min\!\left(1,\; \sum_{h=1}^J l_h - (J-1)\right)\right)
\;\le\;
\Pr[\mathbf{X}_J \in B_k]
\;\le\;
\min\!\left(1,\; \min_{1 \le h \le J} u_h\right),
 \quad \forall \Pr[\mathbf{X}_J \in B_k]=\mu_J(B_k).
\eqnlabel{Pkbounds}
\end{equation}

\section{Detailed Derivation of $\underline{g}_{k}(y)$ and $\overline{g}_{k}(y)$  and Proof of Property}
\applabel{proof}
In this section, we first show how to get the sound $\underline{g}_{k}(y)$ and $\overline{g}_{k}(y)$ mathematically, i.e., bounding the Gaussian density, and prove their properties.
We simplified the notation here for clarity. Considering
$
\mathcal{N}(y; m, v) = \frac{1}{\sqrt{2\pi v}}
\exp\!\left(-\frac{(y-m)^2}{2v}\right),
$
where the mean and variance range over intervals
$
m \in [\underline{m}, \overline{m}],
v \in [\underline{v}, \overline{v}].
$

To get the upper bound of the Gaussian density, we first minimize the squared distance between $y$ and $m$:
\[
d^2(y)
=
\min_{m \in [\underline{m}, \overline{m}]} (y - m)^2.
\]
This yields
\[
d^2(y)
=
\begin{cases}
(\underline{m} - y)^2, & y < \underline{m}, \\
0, & y \in [\underline{m}, \overline{m}], \\
(y - \overline{m})^2, & y > \overline{m}.
\end{cases}
\]
The optimal mean is therefore
\[
m^*(y)=
\begin{cases}
\underline{m}, & y < \underline{m},\\
y, & y \in [\underline{m}, \overline{m}],\\
\overline{m}, & y > \overline{m}.
\end{cases}
\]
Substituting $d^2(y)$, we then maximize
\[
v^{-1/2} \exp\!\left(-\frac{d^2(y)}{2v}\right)
\quad \text{over } v \in [\underline{v}, \overline{v}].
\]
The maximizer occurs at $v = d^2(y)$, but constrained to the interval of $y$, giving
\[
v^*(y)=
\begin{cases}
\underline{v}, & d^2(y) < \underline{v},\\
d^2(y), & \underline{v} \le d^2(y) \le \overline{v},\\
\overline{v}, & d^2(y) > \overline{v}.
\end{cases}
\]
Thus, the upper bound is
\[
\overline{g}(y)
=
\mathcal{N}\bigl(y; m^*(y), v^*(y)\bigr).
\]
In particular, when $y \in [\underline{m}, \overline{m}]$, we have $d^2(y)=0$ and hence
\[
\overline{g}(y)
=
\frac{1}{\sqrt{2\pi \underline{v}}}.
\]
Similarly, to get the lower bound of the Gaussian density, we maximize the squared distance:
\[
D^2(y)
=
\max_{m \in [\underline{m}, \overline{m}]} (y - m)^2
=
\max\{(y - \underline{m})^2,\ (y - \overline{m})^2\}.
\]
Let $m_{\mathrm{far}}(y)$ denote the endpoint attaining this maximum.
We then minimize
\[
v^{-1/2} \exp\!\left(-\frac{D^2(y)}{2v}\right)
\quad \text{over } v \in [\underline{v}, \overline{v}],
\]
which is achieved at the interval endpoints. Therefore,
\[
\underline{g}(y)
=
\min\Bigl\{
\mathcal{N}(y; m_{\mathrm{far}}(y), \underline{v}),
\,
\mathcal{N}(y; m_{\mathrm{far}}(y), \overline{v})
\Bigr\}.
\]

\begin{lemma}[Log-shape of Gaussian bounding cases] 
Let 
\(
g(y;m,v)= \frac{1}{\sqrt{2\pi v}} \exp\!(-\frac{(y-m)^2}{2v}), 
\) 
where 
$
m\in[\underline m,\overline m],  v\in[\underline v,\overline v],  0<\underline v\le \overline v. 
$ 
The upper bound 
$
\overline g(y) = \sup_{m,v} g(y;m,v), 
$
is either log-concave or log-convex in $y$. 
The lower bound 
$
\underline g(y) = \inf_{m,v} g(y;m,v), 
$is log-concave in $y$. 
\lemmalabel{lemma:g}
\end{lemma}
\begin{proof} 
We first prove the property of the upper bound.
From the construction, the optimal value of $m$ is
\[
m^*(y)=\operatorname{clip}(y,\underline m,\overline m),
\]
and we define
\[
d^2(y)=(y-m^*(y))^2.
\]
The remaining maximization over $v$ yields
\[
v^*(y)=
\begin{cases}
\underline v, & d^2(y)<\underline v,\\
d^2(y), & \underline v \le d^2(y) \le \overline v,\\
\overline v, & d^2(y)>\overline v.
\end{cases}
\]
We now analyze the log-shape of $\overline g(y)=g(y;m^*(y),v^*(y))$ by enumerating all possible cases.

\begin{enumerate}
\item \textbf{Case 1: $y < \underline m$.}
Hence $m^*(y)=\underline m$, $d^2(y)=(y-\underline m)^2$. The optimal $v^*(y)$ has the following possibilities:
\begin{itemize}
    \item If $d^2(y) < \underline v$, then
    \[
    \log \overline g(y)
    =
    -\frac12 \log(2\pi \underline v)
    -
    \frac{(y-\underline m)^2}{2\underline v},
    \]
    which is concave.

    \item If $\underline v \le d^2(y) \le \overline v$, then
    \[
    \log \overline g(y)
    =
    -\frac12 \log(2\pi d^2(y))
    - \frac12
    =
    -\frac12 \log(2\pi)
    - \frac12 \log d^2(y)
    - \frac12,
    \]
    which simplifies to
    \[
    -\frac12 \log(2\pi)
    - \log|y-\underline m|
    - \frac12,
    \]
    whose second derivative is positive, and hence it is convex.

    \item If $d^2(y) > \overline v$, then
    \[
    \log \overline g(y)
    =
    -\frac12 \log(2\pi \overline v)
    -
    \frac{(y-\underline m)^2}{2\overline v},
    \]
    which is concave.
\end{itemize}

\item \textbf{Case 2: $y \in [\underline m,\overline m]$.}
Hence $m^*(y)=y$, $d^2(y)=0$, and thus
\[
\log \overline g(y)
=
-\frac12 \log(2\pi v^*(y)),
\]
which is constant, and hence both concave and convex.

\item \textbf{Case 3: $y > \overline m$.}

Hence $m^*(y)=\overline m$, $d^2(y)=(y-\overline m)^2$. The $v^*(y)$ has the following possibilities:

\begin{itemize}
    \item If $d^2(y) < \underline v$, then
    \[
    \log \overline g(y)
    =
    -\frac12 \log(2\pi \underline v)
    -
    \frac{(y-\overline m)^2}{2\underline v},
    \]
    which is concave.

    \item If $\underline v \le d^2(y) \le \overline v$, then
    \[
    \log \overline g(y)
    =
    -\frac12 \log(2\pi)
    - \log|y-\overline m|
    - \frac12,
    \]
    which is convex.

    \item If $d^2(y) > \overline v$, then
    \[
    \log \overline g(y)
    =
    -\frac12 \log(2\pi \overline v)
    -
    \frac{(y-\overline m)^2}{2\overline v},
    \]
    which is concave.
\end{itemize}

\end{enumerate}
Combining all cases, we conclude that $\overline g(y)$ is either log-concave or log-convex.

We then prove the property of the lower bound. For the lower bound, we maximize the squared distance
\[
D^2(y)
=
\max_{m\in[\underline m,\overline m]}(y-m)^2
=
\max\{(y-\underline m)^2,(y-\overline m)^2\},
\]
and let $m_{\mathrm{far}}(y)$ be the corresponding endpoint. The minimization over $v$ is attained at one of the endpoints $\underline v$ or $\overline v$, so $\underline g(y)$ takes the form
\[
\underline g(y)
=
\mathcal N(y;m_0,v_0),
\]
where $m_0\in\{\underline m,\overline m\}$ and $v_0\in\{\underline v,\overline v\}$.
Therefore,
\[
\log \underline g(y)
=
-\frac12 \log(2\pi v_0)
-
\frac{(y-m_0)^2}{2v_0},
\]
whose second derivative is
\[
-\frac{1}{v_0} < 0.
\]
Hence $\log \underline g(y)$ is concave in $y$.
\end{proof}

\section{Approximating the summation for abstract affine transformation}
\applabel{boundsummation}
In this section, we detail the approximation step of the summations in~\eqnref{pybound} to obtain an exponential-affine abstraction over output intervals to construct the output abstract element.
\begin{equation}
\sum_k \underline{P}_k \, \underline{g}_{k}(y)
\;\le\;
f_{Y}(y)
\;\le\;
\sum_k \overline{P}_k \, \overline{g}_{k}(y).
\eqnlabel{pybound}
\end{equation}

Firstly, consider partitioning the domain of $y$ into a finite set of intervals
\(
\{J^r = [l^r, u^r]\}_{r\in [m_y]}.
\) with $m_y\in \mathbb N_{>0}$ partitions.
Note that this partition is not stochastic. Instead, it is constructed such that, for every partition $J^r$, the expressions of $\underline{g}_k(y)$ and $\overline{g}_k(y)$ remain the same for all $y\in J^r$, i.e., no further case distinction is required within a partition. This will significantly simplify the subsequent derivation by avoiding unnecessary case analysis. To satisfy this requirement, we propose a dynamic partition strategy tailored to this setting.
In practice, we first construct a coarse uniform partition of the output domain and then refine it adaptively using the critical switch points induced by the bounds of \(m(\mathbf{x})\) and \(v(\mathbf{x})\) in \(g(y,\mathbf{X}_J)\) over all boxes. The refinement process is designed to ensure that each interval \(J^r\) lies entirely within a single log-shape case (e.g., log-concave or log-convex) in~\appref{proof}. This guarantees that the lower and upper bounds of \(g(y,\mathbf{X}_J)\) can be soundly approximated using tangent or secant affine bounds in the log domain within each interval.

On each interval $J^r$, we then derive exponential-affine upper and lower bounds of the density. 
We first approximate each term of the bounds in~\eqnref{pybound} by exponential-affine functions on each interval $J^r$. As shown in ~\lemmaref{lemma:g}, $\underline{g}_{k}(y)$ always satisfies the log-concavity and $\overline{g}_{k}(y)$ satisfies either the log-concavity property or the log-convexity. Therefore, for bounded interval $J^r$, we are able to bound $\log \underline{g}_{k}(y)$ by an affine function on $J^r$ using a secant line, and bound $\log \overline{g}_{k}(y)$ using the secant line or the tangent line, depending on the property it has. For unbounded intervals, we conservatively set the lower bound of $\underline{g}_{k}(y)$ to zero and omit the corresponding term from the lower-bound summation. Meanwhile, $\overline{g}_{k}(y)$ is log-concave on the unbounded tail intervals, allowing us to construct a sound affine upper bound on $\log \overline{g}_{k}(y)$ using a tangent line at the finite endpoint.
Consequently, we obtain
\[
\overline{g}_{k}(y)
\le
\exp(a_{k,r}^U y + b_{k,r}^U),
\quad
\underline{g}_{k}(y)
\ge
\exp(a_{k,r}^L y + b_{k,r}^L),
\quad \forall y \in J^r.
\]
Multiplying by the constants $\overline{P}_k$ and $\underline{P}_k$, we obtain
\[
\overline{P}_k \, \overline{g}_{k}(y)
\le
\exp(a_{k,r}^U y + b_{k,r}^U + \log \overline{P}_k),
\quad
\underline{P}_k \, 
\underline{g}_{k}(y)
\ge
\exp(a_{k,r}^L y + b_{k,r}^L + \log \underline{P}_k).
\]
Thus, on each interval $J^r$, the density is bounded by sums of exponential-affine functions:
\[
\sum_k \exp(a_{k,r}^L y + b_{k,r}^L + \log \underline{P}_k)
\;\le\;
f_{Y}(y)
\;\le\;
\sum_k \exp(a_{k,r}^U y + b_{k,r}^U + \log \overline{P}_k),
 \forall y \in J^r.
\]

Finally, we approximate these sums. Note that $\sum_k \exp(a_{k,r} y + b_{k,r}+ \log P_k)$ satisfies the log-convexity property,i.e., \(
F(y)
=
\log \left( \sum_k \exp(a_{k,r} y + b_{k,r}+\log P_k) \right)
\) is a convex function.
We provide the complete proof below:
\begin{proof}
Let
\(q_k(y) = \exp(a_{k,r}y+b_{k,r}+\log P_k), \ S(y) = \sum_k q_k(y).\)
Then we have
\(F(y)=\log S(y).\)
Therefore we have 
\[
F''(y) =
\frac{
S''(y)S(y)-S'(y)^2
}{
S(y)^2
},
\]
Since
\(
q_k'(y) = a_{k,r}q_k(y),
\)
we have
\(
S'(y) = \sum_k a_{k,r}q_k(y),
\)
and similarly,
\(
S''(y) = \sum_k a_{k,r}^2q_k(y).
\)
Substituting them into $F''(y)$, we have
\[
F''(y) =
\frac{
\left(\sum_k a_{k,r}^2 q_k(y)\right)
\left(\sum_k q_k(y)\right)
-
\left(\sum_k a_{k,r} q_k(y)\right)^2
}{
\left(\sum_k q_k(y)\right)^2
}.
\]
To show that the numerator is nonnegative, define
\(
u_k = a_{k,r}\sqrt{q_k(y)},
v_k = \sqrt{q_k(y)}.
\)
By the Cauchy--Schwarz inequality,
\[
\left(\sum_k u_k v_k\right)^2
\le
\left(\sum_k u_k^2\right)
\left(\sum_k v_k^2\right).
\]
Substituting the definitions of $u_k$ and $v_k$, we obtain
\[
\left(\sum_k a_{k,r}q_k(y)\right)^2
\le
\left(\sum_k a_{k,r}^2q_k(y)\right)
\left(\sum_k q_k(y)\right).
\]
Therefore, the numerator of $F''(y)$ is nonnegative. Moreover,
$q_k(y)>0$ for all $k$, so
\[
\left(\sum_k q_k(y)\right)^2>0.
\]
Thus,
\(
F''(y)\ge 0,
\)
which proves that $F$ is convex.
\end{proof}

%
As such, we can get the upper bound of $F^U(y)=
\log (\sum_k \exp(a_{k,r}^U y + b_{k,r}^U))$ on bounded $J^r$ using its secant line, obtaining
\(
F^U(y)
\le
a_r^U y + b_r^U.
\)
For an unbounded interval \(J^r\), the upper bound of \(F^U\) cannot be constructed using a secant line, since one endpoint is infinite. Instead, we exploit the slopes of the exponential-affine terms in the summation. Specifically,
For a right-unbounded interval, i.e., \(y\in [l_{m_y},+\infty)\), we define
\[
a_{m_y}^U=\max_k a_{k,r}^U,
\quad
b_{m_y}^U=F^U(l_{m_y})-a_r^U l_{m_y},
\]
which yields \(F^U(y)\le a_{m_y}^U y+b_{m_y}^U\) for all \(y\in [l_{m_y},+\infty)\).
Similarly, for a left-unbounded interval \(y \in (-\infty,u_1]\), we define
\[
a_1^U=\min_k a_{k,r}^U,
\quad
b_1^U=F^U(u_1)-a_r^U u_1,
\]
which yields \(F^U(y)\le a_1^U y+b_1^U\) for all \(y\in (-\infty,u_1]\). We provide the proof in~\appref{unboundedinterval}.
Thus, the resulting bounds remain exponential-affine and are sound over the entire unbounded interval.
Exponentiating both sides yields
\(
f_{Y}(y)
\le
\exp(a_r^U y + b_r^U),
\forall y \in J^r.
\)
Similarly, defining
\(
F^L(y)
=
\log \left( \sum_k \exp(a_{k,r}^L y + b_{k,r}^L) \right),
\)
we can get the lower bound of $F^L(y)$ using a tangent line, yielding
\(
F^L(y)
\ge
a_r^L y + b_r^L,
\)
and hence
\(
\exp(a_r^L y + b_r^L)
\le
f_{Y}(y),
\forall y \in J^r.
\)

\section{Soundness of Summation Bounds on Unbounded Intervals}
\applabel{unboundedinterval}

In this section, we prove that the exponential-affine upper bounds constructed for the summation in the affine transformer in~\secref{affine} are sound over unbounded output intervals.

Let $F^U(y)=\log\left(\sum_k \exp(a_{k,r}^U y+b_{k,r}^U)\right)$.
We first consider the right-unbounded interval, i.e., $y \in [l_{m_y},+\infty)$. We define 
\[
a_{m_y}^U=\max_k a_{k,r}^U,
b_{m_y}^U=F^U(l_{m_y})-a_{m_y}^U l_{m_y}.
\]
For any $y \in [l_{m_y},+\infty)$, we have $y-l_{m_y}\ge 0$. By the definition of \(a_{m_y}^U\), we have \(a_{k,r}^U\le a_{m_y}^U\) for every $k$, and hence
\[
a_{k,r}^U(y-l_{m_y})\le a_{m_y}^U(y-l_{m_y}).
\]
Therefore,
\[
\begin{aligned}
F^U(y)
&=\log\left(
\sum_k
\exp(a_{k,r}^U l_r+b_{k,r}^U)
\exp(a_{k,r}^U(y-l_{m_y}))
\right)\\
&\le
\log\left(
\sum_k
\exp(a_{k,r}^U l_r+b_{k,r}^U)
\exp(a_{m_y}^U(y-l_{m_y}))
\right)\\
&=F^U(l_{m_y})+a_{m_y}^U(y-l_{m_y})\\
&=a_{m_y}^U y+b_{m_y}^U.
\end{aligned}
\]
Thus, \(F^U(y)\le a_{m_y}^U y+b_{m_y}^U\) holds for all \(y\in[l_{m_y},+\infty)\).

We then consider the left-unbounded interval, i.e., $y \in(-\infty,u_1]$. We define
\[
a_1^U=\min_k a_{k,r}^U,
b_1^U=F^U(u_1)-a_r^U u_1.
\]
For any \(y\in (-\infty,u_1]\), we have \(y-u_1\le 0\). Since \(a_{k,r}^U\ge a_1^U\) for every \(k\), it follows that
\[
a_{k,r}^U(y-u_1)\le a_1^U(y-u_1).
\]
Therefore,
\[
\begin{aligned}
F^U(y)
&=\log\left(
\sum_k
\exp(a_{k,r}^U u_r+b_{k,r}^U)
\exp(a_{k,r}^U(y-u_1))
\right)\\
&\le
\log\left(
\sum_k
\exp(a_{k,r}^U u_r+b_{k,r}^U)
\exp(a_1^U(y-u_1))
\right)\\
&=F^U(u_1)+a_1^U(y-u_1)\\
&=a_1^U y+b_1^U.
\end{aligned}
\]
Thus, \(F^U(y)\le a_1^U y+b_1^U\) holds for all \(y\in(-\infty,u_1]\).

\section{Soundness Proof of Affine Abstract Transformer}
\applabel{affinesound}
In this section, we provide the complete proof of the soundness of the affine abstract transformer. The notation is consistent with~\secref{affine}.

\begin{proof}
$\forall a \in \mathcal{A}_n$ we have
\[
\begin{aligned}
&T_f(\gamma_n(a))
=
\left\{
T_f(\mathbf K)
\mid
\mathbf K \in\gamma_n(a)
\right\}
\\
&=
\left\{
T_f(\mathbf K)
\;\middle|\;
\mathbf K=(K_1,\ldots,K_n)\in\mathcal K^n,
\forall t\in[n],\,
\forall A\in\mathcal B(\mathbb R),
\int_L^\#(a_t,A)
\leq K_t(\emptyset,A)
\leq \int_U^\#(a_t,A)
\right\}\\
&=\left\{
T_f(K_1, K_2,\ldots, K_n) \mid
\forall t\in[n],\,
\forall A\in\mathcal B(\mathbb R),
\int_L^\#(a_t,A)
\leq K_t(\emptyset,A)
\leq \int_U^\#(a_t,A)
\right\}\\
&=\{
(K_1, K_2, \dots, K_{i-1}) \land (K_i \in \text{Affine}(\mathbf K_J))\land (K_{i+1},\ldots, K_n)
\mid 
J \in [i-1],
\forall t \in [n], t\neq i, 
\forall A\in\mathcal B(\mathbb R),
\\
&
\int_L^\#(a_t,A)
\leq K_t(\emptyset,A)
\leq \int_U^\#(a_t,A)
\}\end{aligned}
\]
That is, $T_f$ updates only the $i$-th variable through the concrete semantics of the Affine transformation, while
all other variables remain unchanged. 

And we have
\[
\begin{aligned}
&\gamma_n(T_f^\sharp(a))= \gamma_n(a')\\
&=(\gamma(a_1^\prime), \ldots,\gamma(a_{i-1}^\prime),\gamma(a_{i}^\prime),\gamma(a_{i+1}^\prime),\ldots, \gamma(a_{n}^\prime))\\
&=(\gamma(a_1), \ldots,\gamma(a_{i-1}),\gamma(a_{i}^\prime),\gamma(a_{i+1}),\ldots, \gamma(a_{n}))\\
&=\{(K_1,\ldots,K_{i-1})\land \gamma(a_i^\prime)\land(K_{i+1},\ldots, K_n)
\mid 
\forall t \in [n], t\neq i, 
\forall A\in\mathcal B(\mathbb R),
\int_L^\#(a_t,A)
\leq K_t(\emptyset,A)
\leq \int_U^\#(a_t,A)
\}\\
\end{aligned}
\]
Therefore, if we want to prove that $\forall a \in \mathcal{A}_n,T_f(\gamma_n(a))= \gamma_n(T_f^\sharp(a))$, it suffices
to prove that
\[
\left\{
K_i\in \text{Affine}(\mathbf K_J)
\;\middle|\;
\forall j\in[J],\,
\forall A\in\mathcal B(\mathbb R),
\int_L^\#(a_j,A)
\leq K_j(\emptyset,A)
\leq \int_U^\#(a_j,A)
\right\}
\subseteq
\gamma(a_i').
\]
Let $LHS$ and $RHS$ denote the left-hand side
and right-hand side of this inclusion, respectively.
By the definition of the concretization function,
\[
RHS
=
\left\{
K_i'\in\mathcal K
\;\middle|\;
\forall A\in\mathcal B(\mathbb R),
\int_L^\sharp(a_i',A)
\leq K_i'(\emptyset,A)
\leq \int_U^\sharp(a_i',A)
\right\}.
\]

Consider an arbitrary $K_i\in LHS$, which is obtained from a joint input probability measure $\mu_J$ by the concrete semantics of the affine transformation in~\secref{marginalcollecting}.
As established in~\secref{affine}, the output probability measure
$K_i(\emptyset,\cdot)$ is absolutely continuous with respect to the Lebesgue measure $\lambda$, with density denoted as $f_Y$. Thus, for any $A\in\mathcal B(\mathbb R)$,
\(
K_i(\emptyset,A)
=
\int_A f_Y(y)\,\lambda(dy).
\)

By the derivation of the affine abstract transformer in~\secref{affine}, we have 
\(
a_i'
=
\{
\langle
l^r, u^r,
\exp(a_{r}^L y+b_{r}^L),
\exp(a_{r}^U y+b_{r}^U)
\rangle
\}_{r}
\). Therefore for every output interval $J^r$, the derived  bounds satisfy
\(
L_r(y)\leq f_Y(y)\leq U_r(y), \forall y\in J^r.
\) 
which holds for each $f_Y(y)$ derived from each joint input probability measure $\mu_J$. The soundness follows from the derivation steps in~\secref{affine}.
%
Therefore we have
\[
\int_L^\sharp(a_i',A)
\leq
\int_A f_Y(y)\,\lambda(dy)
=
K_i(\emptyset,A)
\leq
\int_U^\sharp(a_i',A).
\]
Therefore, by the definition of $\gamma$,
we have $K_i\in\gamma(a_i')$.
Since $K_i$ was chosen arbitrarily from $LHS$,
we conclude that
\(
LHS\subseteq RHS.
\)
Therefore we obtain
$
T_f(\gamma_n(a))
\subseteq
\gamma_n(T_f^\sharp(a)).
$
Therefore, the affine abstract transformer is sound.
\end{proof}

\section{Extension to Bayesian Inference}
\applabel{extend}
In this section, we demonstrate the generality of our framework by extending it beyond Bayesian neural networks to probabilistic programs with bounded loops and Bayesian inference. We support a broader class of random variables that admit probability densities with respect to the Lebesgue measure, including Uniform, Exponential, Gaussian, Laplace, Logistic, Gamma (with shape parameter $\alpha\ge1$), Beta (with parameters $\alpha\ge1,\beta\ge1$), Student's $t$, and Pareto distributions, all of which admit densities that do not diverge in the tails. We provide sound abstractions for these distributions by exploiting structural properties of their density functions while employing a conservative fallback abstraction for distributions that do not satisfy these properties, 
as detailed in~\appref{app:abstraction}.
Building on this abstraction, our domain supports Bayesian inference~\cite{bayes1,bayes2}, a fundamental mechanism in probabilistic programming where prior distributions are updated using observations to obtain posterior distributions through Bayes' rule.
Bayesian inference involves multiple nonlinear operations over probability distributions, such as integration and normalization, which rarely admit closed-form solutions. As a result, directly reasoning about posterior distributions in the concrete domain is challenging~\cite{Zaiser}, motivating the development of a sound abstract transformer for tractable reasoning. As a simple extension, our current framework supports Bayesian inference consisting of (1) equality observations and (2) likelihood functions depending on a single latent variable.  This allows us to formulate Bayesian inference directly in terms of prior and posterior densities. Specifically, given a prior density \(p(x)\) and a likelihood \(p(y\mid x)\) with the observation $y=c$, we get the posterior density\(p(x \mid y=c)=\frac{p(x)\,p(y=c \mid x)}{\int_{-\infty}^{+\infty} p(x')\,p(y=c \mid x')\,dx'}.\)
Based on this, we construct sound posterior abstractions by combining prior and likelihood abstractions through multiplication and normalization.
Finally, our domain supports probabilistic programs with bounded loops through iterative application of the corresponding abstract transformers.

\subsection{Concrete Semantics}
\appref{semanticsbayes}
Formally, the syntax is defined as:
\[
\begin{array}{rcl}
P & ::= & s \mid s;P \\[3pt]

s & ::= &  x \sim dist(c_1,\dots,c_m)
\mid  
\mathbf{observe}(y \sim dist(x\pm c_1,\dots,c_m), c)
\mid
\mathbf{for}\ i=c_1\ \mathbf{to}\ c_2\ \mathbf{do}\ P  \\[3pt]

E & ::= & x,y
      \mid c \in \mathbb{R}
\\[3pt]
\end{array}
\]
where $dist \in
\{\mathsf{Normal},\mathsf{Uniform},\mathsf{Laplace},\mathsf{Logistic},
\mathsf{Exponential},\mathsf{Gamma},\mathsf{Pareto},\mathsf{Student\mbox{-}t}\}$. This syntax captures a simple setting involving two random variables, $x$ and $y$. Specifically, the sampling statement $x \sim \mathit{dist}(c_1,\ldots,c_m)$ samples $x$ independently from a distribution with $m$ constant parameters. The observation statement $\mathbf{observe}(y \sim \mathit{dist}(x\pm c_1,\ldots,c_m),c)$ consists of two steps. First, it samples $y$ from a distribution whose parameters depend on $x$, with one parameter taking the expression of the form $x \pm c_1$ and the remaining parameters being constants. Second, it conditions the program on observing $y=c, c\in \mathbb R$. 
It states that mathematically, given a prior density \(p(x)\) and a likelihood \(p(y\mid x)\) with the observation $y=c$, we get the posterior density\(p(x \mid y=c)=\frac{p(x)\,p(y=c \mid x)}{\int_{-\infty}^{+\infty} p(x')\,p(y=c \mid x')\,dx'}.\) 
We additionally allow bounded loops to express repeated observations. 
%
We then apply the marginal collecting semantics defined in~\secref{sec:semantics} to support the above probabilistic programs. 
Since the distributions used for sampling and likelihood evaluation are continuous, we adopt the definition defined in~\secref{sec:abstraction}. Let
$\mathcal K_{\mathrm{con}}\subseteq\mathcal K$ denote the set of marginal kernels whose associated probability measures are absolutely continuous with respect to the Lebesgue measure $\lambda$.
For $K\in\mathcal K_{\mathrm{con}}$, let $f_K$ denote its
probability density function, such that
\(
K(\emptyset,A)=\int_A f_K(x)\,dx,
\forall A\in\mathcal B(\mathbb R).
\)
For a constant $c\in\mathbb R$, let $K_c$ denote the Dirac kernel concentrated at $c$, i.e., \(K_c(\emptyset,A)=\delta_c(A).\)
The semantics is shown in~\figref{fig:bayesian-semantics}.
\begin{figure}[t]
\centering
\small
\[
\begin{aligned}
&
\llbracket x\rrbracket_p^0(\sigma_p)
=\{\sigma_p(x)\},
\qquad
\llbracket y\rrbracket_p^0(\sigma_p)
=\{\sigma_p(y)\},
\\[8pt]
&
\llbracket x\sim dist(c_1,\ldots,c_m)
\rrbracket_p^0(\sigma_p)
=
\{\sigma_p[x\mapsto K_x]\},
\quad \text{where }\
K_x(\emptyset,dx)
=
f_{dist(c_1,\ldots,c_m)}(x)\,dx,
\\[8pt]
&
\llbracket
\mathbf{observe}
(y\sim dist(x\pm c_1,c_2,\ldots,c_m),c)
\rrbracket_p^0(\sigma_p)
=
\{\sigma_p[x\mapsto K_x',\,y\mapsto K_c]\},
\\
&\qquad
\text{where }\
K_x'(\emptyset,dx)
=
\frac{
f_{\sigma_p(x)}(x)\,
f_{dist(x\pm c_1,c_2,\ldots,c_m)}(c)
}{
\int_{\mathbb R}
f_{\sigma_p(x)}(u)\,
f_{dist(u\pm c_1,c_2,\ldots,c_m)}(c)\,du
}\,dx,
\\[8pt]
&
\llbracket
\mathbf{for}\ i=c_1\ \mathbf{to}\ c_2\
\mathbf{do}\ P
\rrbracket_p^0(\sigma_p)
=
\llbracket
\underbrace{P;P;\cdots;P}_{c_2-c_1+1}
\rrbracket_p^0(\sigma_p),
\quad
\llbracket s_1;s_2\rrbracket_p(c')
=
\llbracket s_2\rrbracket_p
\bigl(\llbracket s_1\rrbracket_p(c')\bigr).
\end{aligned}
\]
\caption{Marginal collecting semantics for probabilistic programs with Bayesian inference and bounded loops.}
\figlabel{fig:bayesian-semantics}
\vspace{-10pt}
\end{figure}
The expression rules return the marginal kernel associated with each variable.
The sampling rule initializes the marginal distribution of $x$ with the specified distribution, where $f_{dist(c_1,\ldots,c_m)}$ denotes the corresponding probability density function of the distribution parameterized by $c_1,\ldots,c_m$. The observation rule updates the marginal distribution of $x$ and $y$ according to Bayes' rule, where \(f_{dist(x\pm c_1,c_2,\ldots,c_m)}(c)\) denotes the probability density function of \(dist(x\pm c_1,c_2,\ldots,c_m)\) evaluated at \(y=c\), and can therefore be viewed as a function of \(x\), and the denominator is the normalization constant. The resulting posterior distribution of $x$ is represented by $K_x'$, while the observed variable $y$ is assigned the Dirac kernel
$K_c$, concentrated at the observed value $c$. The bounded loop is interpreted by unrolling it into a finite sequence of statements, where each execution of an observation statement represents a fresh observation that is independent of previous observations. 

\subsection{Bayesian inference abstract transformer}
Using the abstract domains defined in~\secref{expo}, we now introduce the design of the Bayesian inference abstract transformer in our setting. Formally, the Bayesian inference abstract transformer is defined as below.
Let
$
f:\mathcal K \rightarrow \mathcal K^2
$
be the function that executes the assignment
$
x_i, x_j \leftarrow \text{observe}(x_j\sim dist(x_i\pm c_1,c_2,\ldots,c_m),c)
$
on two random variables $x_i, x_j$, where the new-sampled random variable $x_j$ depends on $x_i$ through the specified conditional distribution and $c\in \mathbb R$ is the observed value.
The corresponding abstract transformer is defined as
\(
T_f^\sharp(a) = a^\prime\), where $a, a^\prime \in \mathcal{A}_n, a = (a_1,\ldots, a_n), a^\prime = (a_1^\prime, \ldots, a_n^\prime)$. For all $t\notin\{i,j\}$, $a_t^\prime= \{(l_t^{\prime k'},u_t^{\prime k'},L_t^{\prime k'},U_t^{\prime k'})\}_{k'\in[m^\prime _t]}=a_t = \{(l_t^{k},u_t^{k},L_t^{k},U_t^{k})\}_{k\in[m_t]}$. 

Since the observation fixes $x_j$ at the observed value $c$, its marginal distribution becomes a Dirac measure concentrated at $c$. Thus, we trivially have
$
a_j^\prime
=
\{(l_j^{\prime k},u_j^{\prime k},
L_j^{\prime k},U_j^{\prime k})\}_{k\in[m_j^\prime]}
=
\{(c,c,1,1)\},
$
with $m_j^\prime=1$. To construct the new component
\(
a_i^\prime
=
\{(l_i^{\prime k},u_i^{\prime k},
L_i^{\prime k},U_i^{\prime k})\}_{k\in[m_i^\prime]},
\)
we divide the transformer into two steps: (1)likelihood abstraction and (2)posterior transformation.
In the first step, we treat the likelihood of the observation $x_j=c$ given $x_i$ as a nonnegative function of $x_i$ and construct its abstraction with our abstraction function. In the second step, we combine the resulting likelihood abstraction with the prior abstraction $a_i$ and apply normalization to construct the posterior abstraction $a_i^\prime$.
Specifically, we first abstract the observed likelihood function. Given the observation $x_j\sim dist(x_i\pm c_1,c_2,\ldots,c_m)$ with observed value $c$, we have the observed likelihood function $\ell:\mathbb R\rightarrow\mathbb R_{\geq 0}$ as
$
\ell(x_i) = f_{dist(x_i\pm c_1,c_2,\ldots,c_m)}(c).
$
Note that $\ell$ is seen as a function of $x_i$ for the following construction. We construct an abstraction of $\ell(x_i)$ using the piecewise representation introduced in~\secref{expo}.
Formally, let $a_{\mathrm{lik}} \in \mathcal A$ and 
\(
a_{\mathrm{lik}}
=
\alpha(\ell)
=
\{(l_{\mathrm{lik}}^r,u_{\mathrm{lik}}^r,
L_{\mathrm{lik}}^r,U_{\mathrm{lik}}^r)\}_{r\in[m_{\mathrm{lik}}]},
\)
where, for every $A\in\mathcal B(\mathbb R),
\int_L^\sharp(a_{\mathrm{lik}},A)
\leq
\int_A \ell(x)\,dx
\leq
\int_U^\sharp(a_{\mathrm{lik}},A).
$
The lower and upper bounds are constructed using the same techniques described in~\appref{app:abstraction}. 
Then we compute the posterior. Given the prior abstraction $a_i=\{(l_i^{k},u_i^{k},L_i^{k},U_i^{k})\}_{k\in[m_i]}$ and the likelihood abstraction $a_{\mathrm{lik}}=\{(l_{\mathrm{lik}}^r,u_{\mathrm{lik}}^r,
L_{\mathrm{lik}}^r,U_{\mathrm{lik}}^r)\}_{r\in[m_{\mathrm{lik}}]}$, both defined over the variable \(x_i\), we construct the posterior abstraction $a_i^\prime$ by applying Bayes' rule.
We first construct a common refinement of the partitions in $a_i$ and $a_{\mathrm{lik}}$ for the alignment.
For every pair $k\in[m_i]$ and $r\in[m_{\mathrm{lik}}]$ such that
$
[l_i^{k},u_i^{k}]
\cap
[l_{\mathrm{lik}}^r,u_{\mathrm{lik}}^r]
\neq\emptyset,
$
the corresponding refined interval is defined as
\[
[l_i^{\prime k'},u_i^{\prime k'}]
=
\bigl[
\max(l_i^{k},l_{\mathrm{lik}}^r),
\min(u_i^{k},u_{\mathrm{lik}}^r)
\bigr].
\]
By Bayes' rule, the unnormalized posterior density is
$
\widetilde f_{K_i'}(x_i)
=
f_{K_i}(x_i)\ell(x_i).
$
By the soundness of the prior and likelihood abstractions, for every $A\in\mathcal B(\mathbb R)$, we have
$
\int_L^\sharp(a_i,A)
\leq \int_A f_{K_i}(x_i)\,dx_i
\leq \int_U^\sharp(a_i,A),
\int_L^\sharp(a_{\mathrm{lik}},A)
\leq \int_A \ell(x_i)\,dx_i
\leq \int_U^\sharp(a_{\mathrm{lik}},A).
$
Consequently, for each refined partition
$[l_i^{\prime k'},u_i^{\prime k'}]$,
where $k'\in[m_i']$, we have
$
L_i^k(x_i)
\leq f_{K_i}(x_i)
\leq U_i^k(x_i),
L_{\mathrm{lik}}^r(x_i)
\leq \ell(x_i)
\leq U_{\mathrm{lik}}^r(x_i)
$
for almost every
$x_i\in[l_i^{\prime k'},u_i^{\prime k'}]$.
Since all bounds are nonnegative, we obtain bounds for unnormalized posterior of $x_i$:
\[
\widetilde L_i^{\prime k'}(x_i)
=
L_i^{k}(x_i)L_{\mathrm{lik}}^r(x_i)
\leq
\widetilde f_{K_i'}(x_i)
\leq
U_i^{r}(x_i)U_{\mathrm{lik}}^r(x_i)
=
\widetilde U_i^{\prime k'}(x_i).
\]
As the bounds are exponential-affine, their products remain exponential-affine.
We then compute bounds on the normalization constant
$
Z=\int_{\mathbb R}\widetilde f_{K_i'}(x_i)\,dx_i
$
by integrating the unnormalized posterior bounds over all refined intervals:
\[
Z^L
=
\sum_{k'\in[m_i^\prime]}
\int_{l_i^{\prime k'}}^{u_i^{\prime k'}}
\widetilde L_i^{\prime k'}(x_i)\,dx_i,
\qquad
Z^U
=
\sum_{k'\in[m_i^\prime]}
\int_{l_i^{\prime k'}}^{u_i^{\prime k'}}
\widetilde U_i^{\prime k'}(x_i)\,dx_i.
\]
We require $0<Z^L\leq Z^U<\infty$ as a precondition for the Bayesian inference transformer. If $Z^L=0$ or $Z^U=\infty$, a larger partition budget is needed to improve precision and satisfy the precondition for applying the abstract Bayesian inference transformer.
The normalized posterior bounds are defined as
\[
L_i^{\prime k'}(x_i)
=
\frac{\widetilde L_i^{\prime k'}(x_i)}{Z^U},
\qquad
U_i^{\prime k'}(x_i)
=
\frac{\widetilde U_i^{\prime k'}(x_i)}{Z^L}.
\]
Therefore, for every $k'\in[m_i^\prime]$ and
$x_i\in [l_i^{\prime k'},u_i^{\prime k'}]$,
\(
L_i^{\prime k'}(x_i)
\leq
f_{K_i'}(x_i)
\leq
U_i^{\prime k'}(x_i)
\). We therefore have the output $a_i'$: 
\[
a_i' = \{\langle l_i^{\prime k'},u_i^{\prime k'}, L_i^{\prime k'},U_i^{\prime k'} \rangle\}_{k' \in [m_i']}.
\]
\begin{theorem}
The Bayesian inference abstract transformer is sound.
\end{theorem}
\begin{proof}
We prove the soundness of the Bayesian inference abstract transformer by proving $T_f(\gamma_n(a))\sqsubseteq \gamma_n(T_f^\sharp (a)), \forall a \in \mathcal{A}_n$.
Given an $a \in \mathcal{A}_n$ we have
\[
\begin{aligned}
&T_f(\gamma_n(a))
=
\left\{
T_f(\mathbf K)
\mid
\mathbf K \in\gamma_n(a)
\right\}
\\
&=
\left\{
T_f(\mathbf K)
\;\middle|\;
\mathbf K=(K_1,\ldots,K_n)\in\mathcal K^n,
\forall t\in[n],\,
\forall A\in\mathcal B(\mathbb R),
\int_L^\#(a_t,A)
\leq K_t(\emptyset,A)
\leq \int_U^\#(a_t,A)
\right\}\\
&=\left\{
T_f(K_1, K_2,\ldots, K_n) \mid
\forall t\in[n],\,
\forall A\in\mathcal B(\mathbb R),
\int_L^\#(a_t,A)
\leq K_t(\emptyset,A)
\leq \int_U^\#(a_t,A)
\right\}\\
&=\{
(K_1, K_2, \dots, K_{i-1}) \land K_i'\land (K_{i+1},\ldots, K_{j-1}) \land
 K_j' \land (K_{j+1},\ldots, K_{n})
\mid 
\forall t \in [n],
\forall A\in\mathcal B(\mathbb R),
\\
& 
\quad 
\int_L^\#(a_t,A)
\leq K_t(\emptyset,A)
\leq \int_U^\#(a_t,A)
\}
\end{aligned}
\]
That is, $T_f$ updates the $i$-th and $j$-th variables through the concrete semantics of the Bayesian Inference transformation defined in~\applabel{semanticsbayes}. Here $K_i'$ represents the posterior distribution of $x_i$, and \(K_j'\) represents the Dirac measure concentrated at the observed value \(c\). Specifically, based on the marginal collecting semantics, 
\(
K_i' \in
\{
K\in\mathcal K
\;|\;
\forall A\in\mathcal B(\mathbb R),
K(\emptyset,A)
=
\int_A
\frac{p_i(x)\ell(x)}{\int_{\mathbb R}p_i(u)\ell(u)\,du}\,\lambda(dx),
 \forall A \in \mathcal B (\mathbb R),
\int_L^\#(a_i,A)
\leq
\int_A p_i(x)\,\lambda(dx)=K_i(\emptyset,A)
\leq
\int_U^\#(a_i,A)
\},
\)
and
\(
K_j'(\emptyset,A)=\delta_c(A).
\)
Here $p_i$ is the density associated with $K_i$, $\ell(x)$ is the observed likelihood function. All remaining marginal kernels are unchanged.

And we have
\[
\begin{aligned}
&\gamma_n(T_f^\sharp(a))= \gamma_n(a')\\
&=(\gamma(a_1^\prime), \ldots,\gamma(a_{i-1}^\prime),\gamma(a_{i}^\prime),\gamma(a_{i+1}^\prime),\ldots, \gamma(a_{n}^\prime))\\
&=(\gamma(a_1), \ldots,\gamma(a_{i-1}),\gamma(a_{i}^\prime),\gamma(a_{i+1}),\ldots, \gamma(a_{n}))\\
&=\{(K_1,\ldots,K_{i-1})\land \gamma(a_i^\prime)\land(K_{i+1},\ldots, K_{j-1})\land \gamma(a_j^\prime)\land(K_{j+1},\ldots, K_{n})
\mid 
\\
&
\quad
\forall t \in [n], t\notin \{i,j\}, 
\forall A\in\mathcal B(\mathbb R),
\int_L^\#(a_t,A)
\leq K_t(\emptyset,A)
\leq \int_U^\#(a_t,A)
\}\\
\end{aligned}
\]
Therefore, if we want to prove that $\forall a \in \mathcal{A}_n,T_f(\gamma_n(a))\subseteq \gamma_n(T_f^\sharp(a))$, it suffices
to prove that 
\begin{enumerate}
    \item $\{
K_i'\in\mathcal K
\;|\;
\forall A\in\mathcal B(\mathbb R),
K_i'(\emptyset,A)
=
\int_A
\frac{p_i(x)\ell(x)}{\int_{\mathbb R}p_i(u)\ell(u)\,du}\,\lambda(dx),
\forall A \in \mathcal B (\mathbb R),
\int_L^\#(a_i,A)
\leq
\int_A p_i(x)\,\lambda(dx)=K_i(\emptyset,A)
\leq
\int_U^\#(a_i,A)
\}\subseteq \gamma(a_i')$;

\item $\{K_j'\in\mathcal K\mid 
\forall A\in\mathcal B(\mathbb R), K_j'(\emptyset,A)=\delta_c(A)\} \subseteq \gamma(a_j')$.
\end{enumerate}

We first prove (1). Let $\mathrm{LHS}$ and $\mathrm{RHS}$ denote the
left-hand side and right-hand side of (1), respectively.
By the definition of the concretization function,
\[
\mathrm{RHS}
=\gamma(a_i')
=
\left\{
K_i'\in\mathcal K
\;\middle|\;
\forall A\in\mathcal B(\mathbb R),
\int_L^\#(a_i',A)
\leq K_i'(\emptyset,A)
\leq \int_U^\#(a_i',A)
\right\}.
\]
We prove (1) by recalling the step-by-step construction of $a_i'$ based on $a_i$ in terms of $K_i$, from which leads to the conclusion that we have each $K_i'$ in LHS belongs to $\gamma(a_i').$  Specifically, given any $K_i'$,
let
\(
Z=\int_{\mathbb R}p_i(x)\ell(x)\,dx
\)
denote the normalization constant, $p_i'(x)=\frac{p_i(x)\ell(x)}{Z}$
denotes the posterior density associated with $K_i'$.
Since  $\int_L^\#(a_i,A)
\leq
\int_A p_i(x)\,\lambda(dx)=K_i(\emptyset,A)
\leq
\int_U^\#(a_i,A)$, in the continuous setting
of Bayesian inference, this gives the prior density bounds on every non-degenerate partition
$I_i^k=(l_i^k,u_i^k)$:
\(
L_i^k(x)\leq p_i(x)\leq U_i^k(x)
\)
for every $x\in I_i^k$.
By the soundness of the likelihood abstraction
$a^{\mathrm{lik}}=\alpha(\ell)$, on every
partition $I_{\mathrm{lik}}^r$ we have
\(
L_{\mathrm{lik}}^r(x)
\leq \ell(x)
\leq U_{\mathrm{lik}}^r(x)
\)
for every $x$ in its corresponding partition
$I_{\mathrm{lik}}^r$.
Consider an arbitrary nonempty refined interval
\(
I_i^{\prime k'}
=
I_i^k\cap I_{\mathrm{lik}}^r
\)
for some $k\in[m_i]$ and
$r\in[m_{\mathrm{lik}}]$.
We have 
for every $x\in I_i^{\prime k'}$,
the prior and likelihood bounds imply
\(
L_i^k(x)\leq p_i(x)\leq U_i^k(x),
L_{\mathrm{lik}}^r(x)
\leq \ell(x)
\leq U_{\mathrm{lik}}^r(x).
\)
Since all these bounds are nonnegative, multiplying them yields
\(
L_i^k(x)L_{\mathrm{lik}}^r(x)
\leq p_i(x)\ell(x)
\leq U_i^k(x)U_{\mathrm{lik}}^r(x).
\)
Then the unnormalized posterior
bounds over $I_i^{\prime k'}$ are
\(
\widetilde L_i^{\prime k'}(x)
=L_i^k(x)L_{\mathrm{lik}}^r(x),
\widetilde U_i^{\prime k'}(x)
=U_i^k(x)U_{\mathrm{lik}}^r(x).
\)
Therefore,
\(
\widetilde L_i^{\prime k'}(x)
\leq p_i(x)\ell(x)
\leq
\widetilde U_i^{\prime k'}(x).
\)
Since the refined intervals cover the support
of the prior density, integrating these
inequalities over all refined intervals yields
\(
Z^L\leq Z\leq Z^U,
\
\text{where}
\
Z^L
=
\sum_{k'\in[m_i']}
\int_{I_i^{\prime k'}}
\widetilde L_i^{\prime k'}(x)\,dx,
\
Z^U
=
\sum_{k'\in[m_i']}
\int_{I_i^{\prime k'}}
\widetilde U_i^{\prime k'}(x)\,dx.
\)
Obviously we have
$0<Z^L\leq Z^U<\infty$. Therefore we obtain
\(
\frac{1}{Z^U}
\leq \frac{1}{Z}
\leq \frac{1}{Z^L}.
\)
Combining this inequality with the unnormalized
posterior bounds, for
$x\in I_i^{\prime k'}$ we have
\(
\frac{\widetilde L_i^{\prime k'}(x)}{Z^U}
\leq
\frac{p_i(x)\ell(x)}{Z}
\leq
\frac{\widetilde U_i^{\prime k'}(x)}{Z^L}.
\)
Let
\(
L_i^{\prime k'}(x)
=
\frac{\widetilde L_i^{\prime k'}(x)}{Z^U},
U_i^{\prime k'}(x)
=
\frac{\widetilde U_i^{\prime k'}(x)}{Z^L}.
\)
Thus, for 
$x\in I_i^{\prime k'}$,
\(
L_i^{\prime k'}(x)
\leq p_i'(x)
\leq U_i^{\prime k'}(x).
\)
Since the refined intervals cover the posterior
support, integrating the above bounds over any
measurable set $A\in\mathcal B(\mathbb R)$ gives
\(
\int_L^\#(a_i',A)
\leq
\int_A p_i'(x)\,\lambda(dx)=K_i'(\emptyset,A)
\leq
\int_U^\#(a_i',A).
\)
which implies that any $K_i'$ in LHS belongs to $\gamma(a_i)$.Therefore we have $LHS \subseteq RHS$.

We then prove (2).
Considering RHS,
since by the definition of the Bayesian inference
abstract transformer, we have
\(
a_j'=\{\langle c,c,1,1\rangle\}.
\)
Therefore for every
$A\in\mathcal B(\mathbb R)$ we have
\(
\int_L^\#(a_j',A)
=
\int_U^\#(a_j',A)
=
\mathbf 1_A(c).
\)
Consequently, we have
\[
\begin{aligned}
RHS = \gamma(a_j') &= \{K \mid 
\forall A\in\mathcal B(\mathbb R)
\int_L^\#(a_j',A)
\leq K_j'(\emptyset,A)
\leq \int_U^\#(a_j',A)
\}
\\
&
= \{K \mid 
\forall A\in\mathcal B(\mathbb R)
\leq K_j'(\emptyset,A)
=
\mathbf 1_A(c)
\}.
\end{aligned}
\]
Therefore we have LHS =RHS.

\end{proof}

\subsection{Illustrative Example of Bayesian Inference}
\begin{figure}[t]
\centering
\begin{subfigure}{0.32\textwidth}
\centering
\includegraphics[width=\linewidth]{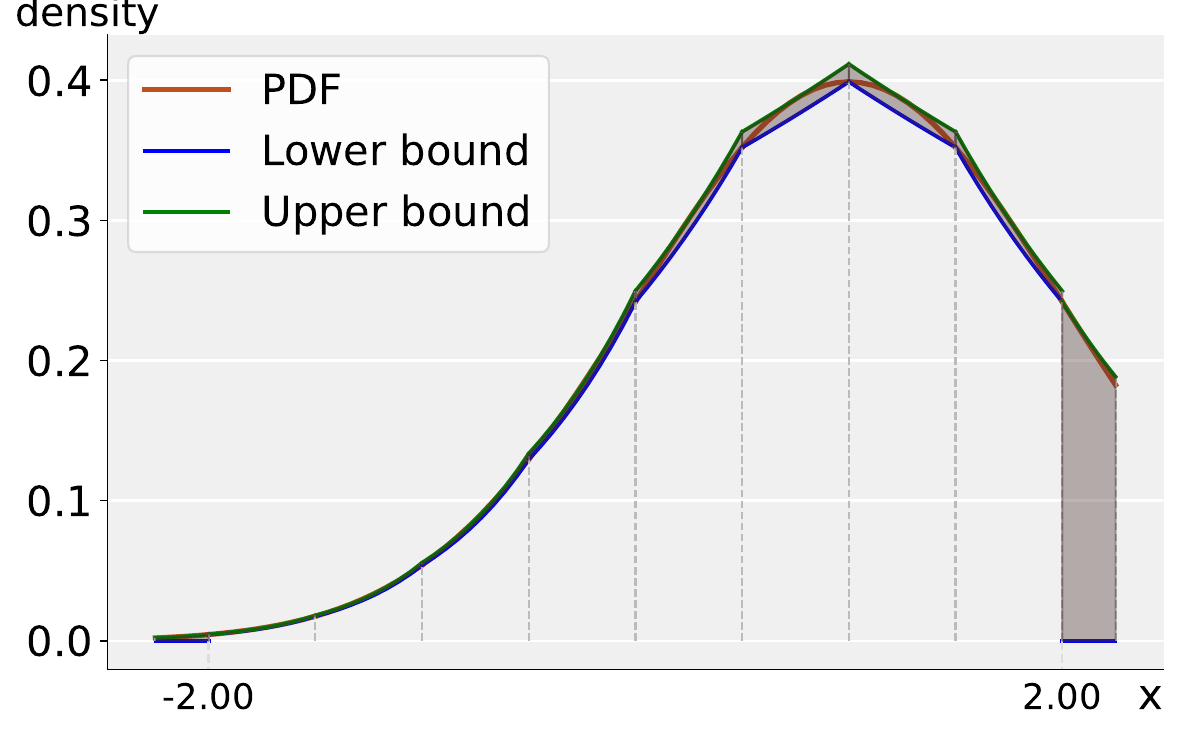}
\caption{Abstraction of the prior $p(x)$ with $x\sim N(1,1)$.}
\figlabel{fig:prior}
\end{subfigure}
\hfill
\begin{subfigure}{0.32\textwidth}
\centering
\includegraphics[width=\linewidth]{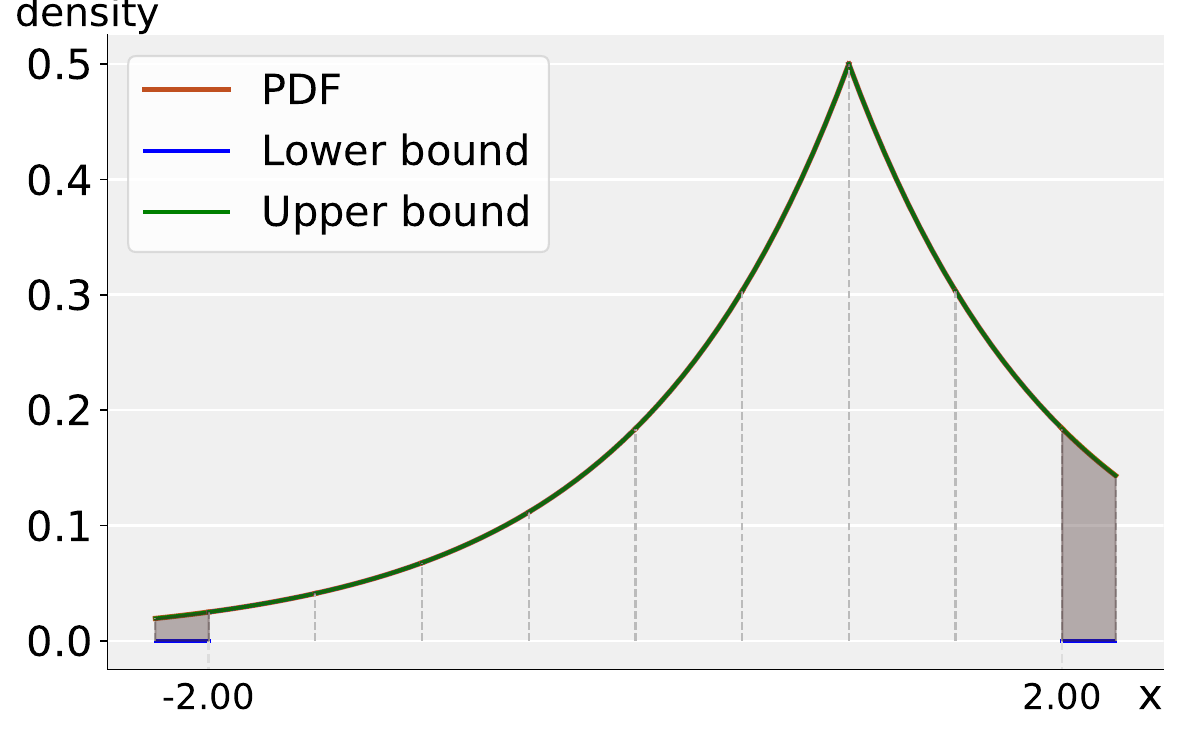}
\caption{Abstraction of the likelihood $l(x)$.}
\figlabel{fig:likelihood}
\end{subfigure}
\hfill
\begin{subfigure}{0.32\textwidth}
\centering
\includegraphics[width=\linewidth]{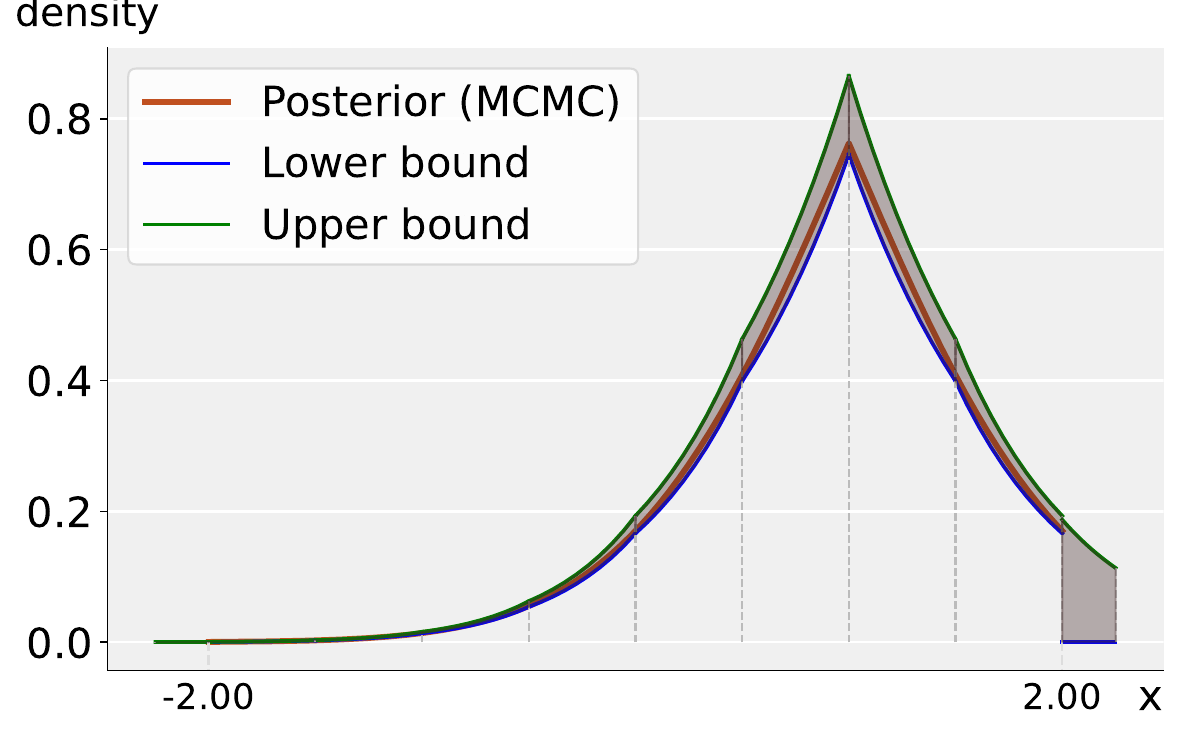}
\caption{Abstraction of the posterior $p(x\mid y=1)$.}
\figlabel{fig:posterior}
\end{subfigure}

\caption{An illustrative example of Bayesian inference in our domain.}
\label{fig:main}
\end{figure}
\begin{wrapfigure}{r}{0.4\textwidth}
\begin{lstlisting}
x ~ Normal(1,1)
observe (y~ Laplace(x,1), 1)
\end{lstlisting}
\vspace{-5pt}
\caption{A simple Bayesian inference example.}
\figlabel{fig:bayesian_example}
\vspace{-10pt}
\end{wrapfigure}

We use a simple probabilistic program with Bayesian inference to illustrate our extension.
Consider the simple probabilistic model in~\figref{fig:bayesian_example}, which defines a Normal prior over the latent variable $x$, a Laplace conditional distribution for $y \mid x$, and performs posterior inference given the observation $y=1$. The original supports of $x$ and $y$ are both $(-\infty, \infty)$. 
The Bayesian inference abstract transformer is built upon three key ideas. First, both the prior and the observed likelihood are viewed as functions over the latent variable $x$ when constructing their abstractions,
therefore the transformer only needs to manipulate different abstractions of the same variable. 
Second, Bayesian inference is performed in the log domain, where complicated multiplication and division in Bayes' rule become simple addition and subtraction, which also naturally matches the structure of our abstraction. Third, the exponential-affine representation enjoys two crucial properties: it is closed under the affine operations arising in the log domain, enabling sound bounds on the unnormalized posterior, and it admits closed-form integration, enabling sound bounds on the normalization constant. Combining these two properties allows us to obtain the sound approximation of the normalized posterior.
Returning to our example, following the abstraction function introduced above, we first construct the abstraction $a_{pr}$ of the prior distribution of $x$. We have $K_x(\emptyset, dx)=p(x)dx$, where $p(x)
=
\frac{1}{\sqrt{2\pi}}
\exp(
-\frac{(x-1)^2}{2})$. The abstarction is reduced to abstarct the density function $p(x)$. Similarly, we construct the abstraction $a_{lik}$ for the observed likelihood function $
l(x)
=
\exp\!\left(
-\frac{|1-x|}{b}
\right)
$. In this example we partition their support both into 10 splits $\{
(-\infty,-2],
[-2,-1.5], \cdots,[1.5,2],[2,\infty)\}$. Therefore we have 
\[
a_{pr} =\left\{\, \bigl(l^k, u^k, L_{pr}^k(x), U_{pr}^k(x)\bigr) \,\right\}_{k=1}^{10},
\quad
a_{lik} =\left\{\, \bigl(l^k, u^k, L_{lik}^k(x), U_{lik}^k(x)\bigr) \,\right\}_{k=1}^{10}.
\]
As visualized in~\figref{fig:prior} and \figref{fig:likelihood}.

As the goal is to obtain the abstraction for the posterior that remains in the same abstract domain, we follow the same procedure used for the prior and likelihood. Specifically, we first derive a partition of the support and then construct sound lower and upper bounds on the posterior density within each partition. The support partition of the posterior is induced by the partitions of the prior and likelihood abstractions. In general, this is achieved by aligning the interval boundaries of the two abstractions and refining the support accordingly. In this running example, the prior and likelihood abstractions already share the same partitioning, and therefore no additional refinement is required. Consequently, the posterior abstraction is constructed over the same collection of intervals $\{
(-\infty,-2],
[-2,-1.5], \cdots,[1.5,2],[2,\infty)\}$. To achieve sound lower and upper bounds on each partition of the normalized posterior, the high-level idea is to first derive interval-wise bounds on the unnormalized posterior, which are then combined to obtain bounds on the normalization constant. Together, these bounds allow us to derive sound lower and upper bounds on the normalized posterior density for each partition and thus obtain a sound abstraction of the normalized posterior.
For example, consider the partition $x\in[0,0.5]$. We have the tuple \(
(0,0.5,L_{pr}(x),U_{pr}(x))\) in $a_{pr}$,
where
$
L_{pr}(x)
=
\exp\!\left(\frac{3}{4}x-\log\sqrt{2\pi}-\frac{1}{2}\right),
\
U_{pr}(x)
=
\exp\!\left(\frac{3}{4}x-\log\sqrt{2\pi}-\frac{15}{32}\right)
$ by computation, and the tuple $(0,0.5,L_{lik}(x),U_{lik}(x))$ in $a_{lik}$,
where
$
L_{lik}(x)
=
U_{lik}(x)
=
\exp(x-1)
$ through the abstraction above.
After mapping both abstractions to the log domain, the lower and upper bounds of the unnormalized log-posterior can be obtained simply by adding the corresponding slopes and intercepts of the affine functions, i.e.,
\[
\frac74x-\log\sqrt{2\pi}-\frac32 \le \phi_{post}(x) \le
\frac74x-\log\sqrt{2\pi}-\frac{47}{32},
\quad x \in [0,0.5].
\]
Expansion yields bounds for the unnormalized posterior:
\[\tilde{L}_{post}(x) \le p(x)\,l(x) \le \tilde{U}_{post}(x) \]
where
\[
\tilde{L}_{post}(x) = \exp(\frac74x-\log\sqrt{2\pi}-\frac32), 
\quad
\tilde{U}_{post}(x) = \exp(\frac74x-\log\sqrt{2\pi}-\frac{47}{32}),
\quad x \in [0,0.5].
\]
The same construction is applied to all bounded partitions of the posterior. Tail intervals are handled separately, since the abstractions of the prior and likelihood in the tail regions employ a conservative lower bound of zero, which cannot be operated on in the log domain. Specifically, for the tail regions $(-\infty,-2]$ and $[2,\infty)$, sound upper bounds on the unnormalized posterior can still be obtained by combining the exponential-affine upper bounds of the prior and likelihood. For the lower bounds, we employ a conservative fallback and simply set the density lower bound to zero. In this way, we obtain interval-wise bounds on the unnormalized posterior density over the entire support. 
We then use the unnormalized posterior abstraction to construct the bounds of the denominator and then derive sound bounds on the normalized posterior density.
Since exponential-affine functions admit closed-form integration, we can soundly integrate the lower and upper bounds on each partition and aggregate the results to obtain sound scalar bounds on the normalization constant. Specifically, \[
Z_{\mathrm{l}} = \sum_{k=1}^{10} \int_{l_k}^{u_k} \tilde{L}_{post}^k(x)\,dx,
\quad
Z_{\mathrm{u}} = \sum_{k=1}^{10}\int_{l_k}^{u_k} \tilde{U}_{post}^k(x)\,dx,
\quad Z_{\mathrm{l}},Z_{\mathrm{u}} \in \mathbb{R}
\]
Then the sound bounds on the normalized posterior can be obtained as:
\[
L_{\text{post}}^k(x)
=
\frac{\tilde{L}_{post}^k(x)}{Z_{\mathrm{u}}},
\quad
U_{\text{post}}^k(x)
=
\frac{\tilde{U}_{post}^k(x)}{Z_{\mathrm{l}}}, \quad \forall k \in [10].
\]
The abstract element outputed is therefore represented as $a_{post} =\left\{\, \bigl(l^k, u^k, L_{post}^k(x), U_{post}^k(x)\bigr) \,\right\}_{k=1}^{10}$, as shown in~\figref{fig:posterior}, which ensures:
\[
L_{\text{post}}^k(x)
\;\le\;
p(x \mid y=1)
\;\le\;
U_{\text{post}}^k(x), \quad \forall x \in [l^k,u^k], \ \forall k \in [10].
\]

\section{Optimizations for Instantiation}
\applabel{optimization}
After establishing the theoretical formulation, three major challenges arise during instantiation: 
(1)Firstly, the use of constant upper bounds over unbounded intervals may lead to infinite upper integrals. Although such bounds are sound pointwise approximations of the probability density, integrating them over unbounded intervals, such as the output tail regions of the affine abstract transformer, yields $+\infty$.
This leads to uninformative probability bounds during subsequent abstract transformations and significantly reduces the precision of the analysis;
(2) Secondly, the affine abstract transformer suffers from the curse of dimensionality. Suppose the input dimension is \(n\), and each dimension is uniformly partitioned into \(m\) intervals. Then the resulting number of high-dimensional boxes becomes \(m^n\). In practical BNN architectures, the hidden dimension is commonly at least \(64\) or \(128\), making the total number of boxes infeasible to compute even for very small values of \(m\). This will limit \(m\) to only a few splits per dimension (e.g., \(1\), \(2\), or \(5\)), significantly restricting the precision of the abstraction; (3) Thirdly, the output abstraction of the affine abstract transformer may suffer from partition explosion. As sound exponential-affine bounds require partitioning the output density into multiple subintervals, the number of partitions can grow rapidly during propagation. In practice, a single affine transformation will generate thousands or even millions of partitions if there is no proper control, causing the abstraction size to increase dramatically across layers; 
Therefore, we propose five optimization strategies to make the approach feasible and scalable in high-dimensional settings. These optimizations include:
\begin{enumerate}[leftmargin=*]
    \item a probability mass clipping strategy, where we exploit the normalization property of probability measures to obtain finite and more precise upper bounds on probability mass, especially when the corresponding density upper bounds are not integrable.
    \item an importance-guided partition strategy for inputs, where we allocate partition budgets according to neuron importance;
    \item a tail-region approximation strategy, where we approximate low-probability tail regions conservatively to avoid explicitly analyzing negligible probability regions;
    \item a greedy interval merging strategy for outputs, where we merge the partitions in the abstraction of output based on a cost function to control partition growth;
    \item the GPU parallelization strategy to accelerate efficient computations.
\end{enumerate}

\subsection{Probability Mass Clipping.}
There are cases where the construction of sound density bounds may introduce constant upper bounds over unbounded intervals, particularly in the tail regions of the affine abstract transformer. Although these bounds are sound pointwise, their integrals over unbounded intervals may diverge to $+\infty$, leading to uninformative probability bounds in subsequent abstract transformations.
To address this issue, we exploit the fact that the total probability mass of any probability distribution is $1$. Specifically, let a partition of the abstract element $\langle l_k, u_k, L_k, U_k\rangle, l_k<u_k,$ be a sound abstraction of a probability density function $f_X$ over the interval $I_k = [l_k, u_k]$.
We have
$
\int_{I_k} L_k(x)\,dx
\leq
\int_{I_k} f_X(x)\,dx
\leq
\int_{I_k} U_k(x)\,dx.
$
Moreover, since $f_X$ is a probability density function, the probability mass over any measurable interval cannot exceed $1$.
Therefore, we can obtain a more precise upper bound through clipping:
\[
\int_{I_k} L_k(x)\,dx
<
\int_{I_k} f_X(x)\,dx
<
\min\left\{
1,\int_{I_k} U_k(x)\,dx
\right\}.
\]
In particular, even when $\int_{I_k} U_k(x)\,dx = +\infty$, the resulting probability upper bound remains finite. This optimization preserves soundness without requiring the upper bounds themselves to be integrable, and can be applied whenever probability mass bounds are computed during abstract transformation.

\subsection{Importance-Guided Partition Strategy for Inputs}
To address the first challenge, we propose an importance-guided partition strategy to mitigate the dimensional explosion caused by high-dimensional probabilistic inputs. Instead of uniformly partitioning every input dimension with the same granularity, we dynamically allocate more partitions to neurons that contribute more significantly to the output distribution. The importance of each neuron is estimated based on the expected sensitivity of the network output with respect to that neuron. Specifically, for the neuron $n_j$, we estimate its importance score as
$
s_{n_j}=\mathbb E_{\mathbf{w}}\left[\left|\frac{\partial y}{\partial n_j}\right|\right],
$
where the expectation is taken over the posterior distribution of the network parameters. In practice, we estimate this expectation using Monte Carlo sampling from the posterior. $s_{n_j}$ measures the expected influence of neuron $j$ on the network output. Neurons with smaller importance scores are assigned coarser abstractions, while more influential neurons receive finer partitions. This importance-guided strategy allocates computational resources to the dimensions that contribute most to the output uncertainty, significantly reducing the number of high-dimensional boxes while preserving the precision of the probabilistic abstraction.

\subsection{Tail Region Approximation}
\applabel{tailregion}
In addition to the importance-guided partition strategy of input neurons, we further exploit the probabilistic structure of the input distribution to avoid unnecessary abstraction. Observe that although the total number of Cartesian-product boxes grows exponentially with the input dimension, the probability mass is typically highly concentrated on a small subset of regions. In particular, boxes involving multiple tail events often have exponentially small probability mass. Therefore, instead of explicitly refining all tail combinations, we separate the input space into a collection of main high-probability regions and a residual low-probability region. The main regions are handled using the standard affine abstraction procedure, while the residual region is conservatively over-approximated using trivially sound bounds.

Formally, for each input dimension \(i\), we distinguish between non-tail intervals and tail intervals. Let
\(
\mathcal{I}_i'
=
\{I_i^{k_i'} \mid I_i^{k_i'} \text{ is not a tail interval}\}
\)
denote the collection of non-tail intervals of the \(i\)-th input dimension. We then define the set of main boxes as
\(
B_{k'}
=
I_1^{k_1'}
\times
I_2^{k_2'}
\times
\cdots
\times
I_J^{k_J'},
I_i^{k_i'} \in \mathcal{I}_i'.
\)
That is, each main box consists entirely of non-tail intervals across all dimensions.
The remaining regions are merged into a residual region defined as
\(
B_{\mathrm{res}}
=
\mathbb{R}^n
\setminus
\bigcup_{k'} B_{k'}.
\)
Therefore, the density function of the output kernel of affine operation can be rewritten as
\begin{equation}
\begin{aligned}
f_Y(y)
=
&
\sum_{k'}
\Pr[\mathbf{X}_J\in B_{k'}]\,
\mathbb{E}\!\left[
g(y,\mathbf{X}_J)
\mid
\mathbf{X}_J\in B_{k'}
\right]+
\Pr[\mathbf{X}_J\in B_{\mathrm{res}}]\,
\mathbb{E}\!\left[
g(y,\mathbf{X}_J)
\mid
\mathbf{X}_J\in B_{\mathrm{res}}
\right].
\end{aligned}
\eqnlabel{eq:py4}
\end{equation}
We next approximate the residual term in~\eqnref{eq:py4}. 
Unlike the main boxes, the residual region is bounded conservatively. Since the Gaussian density is nonnegative, we immediately have
\(
\Pr[\mathbf{X}_J\in B_{\mathrm{res}}]\,
\mathbb{E}\!\left[
g_j(y,\mathbf{X}_J)
\mid
\mathbf{X}_J\in B_{\mathrm{res}}
\right]
\ge 0.
\)
For the upper bound, using the same way we can get the upper bound \(P_{\mathrm{res}}^U\) on the residual probability mass, i.e.,
\(
\Pr[\mathbf{X}_J\in B_{\mathrm{res}}] \le P_{\mathrm{res}}^U.
\)
Then, suppose that for all \(\mathbf{x}_J\in B_{\mathrm{res}}\), the conditional variance satisfies 
\(
v(\mathbf{x})
=
\sum_{j=1}^J \sigma_{W_{j}}^2 X_j^2 + \sigma_{b}^2
\ge
\sigma_{b}^2
>0.
\)
Let $v_{\mathrm{res}}^L = \sigma_{b}^2$,
then for any \(y\), we have
\(
g(y,\mathbf{X}_J)
=
\frac{1}{\sqrt{2\pi v(\mathbf{X}_J)}}
\exp\!\left(
-\frac{(y-m(\mathbf{X}_J))^2}{2v(\mathbf{X}_J)}
\right)
\le
\frac{1}{\sqrt{2\pi v_{\mathrm{res}}^L}},
\)
Therefore,
\(
\mathbb{E}\!\left[
g(y,\mathbf{X}_J)
\mid
\mathbf{X}_J\in B_{\mathrm{res}}
\right]
\le
\frac{1}{\sqrt{2\pi v_{\mathrm{res}}^L}}.
\)
Combining the two bounds, we obtain
\(
0
\le
\Pr[\mathbf{X}\in B_{\mathrm{res}}]\,
\mathbb{E}\!\left[
g_j(y,\mathbf{X})
\mid
\mathbf{X}\in B_{\mathrm{res}}
\right]
\le
P_{\mathrm{res}}^U
\cdot
\frac{1}{\sqrt{2\pi v_{\mathrm{res}}^L}}.
\)
The tail-region approximation does not affect the construction of exponential-affine bounds introduced in~\secref{affine}, since its contribution can be incorporated into the summation as a constant term.
This optimization introduces a tradeoff between scalability and precision. Specifically, by merging low-probability tail regions into a single residual region, we substantially reduce the number of regions that need to be partitioned and analyzed, thereby improving computational efficiency and scalability in high-dimensional settings. However, since the residual region is approximated using coarse conservative bounds, the resulting bounds may become less tight. In practice, this precision loss is often limited, as the residual region mainly consists of low-probability multi-tail configurations whose overall contribution to the output probability can be overlooked.

\subsection{Greedy Interval Merging for Outputs}

When an output abstract element contains a large number of partitions, it may significantly increase the difficulty of analysis and propagation in deeper layers.
To address this issue, we introduce a greedy interval merging strategy that dynamically compresses neighboring intervals while preserving soundness.
The key idea is to iteratively merge adjacent intervals whose combination introduces the smallest precision loss.

\paragraph{Sound Merged Bounds}Consider two neighboring intervals in one abstract element
$
I_1=[l_1,u_1]
$
and
$
I_2=[l_2,u_2]
$
with exponential-affine lower and upper bounds
$
L_i(y)=\exp(a_i^L y+b_i^L),
$
$
U_i(y)=\exp(a_i^U y+b_i^U),
$
\(i\in\{1,2\}\).
We merge them into a single interval
$
I'=[l_1,u_2].
$
To construct a sound merged upper bound $U(y)$, we have
\begin{equation*}
\begin{aligned}
U_{merged}(y) &=
\max(U_1(y),U_2(y))
=
\max(\exp(a_1^L y+b_1^L),\exp(a_2^L y+b_2^L))\\
&=
\exp(\max(a_1^L y+b_1^L,a_2^L y+b_2^L)).
\end{aligned}
\end{equation*}
Let
$
f_1^U(y)=a_1^U y+b_1^U,
$
$
f_2^U(y)=a_2^U y+b_2^U,
$
and
$
g^U(y)
=
\max(f_1^U(y),f_2^U(y)).
$
We can easily prove that \(g^U(y)\) is convex.

\begin{proof}

For any
\(\lambda\in[0,1]\) and \(y_1,y_2\in\mathbb{R}\),
\begin{align*}
g^U(\lambda y_1+(1-\lambda)y_2)
&=
\max(
f_1^U(\lambda y_1+(1-\lambda)y_2),
f_2^U(\lambda y_1+(1-\lambda)y_2)
)
\\
&=
\max(
\lambda f_1^U(y_1)+(1-\lambda)f_1^U(y_2),
\lambda f_2^U(y_1)+(1-\lambda)f_2^U(y_2)
)
\\
&\le
\lambda
\max(f_1^U(y_1),f_2^U(y_1))
+
(1-\lambda)
\max(f_1^U(y_2),f_2^U(y_2))
\\
&=
\lambda g^U(y_1)
+
(1-\lambda)g^U(y_2).
\end{align*}
Therefore, \(g^U(y)\) is convex.
\end{proof}
Therefore, the secant line connecting the endpoint values of \(g^U(y)\) over \(I'\) forms a sound affine upper bound of \(g^U(y)\).
Exponentiating the resulting affine function yields the merged exponential-affine upper bound, i.e.,
\(
U_{\mathrm{merge}}(y)
=
\exp(a_{\mathrm{merge}}^U y+b_{\mathrm{merge}}^U),
\)
where
\(
a_{\mathrm{merge}}^U
=
\frac{g^U(u_2)-g^U(l_1)}{u_2-l_1},
b_{\mathrm{merge}}^U
=
g^U(l_1)-a_{\mathrm{merge}}^U l_1.
\)
Similarly, to construct a sound merged lower bound \(L(y)\), we have
\begin{equation*}
\begin{aligned}
L_{merged}(y)
&=
\min(L_1(y),L_2(y))=
\min(\exp(a_1^L y+b_1^L),\exp(a_2^L y+b_2^L))\\
&=
\exp(\min(a_1^L y+b_1^L,a_2^L y+b_2^L)).
\end{aligned}
\end{equation*}
Let
$
f_1^L(y)=a_1^L y+b_1^L,
f_2^L(y)=a_2^L y+b_2^L,
$
and
$
g^L(y)
=
\min(f_1^L(y),f_2^L(y)).
$
We can easily prove that \(g^L(y)\) is concave.

\begin{proof}
For any \(\lambda\in[0,1]\) and \(y_1,y_2\in\mathbb{R}\),
\begin{align*}
g^L(\lambda y_1+(1-\lambda)y_2)
&=
\min(
f_1^L(\lambda y_1+(1-\lambda)y_2),
f_2^L(\lambda y_1+(1-\lambda)y_2)
)
\\
&=
\min(
\lambda f_1^L(y_1)+(1-\lambda)f_1^L(y_2),
\lambda f_2^L(y_1)+(1-\lambda)f_2^L(y_2)
)
\\
&\ge
\lambda
\min(f_1^L(y_1),f_2^L(y_1))
+
(1-\lambda)
\min(f_1^L(y_2),f_2^L(y_2))
\\
&=
\lambda g^L(y_1)
+
(1-\lambda)g^L(y_2).
\end{align*}
Therefore, \(g^L(y)\) is concave.
\end{proof}

Therefore, the secant line connecting the endpoint values of \(g^L(y)\) over \(I'\) forms a sound affine lower bound of \(g^L(y)\).
Exponentiating the resulting affine function yields the merged exponential-affine lower bound, i.e. 
\(
L_{\mathrm{merge}}(y)
=
\exp(a_{\mathrm{merge}}^L y+b_{\mathrm{merge}}^L),
\)
where
\(
a_{\mathrm{merge}}^L
=
\frac{g^L(u_2)-g^L(l_1)}{u_2-l_1},
b_{\mathrm{merge}}^L
=
g^L(l_1)-a_{\mathrm{merge}}^L l_1.
\)

\paragraph{Cost-based Greedy Merging} 

Given an abstract element consisting of partitions
$
\mathcal{I}=\{I_1,I_2,\dots,I_n\},
$
we first compute the merge cost for every adjacent interval pair
$
(I_i,I_{i+1}).
$
We then greedily select the pair with the smallest merge cost and replace it with the merged interval constructed above.
We define the merge cost as the precision loss introduced by replacing the original bounds with the merged bounds.
For the upper bound, the loss measures the additional probability mass introduced by the merged upper bound:
\(
\Delta_U
=
\int_{I'}
\left(
U_{\mathrm{merge}}(y)-U_{\mathrm{orig}}(y)
\right)\,dy,
\)
where
\(
U_{\mathrm{orig}}(y)
=
\begin{cases}
U_1(y), & y\in I_1,\\
U_2(y), & y\in I_2.
\end{cases}
\)
For the lower bound, the loss measures the probability mass removed by the merged lower bound:
\(
\Delta_L
=
\int_{I'}
\left(
L_{\mathrm{orig}}(y)-L_{\mathrm{merge}}(y)
\right)\,dy,
\)
where
\(
L_{\mathrm{orig}}(y)
=
\begin{cases}
L_1(y), & y\in I_1,\\
L_2(y), & y\in I_2.
\end{cases}
\)
The total merge cost is defined as
\(
\mathrm{Cost}(I_1,I_2)
=
\lambda_U \Delta_U
+
\lambda_L \Delta_L,
\)
where \(\lambda_U,\lambda_L\ge 0\) control the relative importance of preserving the upper and lower bounds. In practice, we set $\lambda_U=\lambda_L=1$.
At each step, we greedily merge the adjacent interval pair with the smallest cost, and repeat this process until the number of partitions reaches the target budget. We observe that this strategy naturally tends to merge tail regions first. 

We evaluate the performance of the greedy interval merging strategy on an illustrative example, as shown in~\figref{fig:merge}. The example consists of a toy Bayesian neural network with two concrete inputs, followed by a hidden layer with ReLU activations and a Gaussian affine output layer. Using our abstraction algorithm, the output distribution is initially represented by 579 partitions. The greedy interval merging algorithm is then applied to reduce the abstraction to a budget of 20 partitions. The abstractions before and after merging are compared against a Monte Carlo estimate of the true output density. As shown in the figure, the merged abstraction closely matches the original one while using substantially fewer partitions, demonstrating that the proposed greedy merging strategy effectively balances precision and scalability.

\begin{figure}[t]
    \centering

    \begin{subfigure}[t]{0.48\linewidth}
        \centering
        \includegraphics[width=\linewidth]{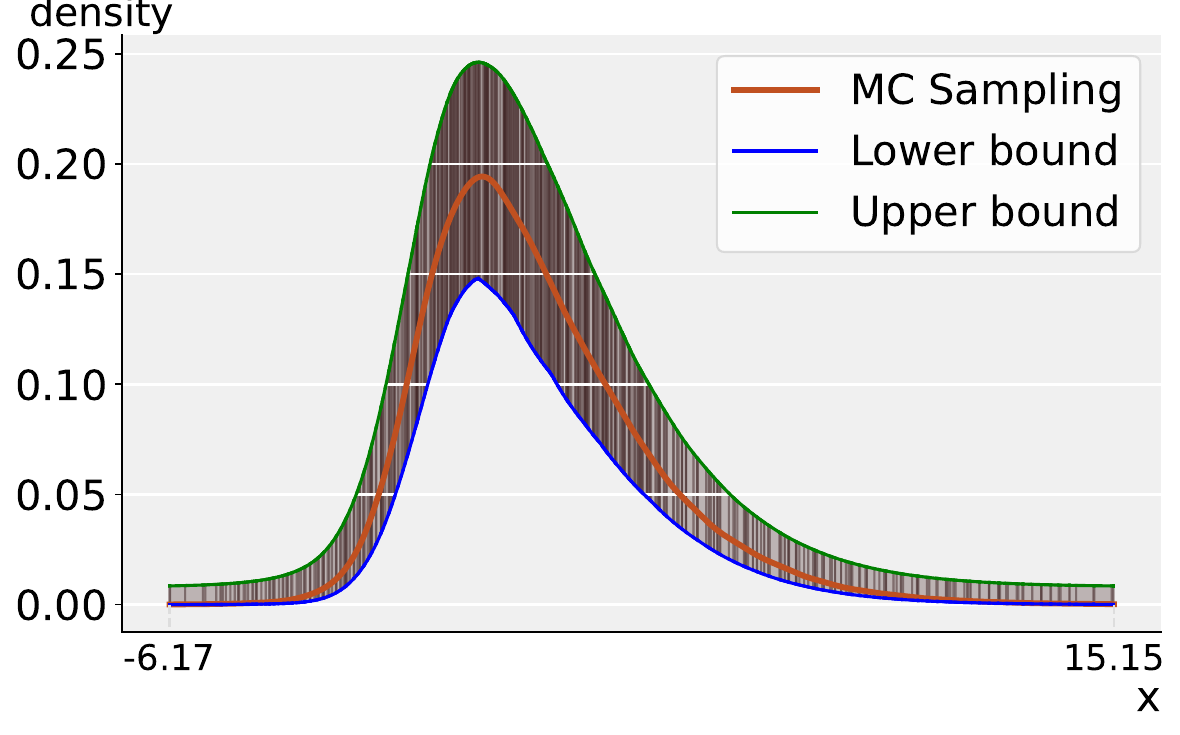}
        \caption{The output abstraction with 579 partitions before merging.}
        \label{fig:1}
    \end{subfigure}
    \hfill
    \begin{subfigure}[t]{0.48\linewidth}
        \centering
        \includegraphics[width=\linewidth]{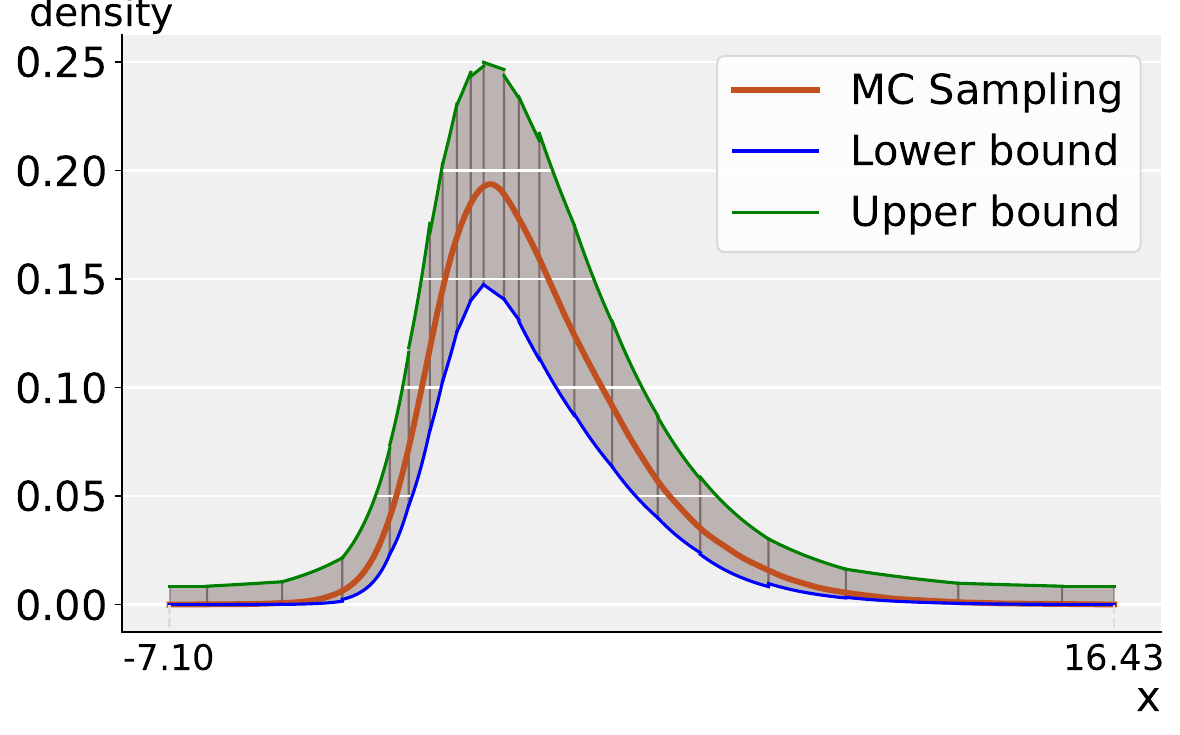}
        \caption{The output abstraction with 20 partitions after merging.}
        \label{fig:2}
    \end{subfigure}

    \caption{Performance of Greedy Interval Merging Strategy.}
    \figlabel{fig:merge}
\end{figure}

\subsection{GPU Parallelization}
In addition to the two adaptive partition strategies above, we further accelerate the affine transformer through GPU-based parallelization. A key observation is that many computations involved in the algorithm above are parallel. For example, the abstraction of conditional expectations over different high-dimensional boxes in the affine transformation, the construction of tangent and secant affine bounds, and the numerical integration of exponential-affine terms can all be performed independently. Therefore, we implement our abstract interpreter using tensor-based computations, batching independent operations into dense tensors for parallel execution on GPUs. This enables efficient vectorized computation by leveraging highly optimized GPU kernels. The speedup becomes increasingly significant as the size of the program grows, especially in the Bayesian neural networks setting, substantially improving the scalability of our tool.

\section{Full Results for Illustrative Evaluation on Small BNNs}
\applabel{fulltoymodel1}

We show the complete the figure results shown in~\secref{sec: eval1} in~\figref{fig:fulltoymodel1} and~\figref{fig:fulltoymodel2}.
\begin{figure}[h]
\centering
\begin{subfigure}[b]{0.48\textwidth}
    \centering
    \includegraphics[width=\linewidth]{fig/eval/toy_bnn_cmp_n1.png}
    \caption{Comparison of abstractions of neuron $x_1$ with different domains.}
    \figlabel{fig:sub1}
\end{subfigure}
\hfill
\begin{subfigure}[b]{0.48\textwidth}
    \centering
    \includegraphics[width=\linewidth]{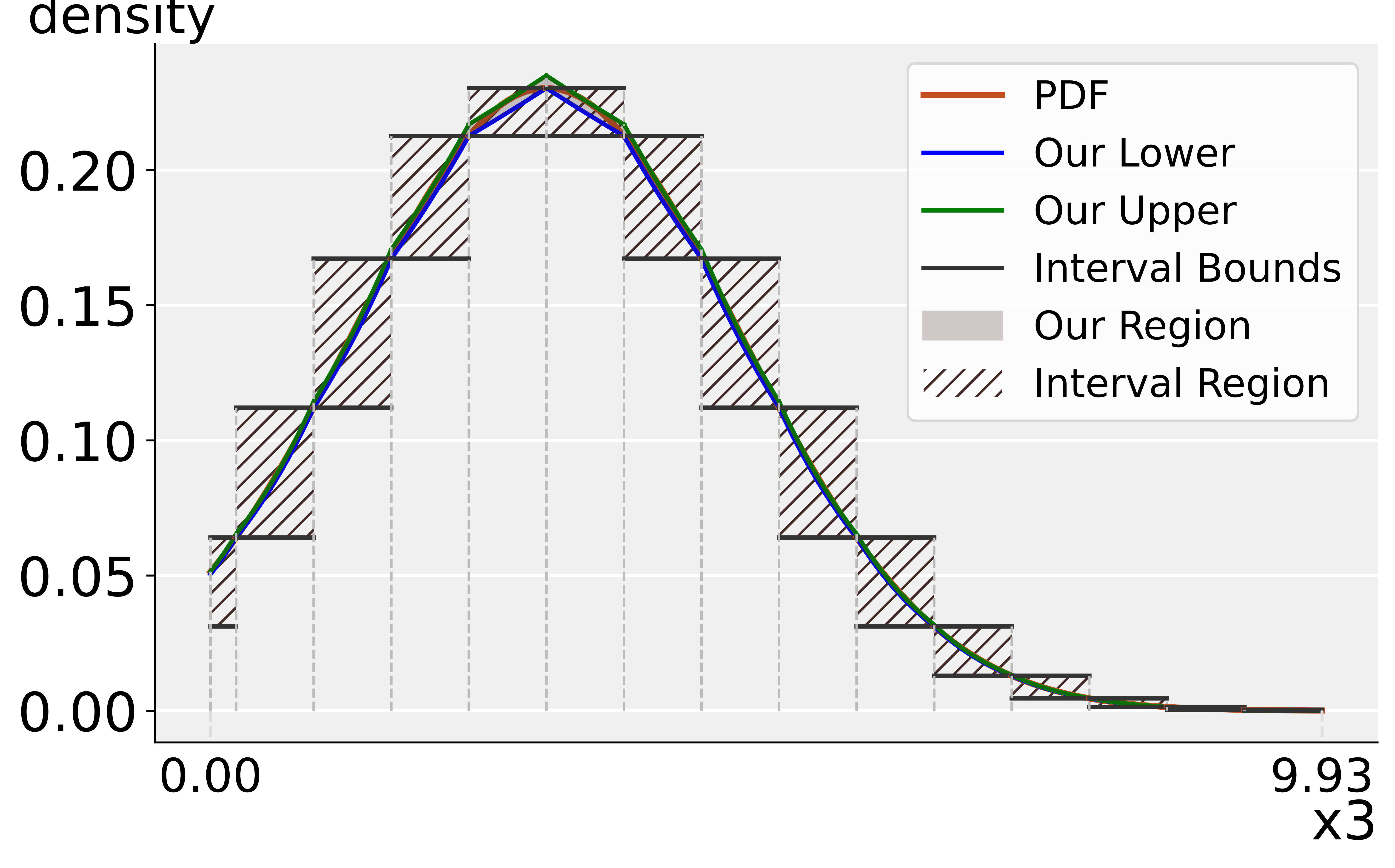}
    \caption{Comparison of abstractions of neuron $x_3$ with different domains.}
    \figlabel{fig:sub2}
\end{subfigure}

\vspace{0.5em}

\begin{subfigure}[b]{0.48\textwidth}
    \centering
    \includegraphics[width=\linewidth]{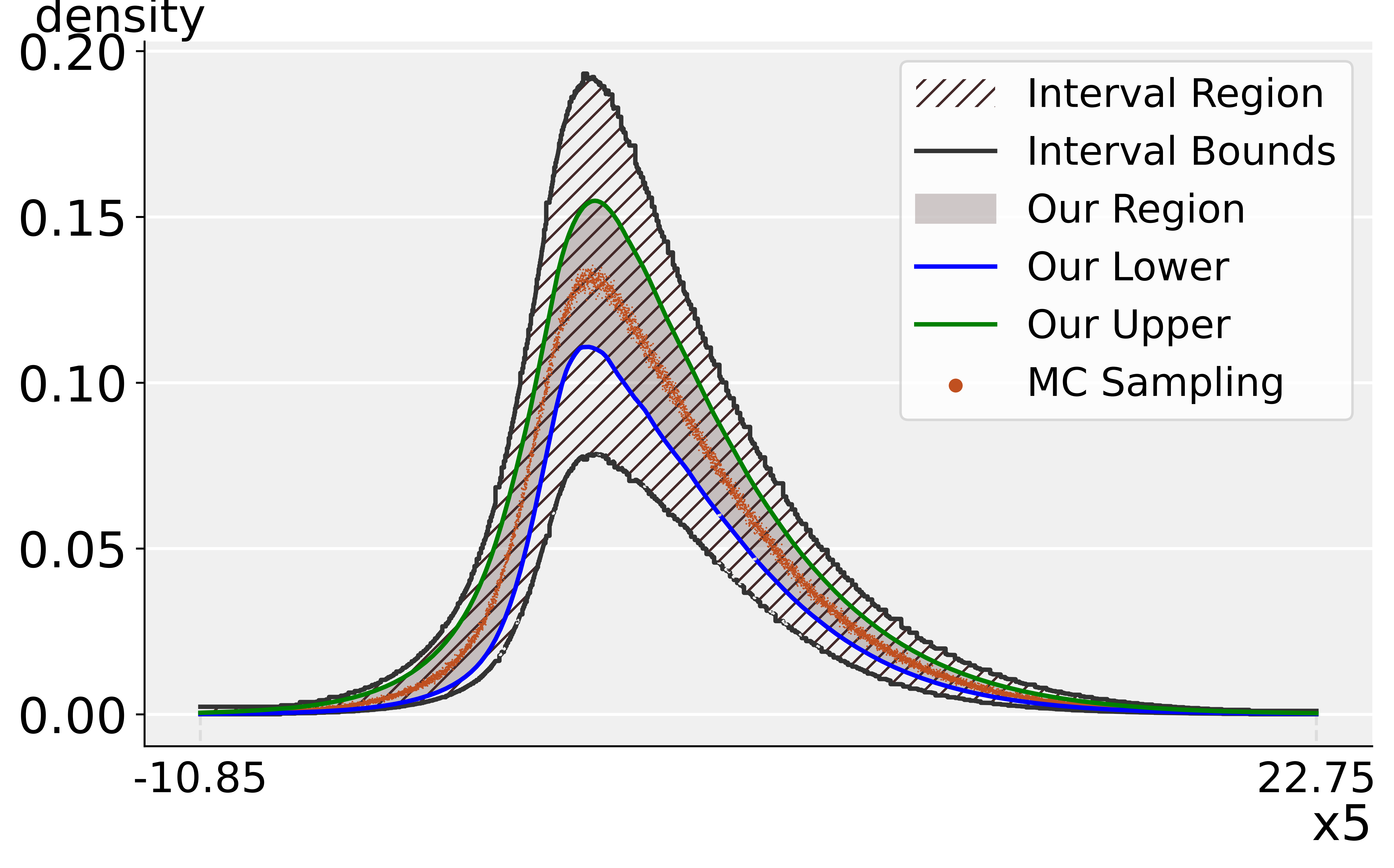}
    \caption{Comparison of abstractions of neuron $x_5$ with different domains.}
    \figlabel{fig:sub3}
\end{subfigure}
\hfill
\begin{subfigure}[b]{0.48\textwidth}
    \centering
    \includegraphics[width=\linewidth]{fig/eval/toy_bnn_cmp_y.png}
    \caption{Comparison of abstractions of neuron $y$ with different domains.}
    \figlabel{fig:sub4}
\end{subfigure}
\vspace{-5pt}
\caption{
Complete comparison between our domain and the interval-based domain on the toy Bayesian neural network~\figref{fig:bnn}  introduced in \secref{overview}.
}
\figlabel{fig:fulltoymodel1}
\end{figure}

\begin{figure}[t]
    \centering
\includegraphics[width=\linewidth]{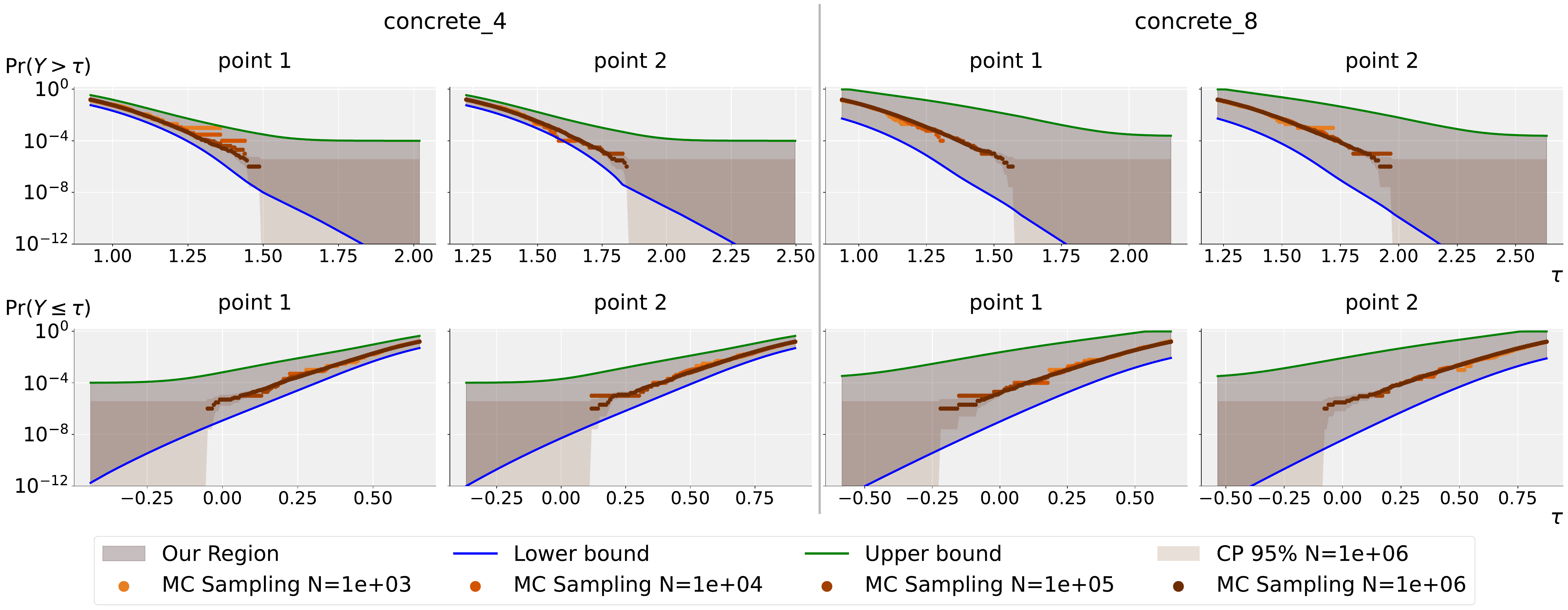}
    \caption{
    Quantifying the probability mass of tail regions for two BNNs (left: 4 hidden units; right: 8 hidden units). Both models are evaluated on the same two test inputs (point 1 and point 2). The top row shows the right-tail probability $\Pr(Y>\tau)$, while the bottom row shows the left-tail probability $\Pr(Y\le\tau)$.
  }
    \figlabel{fig:fulltoymodel2}
\end{figure}

\section{Benchmarks for Bayesian Inference}
\applabel{benchmarksforbayes}

We detail the explanation of benchmarks for Bayesian Inference used in~\secref{sec: eval2} in~\tabref{tab: fullbenchmark}.
\begin{table}[h]
\centering
\small
\setlength{\tabcolsep}{8pt}
\caption{
Benchmark details.
Symbols used:
$\mathcal{N}$: Normal;
$\mathcal{N}_t$: truncated Normal;
$\mathcal N_{\mathrm{lin}}$:
Normal with a linear combination of latent variables in the mean;
$\mathcal N_{\mathrm{prod}}$:
Normal whose mean contains products of latent variables;
$\mathcal{S}$: Logistic;
$\mathcal{G}$: Gamma;
$\mathcal{P}$: Pareto;
$\mathcal{T}$: Student-t.
Operators:
$+$ denotes a mixture distribution;
$^\alpha$ indicates the number of loops.
}
\tablabel{tab:benchmarks}
\begin{tabular}{@{}llll@{}}
\toprule
\textbf{Program} &
\textbf{Prior} &
\textbf{Likelihood} &
\textbf{Description} \\
\midrule

Basic
&
$\mathcal{N}$
&
$\mathcal{N}$
&
Normal prior and Normal likelihood
\\

WealthEstimation
&
$\mathcal{P}$
&
$\mathcal{N}$
&
Estimating latent wealth under measurement noise~\cite{pareto}
\\

RobustBayesian
&
$\mathcal{G}$
&
$\mathcal{T}$
&
Robust estimation under heavy-tailed observations~\cite{studentt}
\\

LearningGaussian
&
$\mathcal{N}$
&
$\mathcal{N}^{20}$
&
Learning the mean of a Gaussian distribution~\cite{InferNET,psi}
\\

ClinicalTrial
&
$\mathcal{N}+\mathcal{N}$
&
$\mathcal{N}^{10}$
&
Treatment effect estimation with a mixture prior~\cite{clinicaltrial}
\\

HumanHeight
&
$\mathcal{N}_{t}+\mathcal{N}_{t}$
&
$\mathcal{N}^{3}$
&
Learning height with a mixture prior~\cite{AURA}
\\

RegLogistic
&
$\mathcal{N}_{t}$
&
$\mathcal{S}^{30}$
&
Linear regression with logistic likelihood~\cite{AURA}
\\

Privacy
&
$\mathcal{N}_{\mathrm{lin}}$
&
$\mathcal{N}$
&
Inferring private attributes from aggregate statistics~\cite{AURA}
\\

ProbabilisticPCA
&
$\mathcal{N}_{\mathrm{prod}}$
&
$\mathcal{N}$
&
Learning latent factors from noisy observations~\cite{PPCA}
\\

\bottomrule
\end{tabular}
\tablabel{tab: fullbenchmark}
\end{table}

\section{Comparison to AURA}
\applabel{aura}

As discussed in~\secref{sec: eval2}, AURA is not directly applicable to our setting because it does not support compositional abstract transformers over abstract inputs, as it operates directly on the final symbolic expression with concrete inputs. It also defines soundness with respect to a semantics that discards tail regions. Consequently, AURA avoids approximation over intermediate abstract inputs and tail regions, both of which generally lead to more precise results. Nevertheless, we still compare against AURA under a setting that favors AURA while keeping the remaining experimental configuration identical. Despite these advantages, the results demonstrate that our method still substantially outperforms AURA in both precision and runtime. 

\paragraph{Benchmark.} Although AURA can handle loops, it does so by fully unrolling the computation and deriving a symbolic posterior before performing approximation. In contrast, our goal is to develop general abstract transformers that can be applied iteratively over abstract states. Therefore, to enable a meaningful comparison, we simplify the loop benchmarks in~\tabref{tab:benchmarks} by removing the loops so that each benchmark consists of a single Bayesian affine update, as shown in~\tabref{tab:benchmarks_aura}.

\paragraph{Metrics.} Same as metrics used in~\secref{sec: eval2}.

\paragraph{Experimental Setup.}Both AURA and our approach are evaluated on the same hardware using the same implementation configuration and the same partitioning of the support. In particular, although our implementation supports tensor-core acceleration, this optimization is disabled because it is unavailable in AURA. Besides, as AURA discards unbounded tail regions, we evaluate all benchmarks under the bounded-support setting. For completeness, we additionally report the results of our full framework, which preserves the original support and soundly handles unbounded tails.

{
\small
\setlength{\tabcolsep}{8pt}
\centering
\begin{longtable}{@{}llll@{}}
\caption{
Benchmark details.
Symbols used:
$\mathcal{N}$: Normal;
$\mathcal{N}_t$: truncated Normal;
$\mathcal N_{\mathrm{lin}}$:
Normal with a linear combination of latent variables in the mean.
$\mathcal N_{\mathrm{prod}}$:
Normal whose mean contains products of latent variables.
$\mathcal{L}$: Laplace;
$\mathcal{U}$: Uniform;
$\mathcal{S}$: Logistic.
$\mathcal{G}$: Gamma.
$\mathcal{P}$: Pareto.
$\mathcal{T}$: Student-t.
Operators:
$+$ denotes a mixture distribution.
$^\alpha$ indicates the number of latent variables or observations.
}
\tablabel{tab:benchmarks_aura}\\[4pt]
\toprule
\textbf{Program} &
\textbf{Prior} &
\textbf{Likelihood} &
\textbf{Description}
\\
\midrule
\endfirsthead

Basic
&
$\mathcal{N}$
&
$\mathcal{N}$
&
Normal prior and Normal likelihood
\\
WealthEstimation
&
$\mathcal{P}$
&
$\mathcal{N}$
&
Estimating latent wealth under measurement noise~\cite{pareto}
\\
RobustBayesian
&
$\mathcal{G}$
&
$\mathcal{T}$
&
Robust estimation under heavy-tailed observations~\cite{studentt}
\\
LearningGaussian
&
$\mathcal{N}$
&
$\mathcal{N}$
&
Learning the mean of a Gaussian distribution~\cite{InferNET,psi}
\\

ClinicalTrial
&
$\mathcal{N}+\mathcal{N}$
&
$\mathcal{N}$
&
Treatment effect estimation with a mixture prior~\cite{clinicaltrial}
\\

HumanHeight
&
$\mathcal{N}_{t}+\mathcal{N}_{t}$
&
$\mathcal{N}$
&
Learning height with a mixture prior~\cite{AURA}
\\

RegLogistic
&
$\mathcal{N}_{t}$
&
$\mathcal{S}$
&
Linear regression with logistic likelihood~\cite{AURA}
\\

Privacy
&
$\mathcal{N}_{\mathrm{lin}}$
&
$\mathcal{N}$
&
Inferring private attributes from aggregate statistics~\cite{AURA}
\\
ProbabilisticPCA
&
$\mathcal{N}_{\mathrm{prod}}$
&
$\mathcal{N}$
&
Learning latent factors from noisy observations~\cite{PPCA}
\\

\bottomrule
\end{longtable}
}

\paragraph{Observation.}
As shown in~\tabref{tab:prec_aura}, our approach consistently outperforms AURA in both precision and runtime across all benchmarks where AURA successfully completes the analysis. Moreover, even the full version of our approach, which soundly preserves the original support and explicitly reasons about unbounded tail regions, achieves better performance than AURA on all comparable benchmarks. Furthermore, AURA fails on \textit{WealthEstimation}, \textit{RobustBayesian}, and \textit{ProbabilisticPCA}, whereas our framework successfully analyzes all benchmarks. The result demonstrates that our domain provides substantially better precision, scalability, and generality.

{
\small
\setlength{\tabcolsep}{8pt}
\centering
\begin{longtable}{@{}lrrrrrr@{}}
\caption{Precision(TVD) and computation time comparison between our domain and AURA.}
\tablabel{tab:prec_aura}\\[4pt]

\toprule

\multirow{2}{*}{\textbf{Benchmark}} &
\multicolumn{3}{c}{\textbf{Precision}} &
\multicolumn{3}{c}{\textbf{Time (s)}} \\

\cmidrule(lr){2-4}
\cmidrule(lr){5-7}

& \textbf{AURA} & \textbf{Ours$^\star$} & \textbf{Ours}
& \textbf{AURA} & \textbf{Ours$^\star$} & \textbf{Ours}
\\

\midrule

Basic
& 0.029969 & 0.000176 & 0.001410
& 0.74 & 0.11 & 1.07 \\

WealthEstimation
& \xmark & 0.001271 & 0.006025
& \xmark & 0.10 & 0.23 \\

RobustBayesian
& \xmark & 0.000901 & 0.000932
& \xmark & 0.17 & 0.23 \\

LearningGaussian
& 0.040994 & 0.000327 & 0.000329
& 0.81 & 0.11 & 0.14 \\

ClinicalTrial
& 0.038450 & 0.000364 & 0.031287
& 0.77 & 0.16 & 0.19 \\

HumanHeight
& 0.041458 & 0.000327 & 0.000327
& 0.91 & 0.17 & 0.21 \\

RegLogistic
& 0.032574 & 0.000203 & 0.000298
& 0.75 & 0.15 & 0.22 \\

Privacy
& 0.031531 & 0.000194 & 0.000282
& 0.67 & 0.13 & 0.17 \\

ProbabilisticPCA
& \xmark & 0.002904 & 0.002919
& \xmark & 15.99 & 13.00 \\

\bottomrule
\end{longtable}
}